[ english version / português abaixo ]

# Apocalypse, survivalism, occultism and esotericism communities on Brazilian Telegram:

## when faith is used to sell quantum courses and open doors to harmful conspiracy theories

*Ergon Cugler de Moraes Silva*

Brazilian Institute of Information in Science and Technology (IBICT)
Brasília, Federal District, Brazil

contato@ergoncugler.com
www.ergoncugler.com

### Abstract

Brazilian communities on Telegram have increasingly turned to apocalyptic and survivalist theories, especially in times of crisis such as the COVID-19 pandemic, where narratives of occultism and esotericism find fertile ground. Therefore, this study aims to address the research question: **how are Brazilian conspiracy theory communities on apocalypse, survivalism, occultism and esotericism topics characterized and articulated on Telegram?** It is worth noting that this study is part of a series of seven studies whose main objective is to understand and characterize Brazilian conspiracy theory communities on Telegram. This series of seven studies is openly and originally available on arXiv at Cornell University, applying a mirrored method across the seven studies, changing only the thematic object of analysis and providing investigation replicability, including with proprietary and authored codes, adding to the culture of free and open-source software. Regarding the main findings of this study, the following were observed: Occult and esoteric communities function as gateways to apocalypse theories; Conspiracies about the New World Order are amplified by apocalyptic discussions; Survivalist narratives grew significantly during the Pandemic; Occultism and esotericism are sources of invitations to off-label drug communities, reinforcing scientific disinformation; Discussions about the apocalypse serve as a start for other conspiracy theories, expanding their reach.

### Key findings

- ➔ Occultism and esotericism communities serve as gateways to extreme survivalism and apocalypse theories, with 2,527,833 interactions connecting these themes to other conspiracies, highlighting how the search for mystical answers fuels catastrophic concerns and reinforces narratives like the New World Order;
- ➔ New World Order conspiracies are amplified by apocalyptic discussions, with 3,488,686 interactions reinforcing the idea of global control, especially during crises like the COVID-19 pandemic. During such times, these narratives connect to disinformation about vaccines and public health, totaling 1,345,060 interactions;
- ➔ Survivalism narratives grew by 1,500% during the pandemic, reflecting how crises drive apocalyptic theories, with mentions rising from 1,000 to 35,000 between 2020 and 2021. These groups have become catalysts for the acceptance of an impending collapse;

- ➔ Occultism and esotericism are the largest sources of invitations to communities promoting off-label drugs and illicit chemicals, leading to vaccine denialism, with 7,367 links. This shows a dangerous intersection between mysticism and scientific disinformation, which reinforces the spread of alternative practices and rejection of conventional science. Additionally, monetization through the sale of courses on quantum magnetism is evident;
- ➔ Apocalypse and survivalism communities operate as central hubs in the disinformation network, with 4,371,125 interactions, connecting and amplifying different conspiracy theories, creating a continuous cycle of mutual reinforcement that strengthens these narratives;
- ➔ The overlap of esoteric and survivalist themes creates a cohesive ideological bubble, resistant to scientific information, with 3,026,065 interactions linked to occultism and esotericism. This perpetuates distrust in science and promotes an alternative worldview;
- ➔ The false narrative associating vaccines with global control persists as one of the most discussed topics in apocalyptic and esoteric communities, reinforcing distrust in medicine even after debunking, particularly in groups focusing on the NWO and survivalism theories;
- ➔ The interconnectivity between occultism, NWO, and apocalypse communities reveals a pattern of mutual reinforcement that amplifies disinformation, with 5,421 links between them, creating an environment where different forms of disinformation become interdependent;
- ➔ Discussions about the apocalypse often serve as a starting point for accepting other conspiracy theories, with survivalism communities acting as gateways to theories about the New World Order, globalism, and other conspiracies, broadening their reach;
- ➔ The COVID-19 pandemic catalyzed the growth of esoteric and survival communities, with peaks of 1,200% in interactions related to occultism and esotericism—underpinning vaccine disinformation and promoting quackery—reflecting the response to global uncertainties and encouraging the search for alternative explanations linked to conspiracy theories.

## 1. Introduction

After analyzing thousands of Brazilian conspiracy theory communities on Telegram and extracting tens of millions of content pieces from these communities, created and/or shared by millions of users, this study aims to compose a series of seven studies that address the phenomenon of conspiracy theories on Telegram, focusing on Brazil as a case study. Through the identification approaches implemented, it was possible to reach a total of 72 Brazilian conspiracy theory communities on Telegram on apocalypse, survivalism, occultism and esotericism topics, summing up 4,371,125 content pieces published between May 2016 (initial publications) and August 2024 (date of this study), with 192,138 users aggregated from within these communities. Thus, this study aims to understand and characterize the communities focused on apocalypse, survivalism, occultism and esotericism present in this Brazilian network of conspiracy theories identified on Telegram.

To this end, a mirrored method will be applied across all seven studies, changing only the thematic object of analysis and providing investigation replicability. In this way, we will adopt technical approaches to observe the connections, temporal series, content, and overlaps of themes within the conspiracy theory communities. In addition to this study, the other six are openly and originally available on arXiv at Cornell University. This series paid particular attention to ensuring data integrity and respecting user privacy, as provided by Brazilian legislation (Law No. 13,709/2018 / Brazilian law from 2018).

Therefore, the question arises: **how are Brazilian conspiracy theory communities on apocalypse, survivalism, occultism and esotericism topics characterized and articulated on Telegram?**

## 2. Materials and methods

The methodology of this study is organized into three subsections: **2.1. Data extraction**, which describes the process and tools used to collect information from Telegram communities; **2.2. Data processing**, which discusses the criteria and methods applied to classify and anonymize the collected data; and **2.3. Approaches to data analysis**, which details the techniques used to investigate the connections, temporal series, content, and thematic overlaps within conspiracy theory communities.

### 2.1. Data extraction

This project began in February 2023 with the publication of the first version of TelegramScrap (Silva, 2023), a proprietary, free, and open-source tool that utilizes Telegram's Application Programming Interface (API) by Telethon library and organizes data extraction cycles from groups and open channels on Telegram. Over the months, the database was expanded and refined using four approaches to identifying conspiracy theory communities:

**(i) Use of keywords:** at the project's outset, keywords were listed for direct identification in the search engine of Brazilian groups and channels on Telegram, such as "apocalypse", "survivalism", "climate change", "flat earth", "conspiracy theory", "globalism", "new world order", "occultism", "esotericism", "alternative cures", "qAnon" "reptilians", "revisionism", "aliens", among others. This initial approach provided some communities whose titles and/or descriptions of groups and channels explicitly contained terms related to conspiracy theories. However, over time, it was possible to identify many other communities that the listed keywords did not encompass, some of which deliberately used altered characters to make it difficult for those searching for them on the network.

**(ii) Telegram channel recommendation mechanism:** over time, it was identified that Telegram channels (except groups) have a recommendation tab called "similar channels", where Telegram itself suggests ten channels that have some similarity with the channel being observed. Through this recommendation mechanism, it was possible to find more Brazilian conspiracy theory communities, with these being recommended by the platform itself.

**(iii) Snowball approach for invitation identification:** after some initial communities were accumulated for extraction, a proprietary algorithm was developed to identify URLs containing "t.me/", the prefix for any invitation to Telegram groups and channels. Accumulating a frequency of hundreds of thousands of links that met this criterion, the unique addresses were listed, and their repetitions counted. In this way, it was possible to investigate new Brazilian groups and channels mentioned in the messages of those already investigated, expanding the network. This process was repeated periodically to identify new communities aligned with conspiracy theory themes on Telegram.

**(iv) Expansion to tweets published on X mentioning invitations:** to further diversify the sources of Brazilian conspiracy theory communities on Telegram, a proprietary search query was developed to identify conspiracy theory-themed keywords using tweets published on X (formerly Twitter) that, in addition to containing one of the keywords, also included "t.me/", the prefix for any invitation to Telegram groups and channels, "https://x.com/search?q=lang%3Apt%20%22t.me%2F%22%20SEARCH-TERM".

With the implementation of community identification approaches for conspiracy theories developed over months of investigation and method refinement, it was possible to build a project database encompassing a total of 855 Brazilian conspiracy theory communities on Telegram (including other themes not covered in this study). These communities have collectively published 27,227,525 pieces of content from May 2016 (the first publications) to August 2024 (the period of this study), with a combined total of 2,290,621 users across the Brazilian communities. It is important to note that this volume of users includes two elements: first, it is a variable figure, as users can join and leave communities daily, so this value represents what was recorded on the publication extraction date; second, it is possible that the same user is a member of more than one group and, therefore, is counted more than once. In this context, while the volume remains significant, it may be slightly lower when considering the deduplicated number of citizens within these Brazilian conspiracy theory communities.

### 2.2. Data processing

With all the Brazilian conspiracy theory groups and channels on Telegram extracted, a manual classification was conducted considering the title and description of the community. If there was an explicit mention in the title or description of the community related to a specific theme, it was classified into one of the following categories: (i) "Anti-Science"; (ii) "Anti-Woke and Gender"; (iii) "Antivax"; (iv) "Apocalypse and Survivalism"; (v) "Climate Changes"; (vi) "Flat Earth"; (vii) "Globalism"; (viii) "New World Order"; (ix) "Occultism and Esotericism"; (x) "Off Label and Quackery"; (xi) "QAnon"; (xii) "Reptilians and Creatures"; (xiii) "Revisionism and Hate Speech"; (xiv) "UFO and Universe". If there was no explicit mention related to the themes in the title or description of the community, it was classified as (xv) "General Conspiracy". In the following table, we can observe the metrics related to the classification of these conspiracy theory communities in Brazil.

**Table 01.** Conspiracy theory communities in Brazil (metrics up to August 2024)

| Categories | Groups | Users | Contents | Comments | Total |
|---|---|---|---|---|---|
| **Anti-Science** | 22 | 58,138 | 187,585 | 784,331 | 971,916 |
| **Anti-Woke and Gender** | 43 | 154,391 | 276,018 | 1,017,412 | 1,293,430 |
| **Antivax** | 111 | 239,309 | 1,778,587 | 1,965,381 | 3,743,968 |
| **Apocalypse and Survivalism** | 33 | 109,266 | 915,584 | 429,476 | 1,345,060 |
| **Climate Changes** | 14 | 20,114 | 269,203 | 46,819 | 316,022 |
| **Flat Earth** | 33 | 38,563 | 354,200 | 1,025,039 | 1,379,239 |
| **General Conspiracy** | 127 | 498,190 | 2,671,440 | 3,498,492 | 6,169,932 |
| **Globalism** | 41 | 326,596 | 768,176 | 537,087 | 1,305,263 |
| **NWO** | 148 | 329,304 | 2,411,003 | 1,077,683 | 3,488,686 |
| **Occultism and Esotericism** | 39 | 82,872 | 927,708 | 2,098,357 | 3,026,065 |
| **Off Label and Quackery** | 84 | 201,342 | 929,156 | 733,638 | 1,662,794 |
| **QAnon** | 28 | 62,346 | 531,678 | 219,742 | 751,420 |
| **Reptilians and Creatures** | 19 | 82,290 | 96,262 | 62,342 | 158,604 |
| **Revisionism and Hate Speech** | 66 | 34,380 | 204,453 | 142,266 | 346,719 |
| **UFO and Universe** | 47 | 58,912 | 862,358 | 406,049 | 1,268,407 |
| **Total** | **855** | **2,296,013** | **13,183,411** | **14,044,114** | **27,227,525** |

Source: Own elaboration (2024).

With this volume of extracted data, it was possible to segment and present in this study only communities and content related to apocalypse, survivalism, occultism and esotericism themes. In parallel, other themes of Brazilian conspiracy theory communities were also addressed with studies aimed at characterizing the extent and dynamics of the network, which are openly and originally available on arXiv at Cornell University.

Additionally, it should be noted that only open communities were extracted, meaning those that are not only publicly identifiable but also do not require any request to access the content, being available to any user with a Telegram account who needs to join the group or channel. Furthermore, in compliance with Brazilian legislation, particularly the General Data Protection Law (LGPD), or Law No. 13,709/2018 (Brazilian law from 2018), which deals with privacy control and the use/treatment of personal data, all extracted data were anonymized for the purposes of analysis and investigation. Therefore, not even the identification of the communities is possible through this study, thus extending the user's privacy by anonymizing their data beyond the community itself to which they submitted by being in a public and open group or channel on Telegram.

### 2.3. Approaches to data analysis

A total of 72 selected communities focused on apocalypse, survivalism, occultism and esotericism themes, containing 4,371,125 publications and 192,138 combined users, will be analyzed. Four approaches will be used to investigate the conspiracy theory communities selected for the scope of this study. These metrics are detailed in the following table:

**Table 02.** Selected communities for analysis (metrics up to August 2024)

| Categories | Groups | Users | Contents | Comments | Total |
|---|---|---|---|---|---|
| **Apocalypse and Survivalism** | 33 | 109,266 | 915,584 | 429,476 | 1,345,060 |
| **Occultism and Esotericism** | 39 | 82,872 | 927,708 | 2,098,357 | 3,026,065 |
| **Total** | **72** | **192,138** | **1,843,292** | **2,527,833** | **4,371,125** |

Source: Own elaboration (2024).

**(i) Network:** by developing a proprietary algorithm to identify messages containing the term "t.me/" (inviting users to join other communities), we propose to present volumes and connections observed on how **(a)** communities within the apocalypse, survivalism, occultism and esotericism theme circulate invitations for their users to explore more groups and channels within the same theme, reinforcing shared belief systems; and how **(b)** these same communities circulate invitations for their users to explore groups and channels dealing with other conspiracy theories, distinct from their explicit purpose. This approach is valuable for observing whether these communities use themselves as a source of legitimation and reference and/or rely on other conspiracy theory themes, even opening doors for their users to explore other conspiracies. Furthermore, it is worth mentioning the study by Rocha *et al.* (2024), where a network identification approach was also applied in Telegram communities, but by observing similar content based on an ID generated for each unique message and its similar ones;

**(ii) Time series:** the "Pandas" library (McKinney, 2010) is used to organize the investigation data frames, observing **(a)** the volume of publications over the months; and **(b)** the volume of engagement over the months, considering metadata of views, reactions, and comments collected during extraction. In addition to volumetry, the "Plotly" library (Plotly Technologies Inc., 2015) enabled the graphical representation of this variation;

**(iii) Content analysis:** in addition to the general word frequency analysis, time series are applied to the variation of the most frequent words over the semesters—observing from May 2016 (initial publications) to August 2024 (when this study was conducted). With the "Pandas" (McKinney, 2010) and "WordCloud" (Mueller, 2020) libraries, the results are presented both volumetrically and graphically;

**(iv) Thematic agenda overlap:** following the approach proposed by Silva & Sátiro (2024) for identifying thematic agenda overlap in Telegram communities, we used the "BERTopic" model (Grootendorst, 2020). BERTopic is a topic modeling algorithm that facilitates the generation of thematic representations from large amounts of text. First, the algorithm extracts document embeddings using sentence transformer models, such as “all-MiniLM-L6-v2”. These embeddings are then reduced in dimensionality using techniques like “UMAP”, facilitating the clustering process. Clustering is performed using “HDBSCAN”, a density-based technique that identifies clusters of different shapes and sizes, as well as outliers. Subsequently, the documents are tokenized and represented in a bag-of-words structure, which is normalized (L1) to account for size differences between clusters. The topic representation is refined using a modified version of “TF-IDF”, called “Class-TF-IDF”, which considers the importance of words within each cluster (Grootendorst, 2020). It is important to note that before applying BERTopic, we cleaned the dataset by removing Portuguese “stopwords” using “NLTK” (Loper & Bird, 2002). For topic modeling, we used the “loky” backend to optimize performance during data fitting and transformation.

In summary, the methodology applied ranged from data extraction using the own tool TelegramScrap (Silva, 2023) to the processing and analysis of the collected data, employing various approaches to identify and classify Brazilian conspiracy theory communities on Telegram. Each stage was carefully implemented to ensure data integrity and respect for user privacy, as mandated by Brazilian legislation. The results of this data will be presented below, aiming to reveal the dynamics and characteristics of the studied communities.

## 3. Results

The results are detailed below in the order outlined in the methodology, beginning with the characterization of the network and its sources of legitimation and reference, progressing to the time series, incorporating content analysis, and concluding with the identification of thematic agenda overlap among the conspiracy theory communities.

### 3.1. Network

The analysis of networks involving apocalypse, survivalism, occultism, and esotericism reveals an intricate and interconnected structure of communities that promote and perpetuate narratives of fear, distrust, and preparation for imminent catastrophes. In the first figure, the internal network among these themes shows how survivalism and apocalypse communities intertwine with those focused on occultism and esotericism. The presence of two large hubs in the network indicates points of centralization where these narratives converge and mutually reinforce each other. The interdependence of these communities suggests that the fear of cataclysmic events and the search for esoteric answers go hand in hand, creating an environment in which uncertainty is amplified and adherence to these beliefs is intensified.

The second figure focuses on communities that serve as gateways to apocalyptic and survivalist narratives. The strong connection of these communities with themes such as occultism and general conspiracies indicates that survivalism not only attracts new members interested in end-of-the-world scenarios but also introduces them to a broader network of conspiracy theories. These communities act as central hubs, capturing the attention of individuals who, from an initial concern with survival, end up being exposed to a wide range of conspiratorial and esoteric ideas.

In the third figure, the network of communities that operate as exit points demonstrates how discussions about the apocalypse and survivalism often lead individuals to explore other areas of disinformation, such as the New World Order and occultism. The centrality of occultism communities in the network suggests that these discussions about extreme survivalism often serve as a starting point for the acceptance of narratives about secret orders and global conspiracies. This reflects how concerns about cataclysmic events can be manipulated to introduce and perpetuate other conspiracy theories, expanding the breadth and complexity of followers' beliefs.

The flowchart of invitation links between apocalypse and survival communities highlights the strong connection of these narratives with the NWO and anti-vaccine movements, suggesting that these themes not only attract but also amplify the spread of disinformation related to global conspiracies and public health. Apocalypse functions as catalysts for the acceptance of a worldview where social collapse is inevitable, fueling a cycle of fear and distrust that predisposes individuals to embrace other forms of disinformation.

Finally, the flowchart of links between occultism and esotericism communities reveals how these themes are deeply interconnected with other conspiracy narratives, such as general conspiracies and UFO and universe topics. Occultism, often associated with a rejection of modern science, also sends significant invitations to off-label drug use, indicating a tendency to combine mysticism with distrust in conventional medicine. This interconnection reflects a pattern of mutual reinforcement between esoteric beliefs and scientific disinformation, creating a vicious cycle of institutional distrust that perpetuates a worldview where alternative knowledge is preferred over established scientific authority.

[ english version / português abaixo ]

**Figure 01.** Internal network between apocalypse, survivalism and esotericism communities

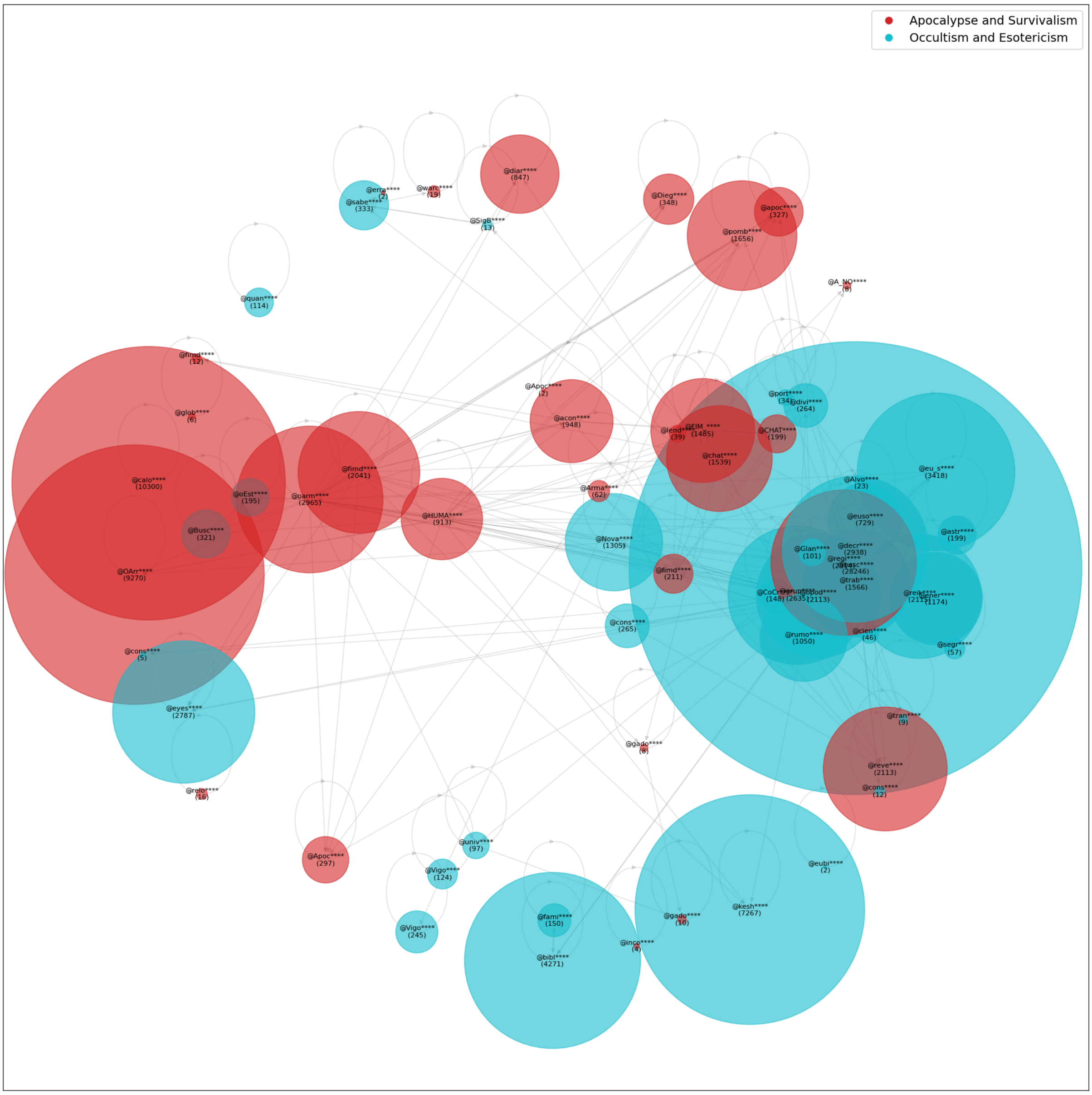

Source: Own elaboration (2024).

The figure displays a network that connects communities on apocalypse, survivalism, occultism, and esotericism. The network, characterized by two large hubs, reflects an interconnection between preparation for catastrophic events and occult beliefs. These themes, though distinct, frequently intersect, creating a cycle where the fear of the apocalypse is fueled by esoteric narratives and vice versa. The large nodes in the graph indicate communities that act as centers of influence, where ideas about survival in a post-apocalyptic world are reinforced by occult practices. The existence of several peripheral communities suggests that these ideas are spreading beyond the central core but remain strongly influenced by the primary disseminators of these beliefs. The interdependence between survivalism and occultism creates an environment where uncertainty and fear are continuously exacerbated, fueling the need to belong to these communities to find solutions and answers.

[ english version / português abaixo ]

**Figure 02.** Network of communities that open doors to the theme (gateway)

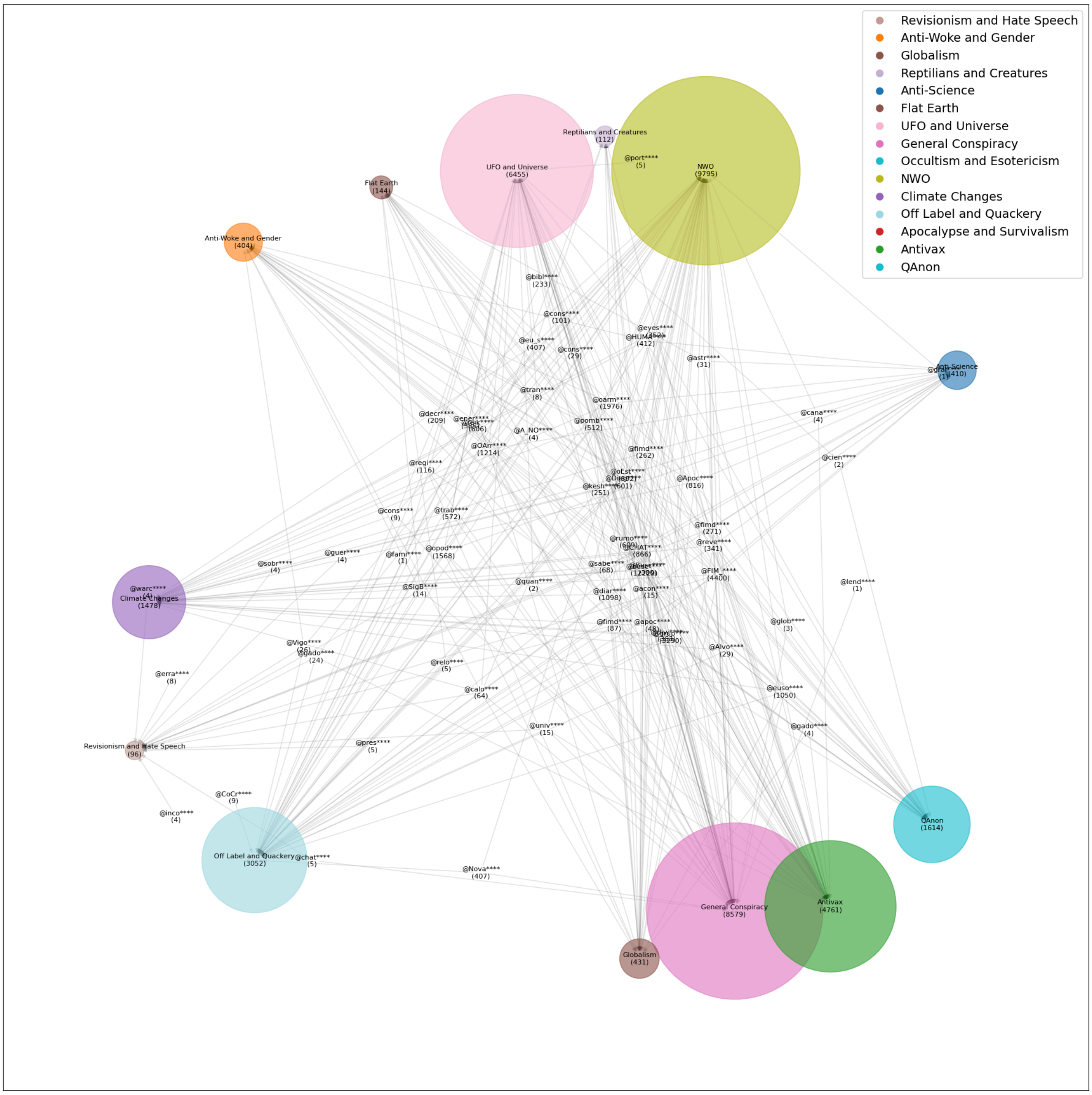

Source: Own elaboration (2024).

This graph highlights the communities that serve as gateways to apocalyptic and survivalist narratives. The large red and blue spheres represent central communities that attract individuals interested in end-of-the-world scenarios and survival practices. The interconnection with other communities, such as occultism and general conspiracy theories, suggests that these ideas feed off each other. Thus, survivalism not only attracts new members with an interest in the apocalypse but can also serve as an entry point for other theories, such as the belief in hidden powers and global conspiracy. The structure of this network reflects how narratives of fear and extreme preparedness can expand individuals' conspiracy horizons, connecting them to a broader set of theories.

[ english version / português abaixo ]

**Figure 03.** Network of communities whose theme opens doors (exit point)

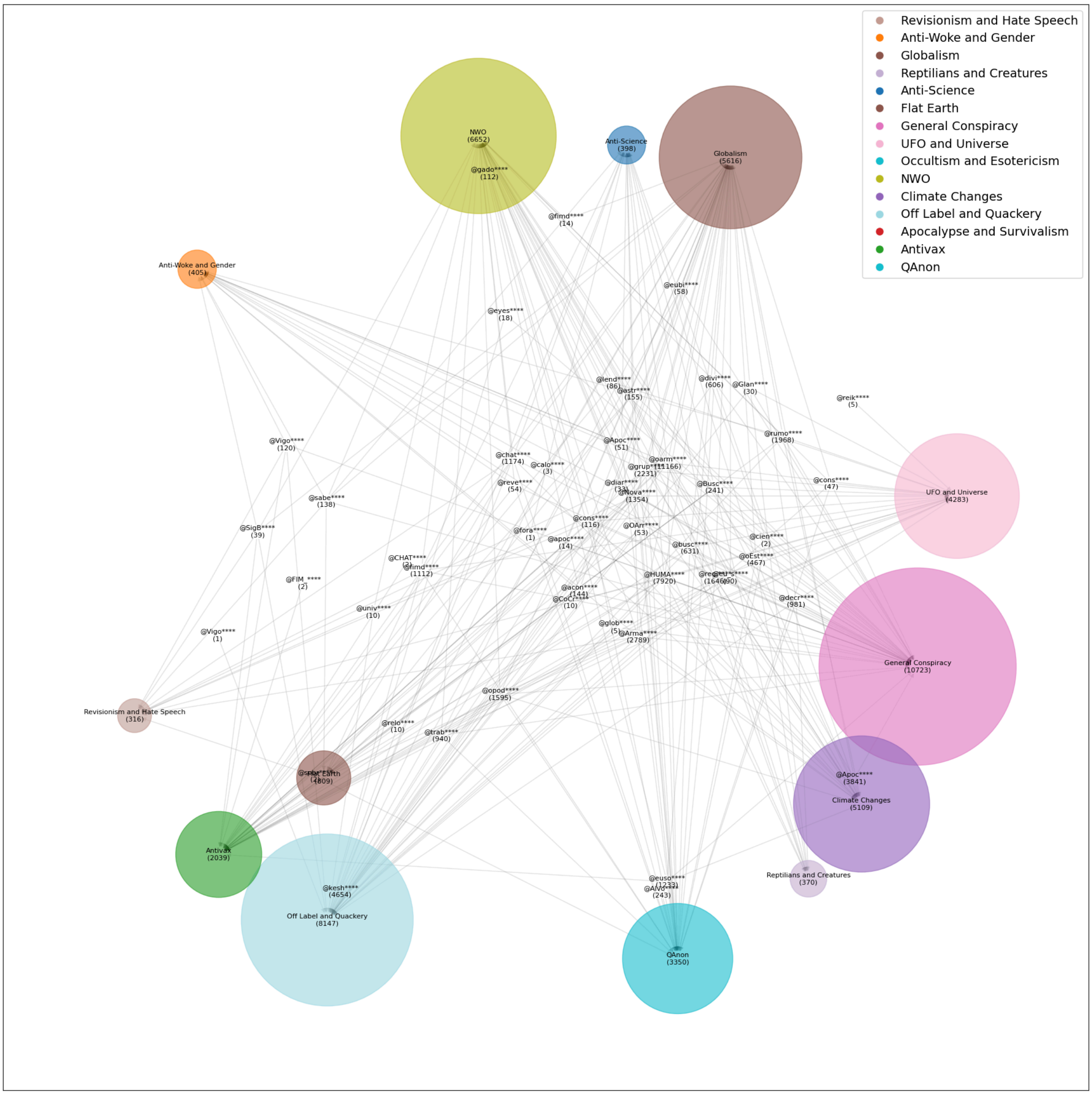

Source: Own elaboration (2024).

his graph explores the connections between communities focused on the apocalypse, survivalism, occultism, and esotericism, revealing how these themes interact within the conspiratorial ecosystem. The large Occultism and New World Order bubbles stand out as central points, suggesting that discussions about survivalism and the apocalypse often lead individuals to broader and interconnected themes, such as global conspiracies. The figure illustrates the fluid transition from concern with cataclysmic events and survival to the acceptance of narratives about secret orders. This reflects how concerns about the future and survival can be manipulated to introduce and perpetuate other conspiracy theories, thus expanding the reach and complexity of conspiracy beliefs among followers.

[ english version / português abaixo ]

**Figure 04.** Flow of invitation links between apocalypse and survivalism communities

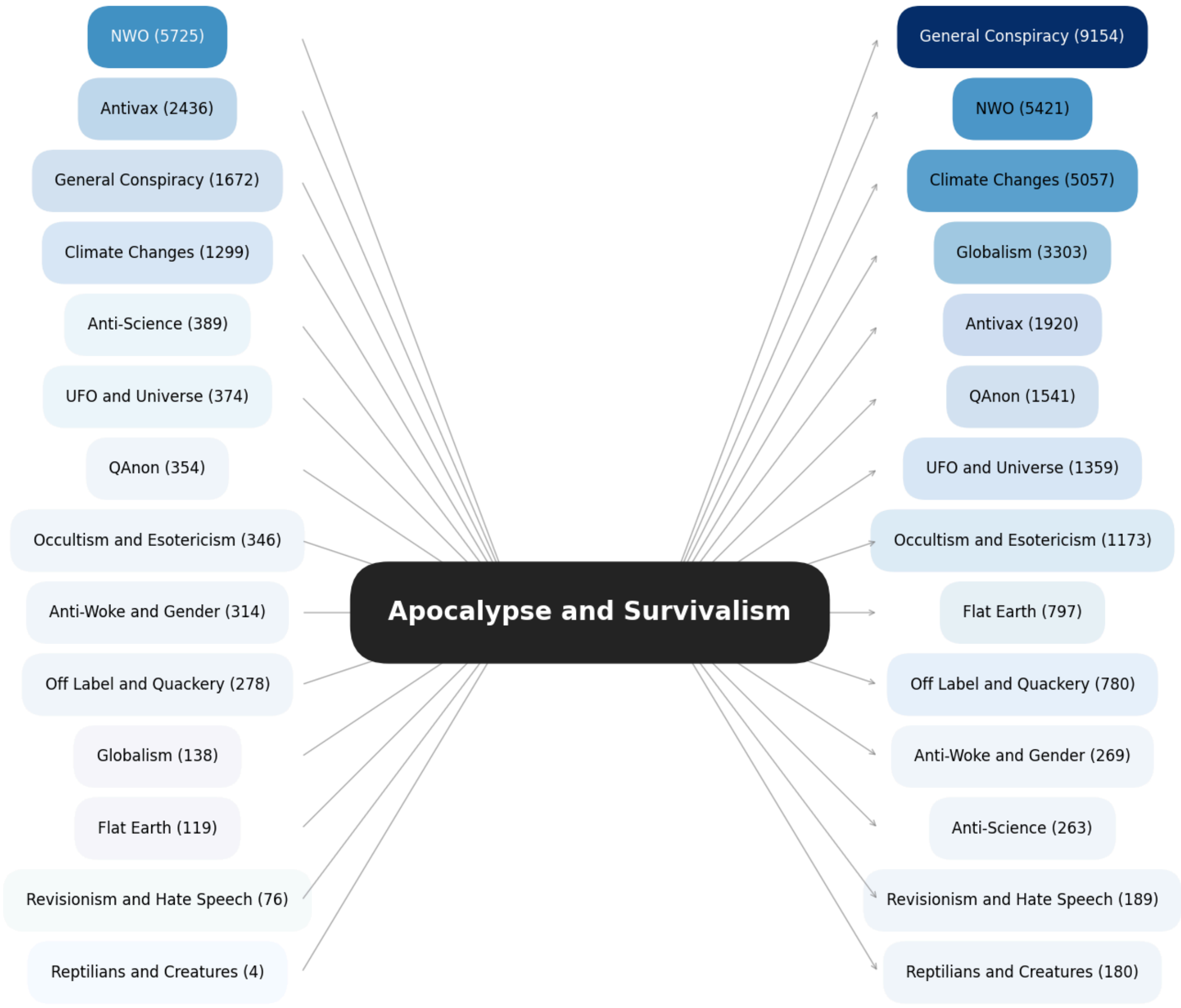

Source: Own elaboration (2024).

The Apocalypse and Survival graph presents a strong connection with the NWO (5,725 links) and anti-vaccine movements (2,436 links), suggesting that these narratives of extreme survivalism and preparation for catastrophes are often associated with a worldview where social collapse is imminent and inevitable. These data indicate that the apocalyptic narrative serves as an effective means of engaging individuals in a constant state of fear and alertness, predisposing them to accept and propagate other forms of disinformation. By analyzing the invitations sent to General Conspiracies (9,154 links) and NWO (5,421 links), it is possible to observe how these communities function as amplifiers of a worldview where global crises are seen not as isolated events but as part of a larger plan of domination and control. Apocalypse and Survivalism, therefore, serve as catalysts for the acceptance of extreme narratives, promoting a “us versus them” mentality that is fundamental to the perpetuation and dissemination of conspiracy theories.

[ english version / português abaixo ]

**Figure 05.** Flow of invitation links between occultism and esotericism communities

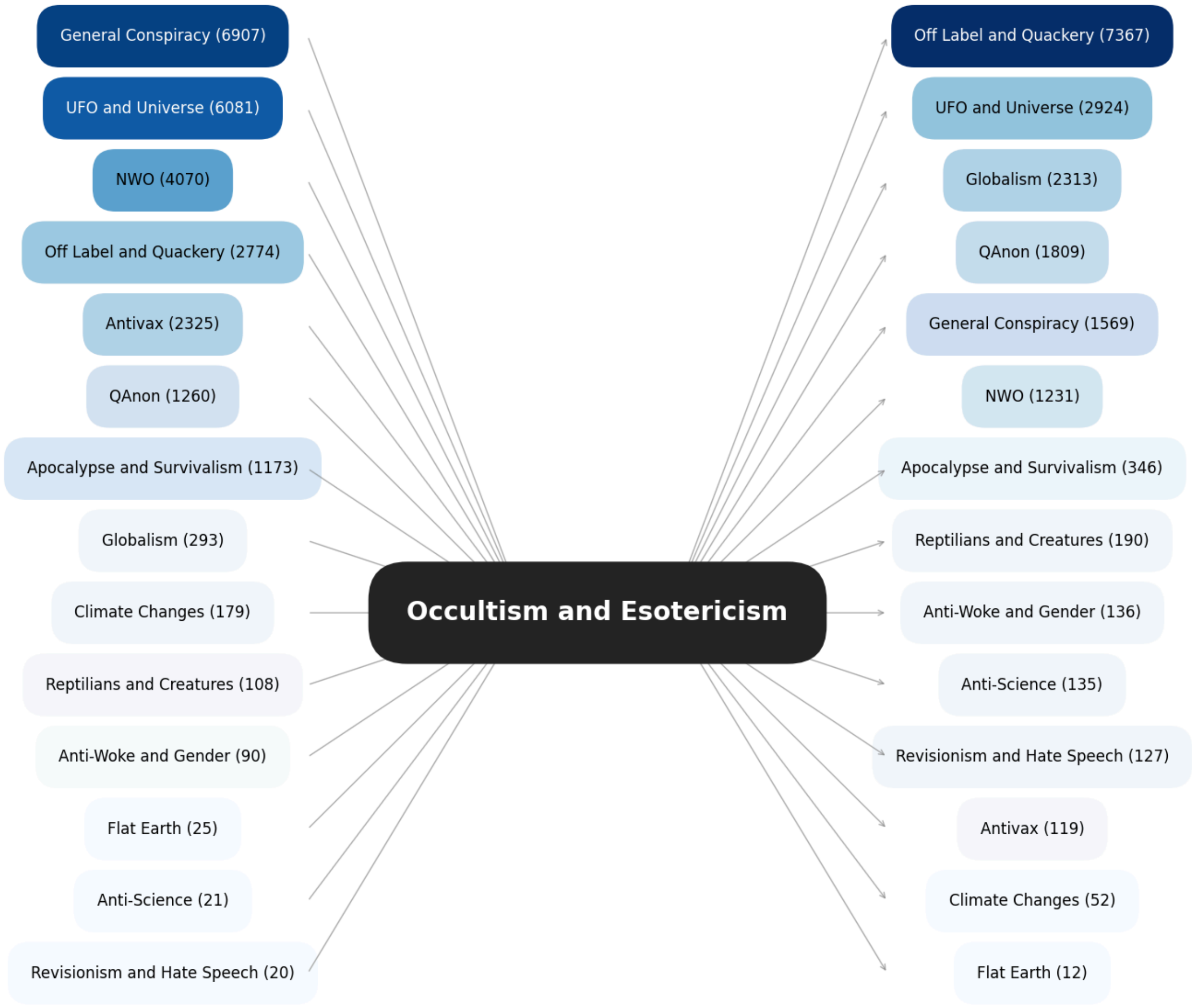

Source: Own elaboration (2024).

The Occultism and Esotericism graph reveals that these communities are deeply intertwined with General Conspiracies (6,907 links) and UFO and Universe (6,081 links), suggesting that the interest in the occult and esoteric often serves as a gateway to broader conspiracy narratives. This interconnection is not just an overlap of interests but rather a fusion of ideologies that feed off each other, creating an environment where belief in hidden powers and global conspiracies reinforce and legitimize each other. On the other hand, the fact that Occultism and Esotericism issue significant invitations to off-label drugs (7,367 links) and UFO and Universe (2,924 links) indicates a trend of expanding the boundaries of belief, uniting mysticism with distrust in conventional medicine and modern science. This convergence reflects a pattern where esotericism and scientific disinformation go hand in hand, forming an alliance based on the rejection of authority and the search for alternative truths. Thus, Occultism and Esotericism are not just isolated themes but a core that connects and expands different forms of disinformation, creating a vicious cycle of beliefs that perpetuate institutional distrust.

### 3.2. Time series

In the next graph, we will analyze how mentions of Apocalypse and Survival, as well as Occultism and Esotericism, have evolved over the past few years. It is observed that with the advent of the COVID-19 pandemic and subsequent global crises, there was an exponential increase in mentions of these themes, especially between 2020 and 2021. This period was marked by a rise in global uncertainties, which may have contributed to the proliferation of apocalyptic theories and the search for mystical explanations. Although both themes show a slight decline after the peaks, interest levels remain high, suggesting that in times of crisis, apocalyptic and esoteric narratives continue to grow.

**Figure 06.** Line graph over the period

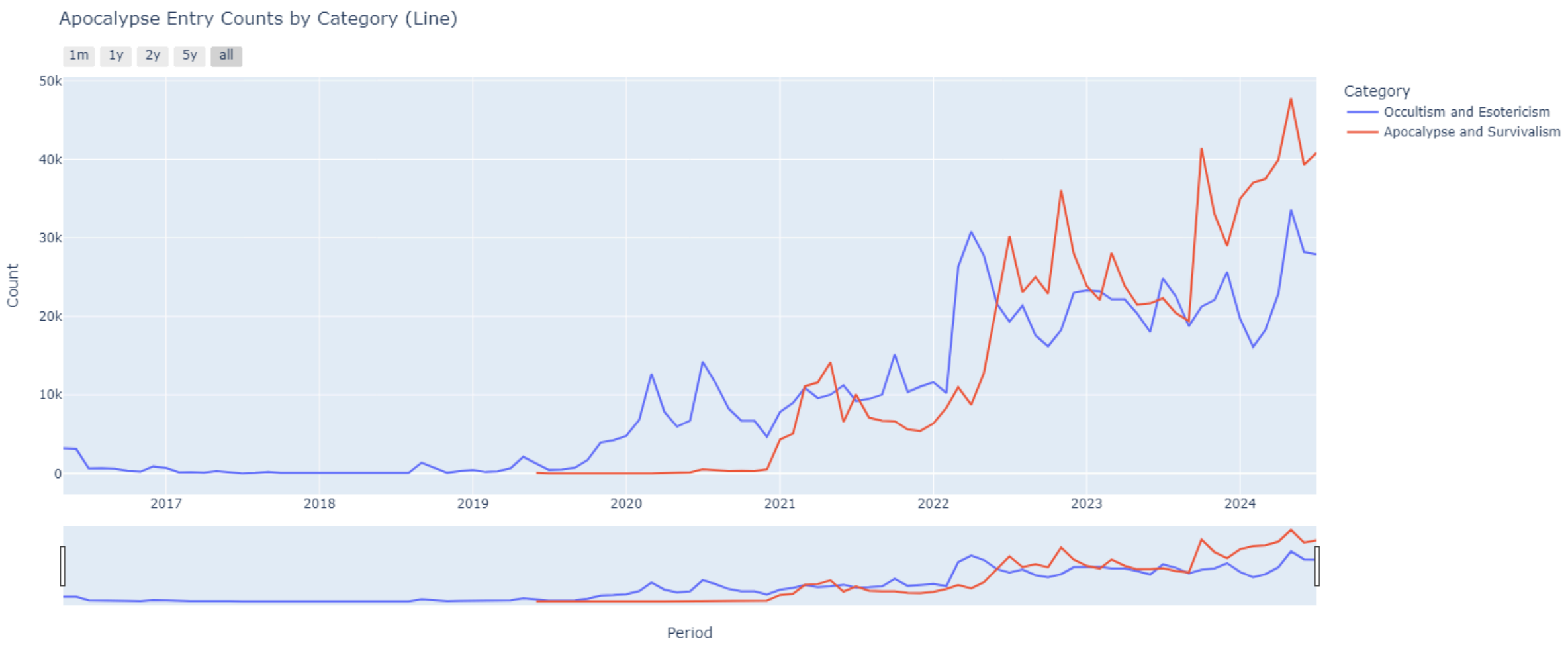

Source: Own elaboration (2024).

For Apocalypse and Survival, mentions grew by 1,500% between 2020 and the peak in April 2021, from approximately 1,000 to 35,000 mentions. This significant increase is correlated with the rise in global uncertainties, such as new COVID-19 variants and global economic crises, which spurred discussions about apocalyptic scenarios. Occultism and Esotericism saw a 1,200% increase over the same period, rising from around 2,000 to 28,000 mentions in November 2021. This shows how the search for alternative and mystical explanations also grew during periods of uncertainty, fueling apocalyptic theories. After the peaks, both themes show a decline of approximately 25%, indicating a slight deceleration but still with elevated interest levels compared to the pre-pandemic period.

[ english version / português abaixo ]

**Figure 07.** Absolute area chart over the period

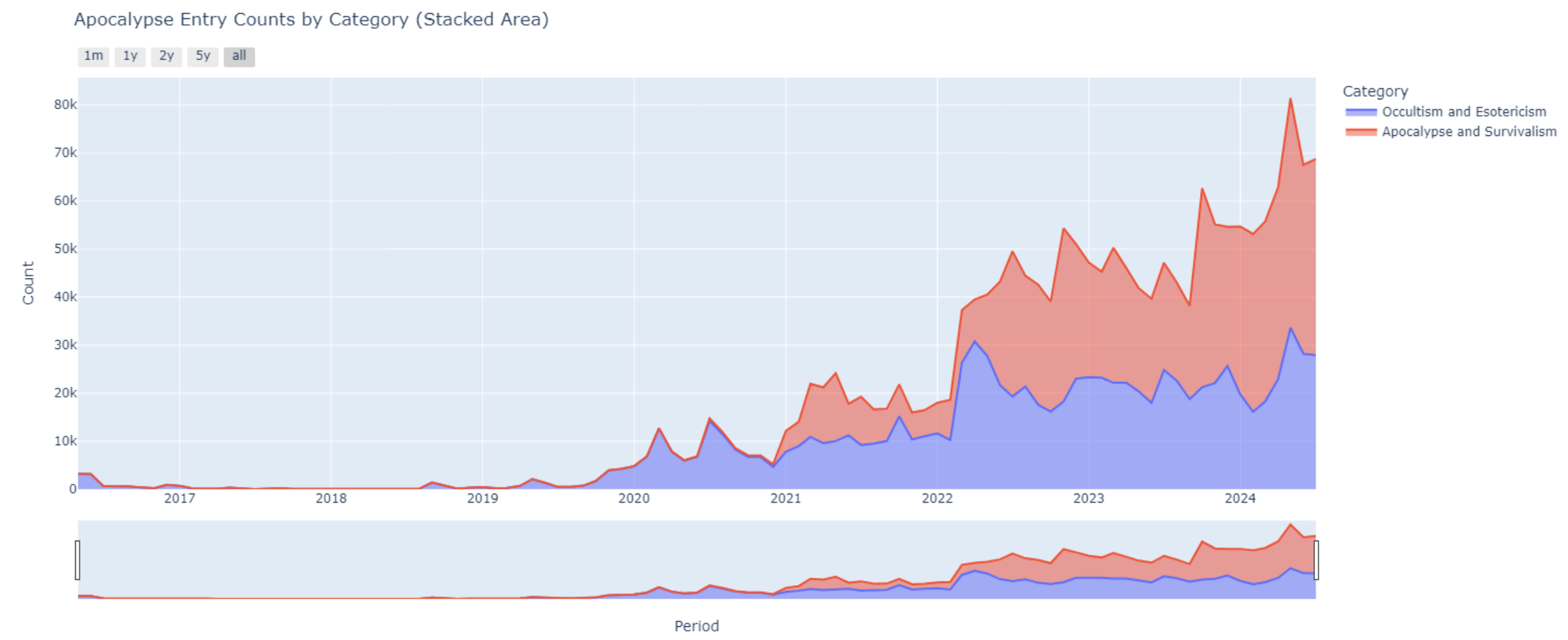

Source: Own elaboration (2024).

The absolute area graph highlights the continued increase in discussions surrounding Occultism and Esotericism and Apocalypse and Survival. Starting in 2020, mentions of Apocalypse and Survival grew substantially, peaking in 2022 with more than 50,000 entries. This reflects the increased fear and uncertainty caused by the COVID-19 pandemic, which fueled apocalyptic narratives and theories about the need for catastrophe preparedness. The pandemic, along with global economic and environmental crises, contributed to the proliferation of these theories. The simultaneous growth of Occultism and Esotericism is also significant, suggesting that in times of crisis, people seek supernatural or alternative explanations, which often connect with apocalyptic survival narratives. These two themes seem to reinforce each other, with adherents of esoteric theories being attracted to apocalyptic discourses and vice versa.

**Figure 08.** Relative area chart over the period

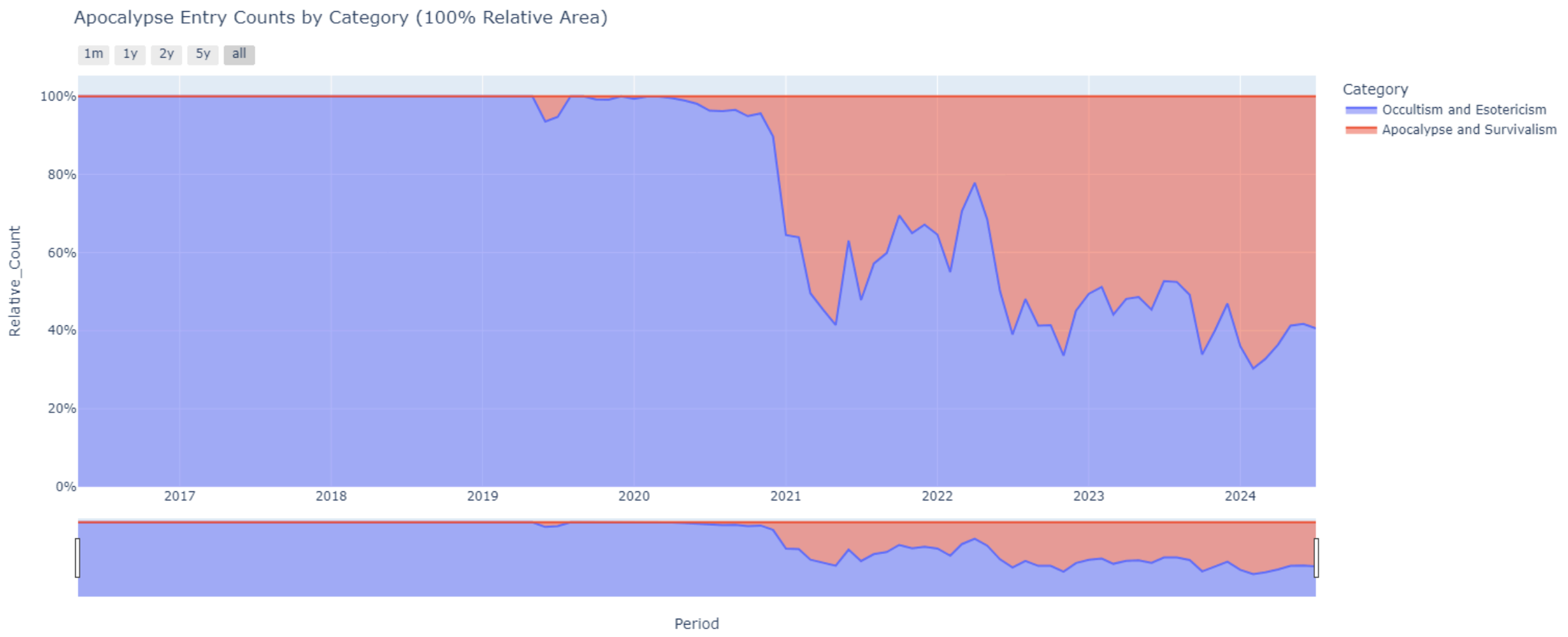

Source: Own elaboration (2024).

The relative area graph reveals that while Occultism and Esotericism were more dominant before 2020, Apocalypse and Survival began to dominate the discussions from 2021 onwards. This can be interpreted as a direct reflection of the pandemic and the perception that global catastrophic events require extreme responses. The graph also indicates that while Occultism and Esotericism maintain a significant presence, they are increasingly absorbed within the apocalyptic context, suggesting a convergence of narratives. The shift in focus suggests that in times of crisis, concerns about survival and disaster preparedness become more prominent, while esotericism serves as explanatory bases for fears and preparations.

### 3.3. Content analysis

The content analysis of apocalypse, occultism, and esotericism communities, through word clouds, reveals how certain concepts and terms stand out and transform over time within these discussions. The most frequent words, such as “life”, “light”, “God”, and “world”, indicate a strong intersection between spiritual and esoteric themes, often linked to perceptions of personal and global transformation. The presence of terms like “energy”, “soul”, and “consciousness” suggests a search for understanding beyond material reality, reflecting an attempt to connect apocalyptic events to a greater purpose or spiritual evolution. Additionally, the focus on words like “now” and “truth” demonstrates the perceived urgency of these groups regarding these changes, pointing to a narrative that blends the imminent with the eternal, the physical with the metaphysical. This analysis allows us to understand not only the central topics of these communities but also how these ideas are disseminated and reformulated in response to contemporary events and cultural shifts.

[ english version / português abaixo ]

**Figure 09.** Consolidated word cloud for apocalypse, survivalism, occultism and esotericism

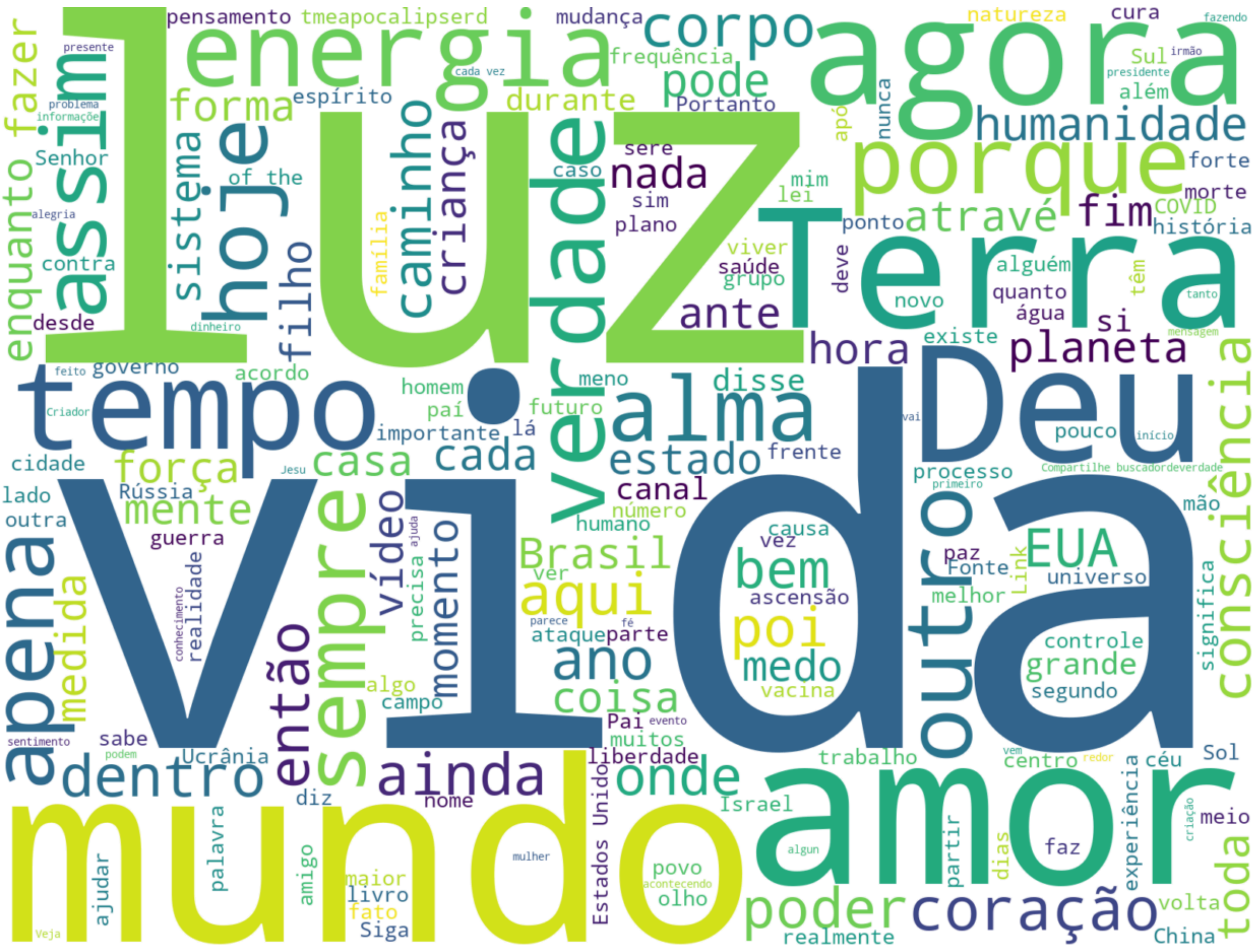

Source: Own elaboration (2024).

The consolidated word cloud of discussions on apocalypse, occultism, and esotericism reveals the centrality of terms deeply loaded with spiritual and existential meaning, such as “life”, “light”, “God”, and “world”. The prominence of the word “life” suggests a fundamental concern with existence and purpose, themes that permeate discussions within these communities. “Light” and “energy” indicate an emphasis on esoteric concepts of illumination and vital force, often associated with the idea of transformation or spiritual ascension. The recurrence of “God” and “soul” points to an attempt to connect apocalyptic events with a narrative of redemption or divine judgment, while “world” and “truth” reflect a dichotomous vision between the known and the unknown, the visible and the invisible. The word “now” suggests an urgency in the discussions, as if these events were imminent and required an immediate response. On the other hand, “fear” and “power” indicate the emotions and forces at play in these narratives, where the apocalypse is seen both as a time of destruction and potential rebirth.

[ english version / português abaixo ]

**Chart 01.** Temporal word cloud series for apocalypse and survivalism

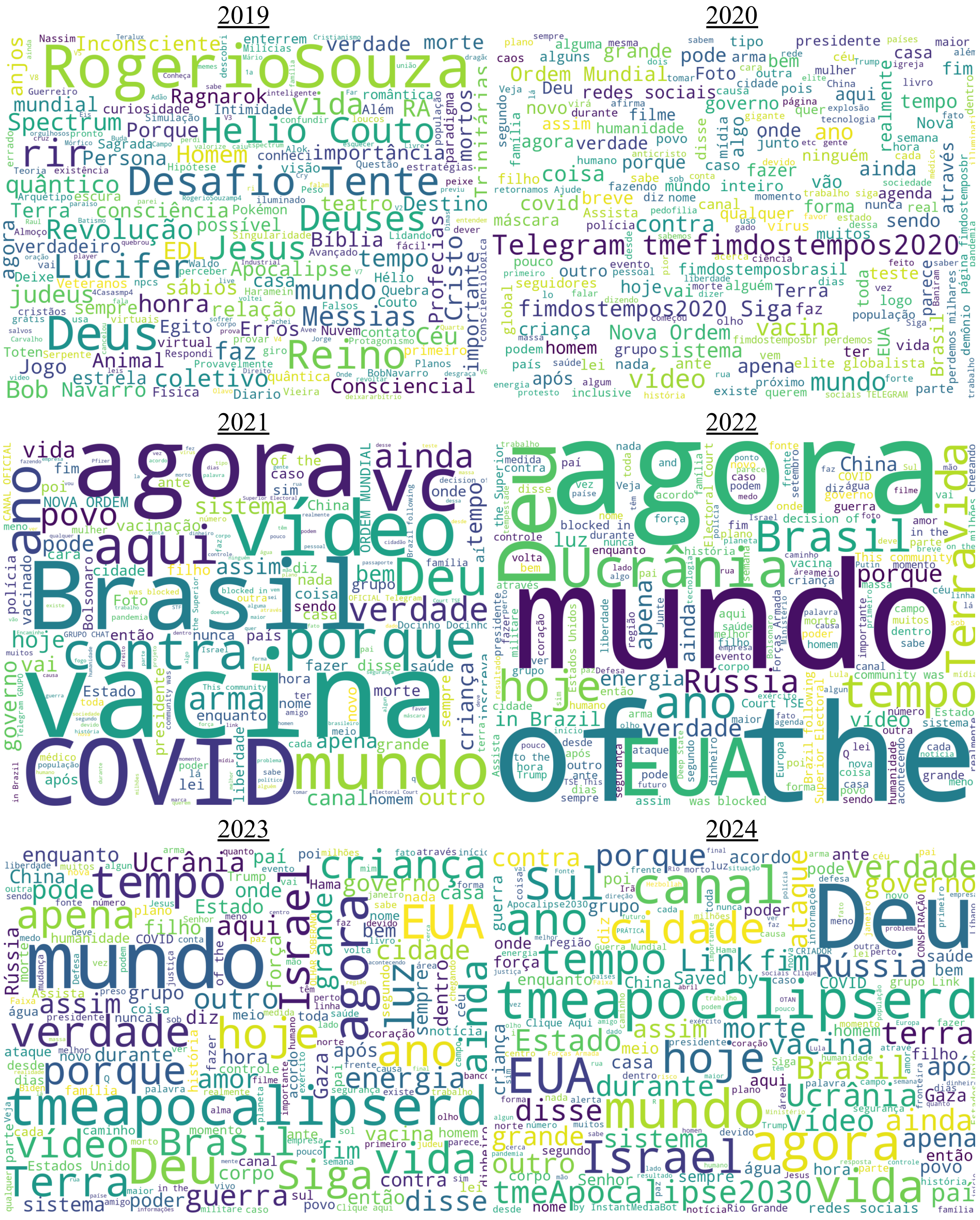

Source: Own elaboration (2024).

The temporal word cloud series on apocalypse and survival reveals how the themes and concerns of these communities evolved from 2019 to 2024. In 2019, terms like “God” and “Kingdom” suggest a strong connection with religious end-times narratives, while “life” and “world” indicate the duality between the present and what is to come. In 2020, with the advent of the COVID-19 pandemic, words like “end” and “world” gain prominence, reflecting the fear and uncertainty brought by the global crisis. In 2021 and 2022, “vaccine” and “COVID” appear frequently, indicating that the pandemic became a central point in apocalyptic discussions, with the vaccine being interpreted in some cases as part of a larger conspiracy. In the following years, the focus seems to expand to include themes like “energy” and “system”, suggesting a growing concern with survival and self-sufficiency in a world perceived as increasingly unstable and dangerous. The continuity of terms like “God” and “apocalypse” over the years shows how these communities maintain a consistent narrative of preparation for an imminent end, adapting it to changes in the global context.

**Chart 02.** Temporal word cloud series for occultism and esotericism

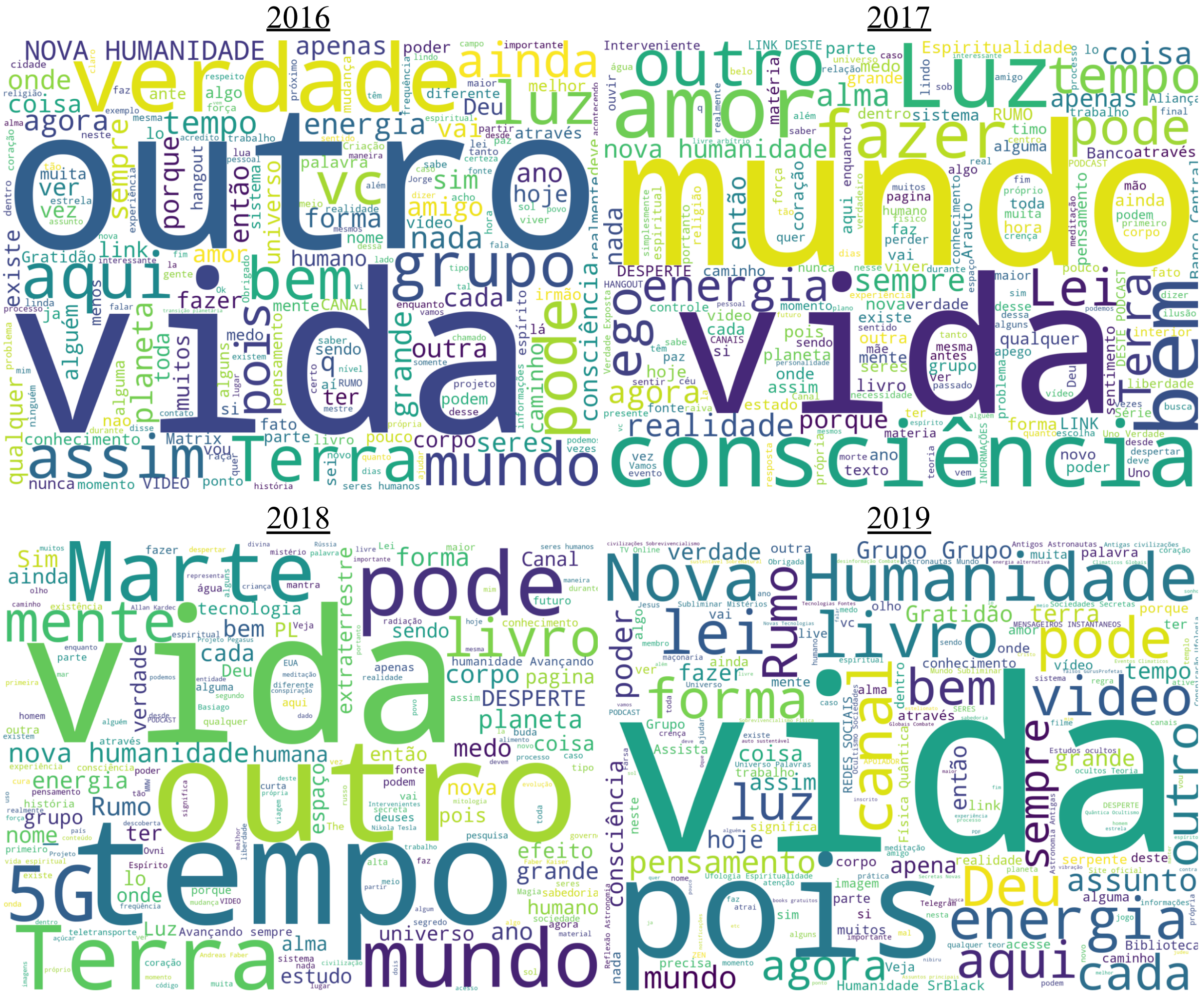

[ english version / português abaixo ]

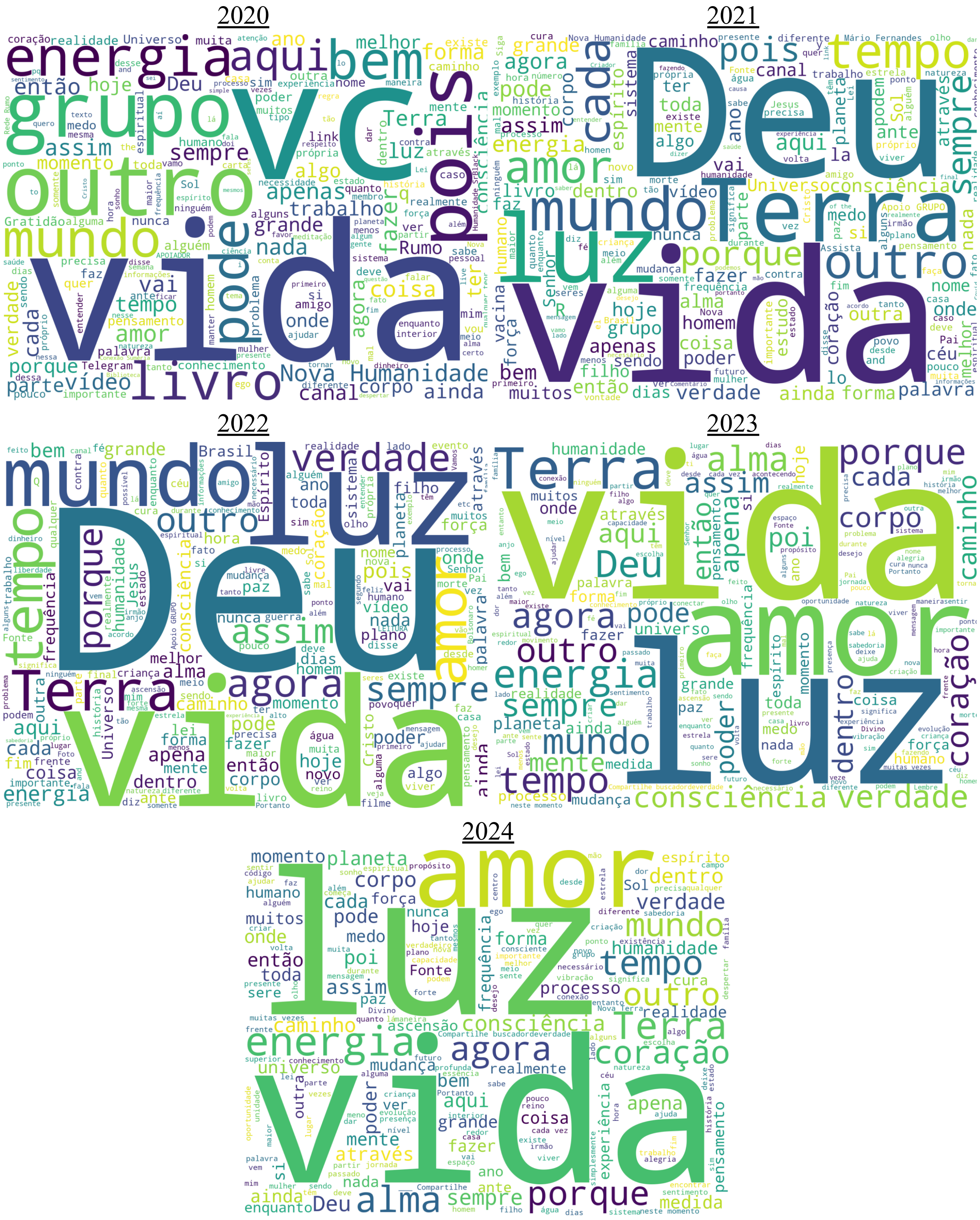

Source: Own elaboration (2024).

The temporal word cloud series on occultism and esotericism highlights the persistence and evolution of spiritual and mystical themes between 2016 and 2024. In 2016, terms like “life” and “world” are predominant, suggesting a search for meaning and understanding of the universe, often through an esoteric lens. Over the years, words like

"light" and "energy" gain importance, reflecting an increasing emphasis on concepts of illumination and vital force. In 2020, during the peak of the pandemic, "God" and "truth" appear prominently, perhaps indicating a search for spiritual answers amid global chaos. In 2021 and 2022, there is a diversification of themes, with the emergence of terms like "consciousness" and "earth", suggesting an expansion of discussions to include issues of planetary consciousness and connection with nature. In 2023 and 2024, "love" and "life" remain central, pointing to a narrative that emphasizes the importance of spiritual transformation and self-knowledge as responses to external crises. The recurrence of terms over time suggests that while the context changes, the fundamental concerns of these communities remain focused on the search for understanding and spiritual enlightenment.

### 3.4. Thematic agenda overlap

The following figures explore how conspiracy theory communities related to Apocalypse and Survivalism connect with other narratives, forming a cohesive discourse that reinforces the beliefs of these communities. By analyzing the topics discussed, it is possible to observe the centrality of these themes in the dissemination of disinformation, where esoteric beliefs, globalism theories, and anti-vaccination sentiments are integrated into a narrative of preparation for an apocalyptic future. This analysis highlights how the overlap of themes amplifies the reach of these communities, making evidence-based interventions challenging.

**Figure 10.** Energy and esoteric enlightenment themes

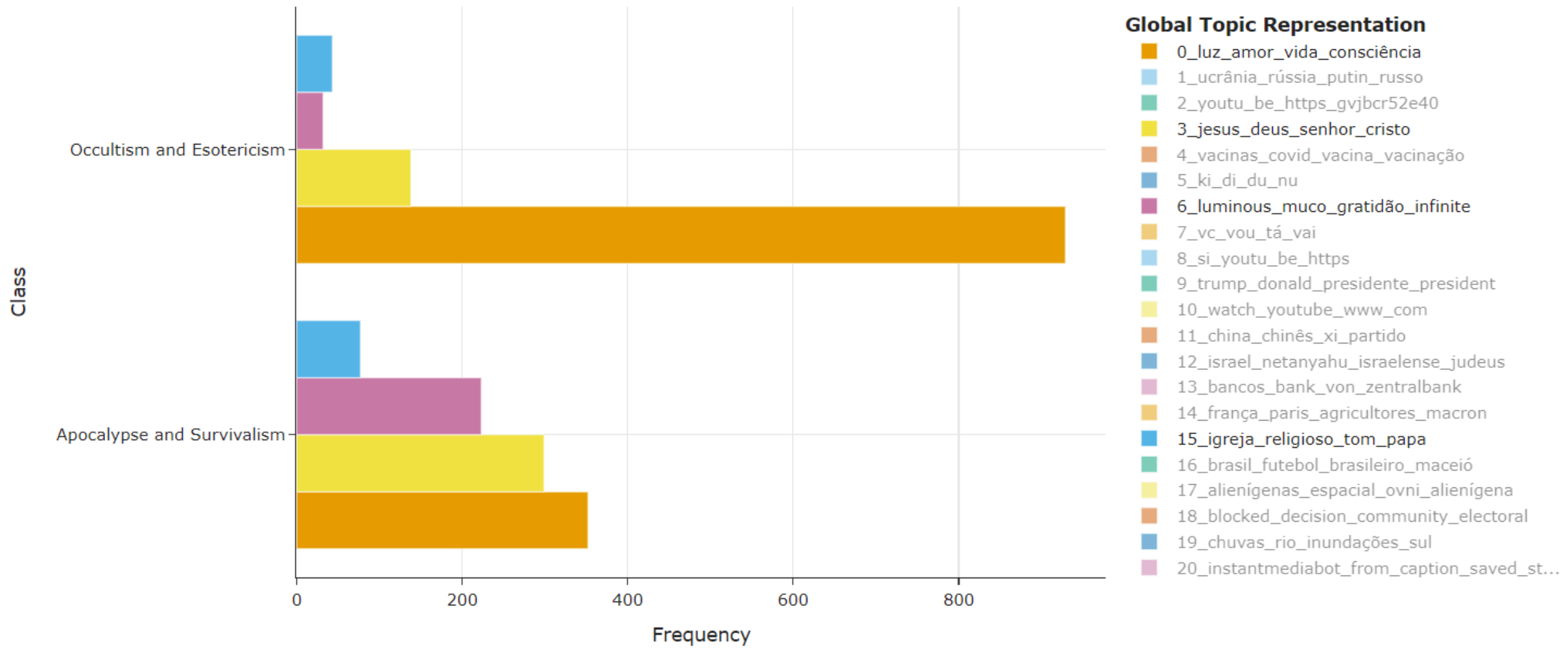

Source: Own elaboration (2024).

Figure 10 highlights the predominance of discussions on "energy" within the esoteric context, contrasting with the more abrupt discussions on apocalypse and survivalism seen in the communities. Topics such as "light", "love", and "life" are widely discussed, suggesting that esoteric communities integrate these concepts as part of a spiritual preparation for catastrophic events. This overlap reflects a search for mystical meanings and an attempt to find spiritual guidance amidst the anticipation of apocalyptic crises.

**Figure 11.** Globalism and geopolitical disputes themes

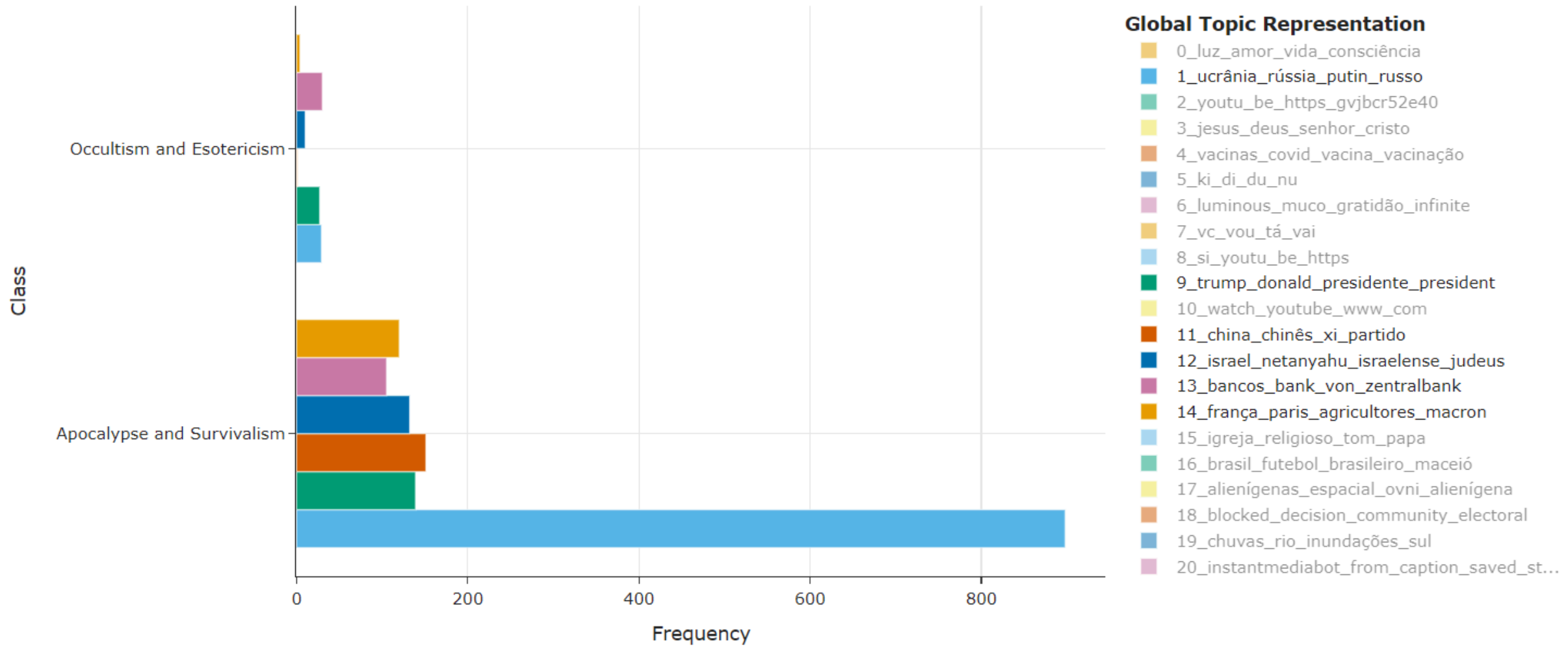

Source: Own elaboration (2024).

In Figure 11, discussions of apocalypse and survivalism dominate, focusing on themes of globalism and geopolitical disputes. Topics such as “Russia”, “Ukraine”, and “Putin” are widely addressed, suggesting that these communities use recent geopolitical events to reinforce their narratives of imminent global control. The overlap with survivalism themes indicates that these theories are used to justify preparation for a global collapse, where international disputes are seen as precursors to an apocalypse. This results in concerns that reverberate in daily alerts issued by communities, including citizens worried about safety.

**Figure 12.** Anti-vaccine denialism, aliens, and QAnon themes

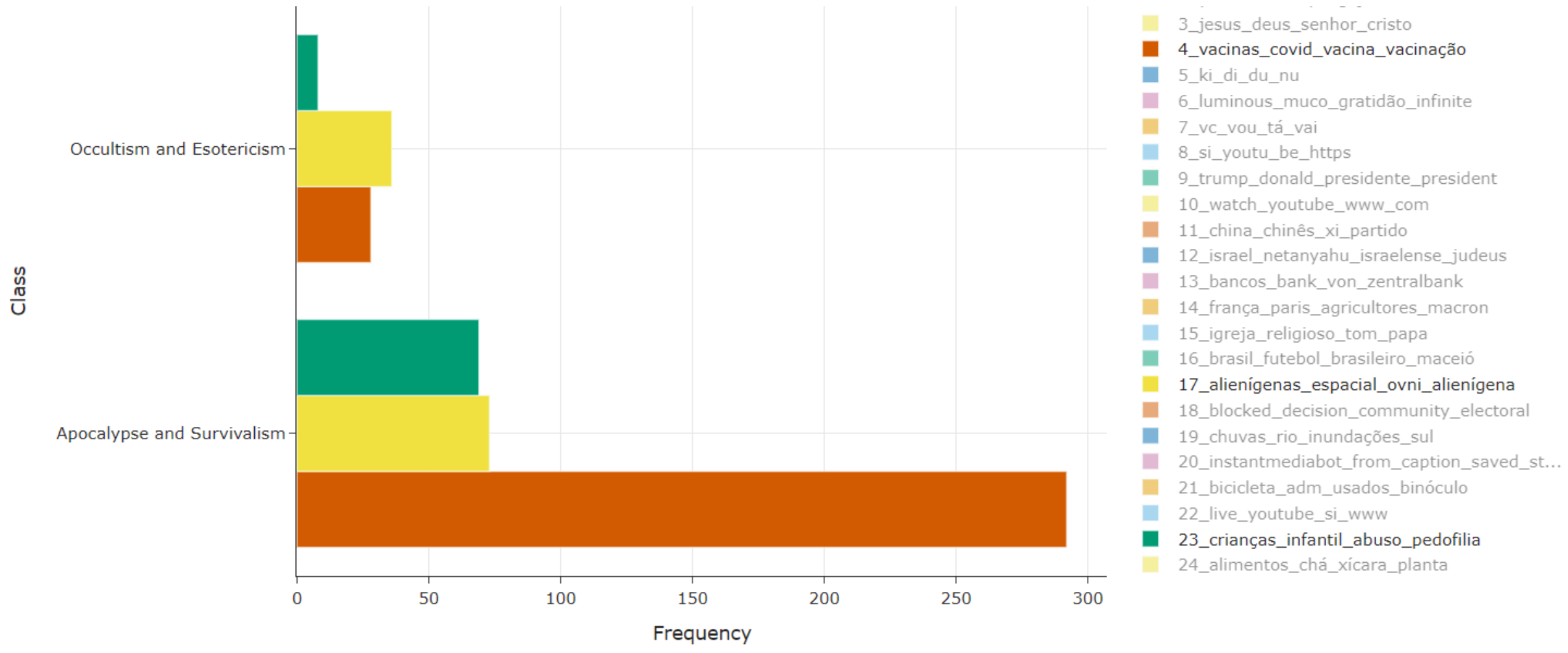

Source: Own elaboration (2024).

Figure 12 shows a strong interconnection between anti-vaccine denialism, alien theories, and QAnon. Topics such as “vaccines”, “COVID”, and “aliens” suggest that these communities view vaccination and the pandemic as parts of a larger conspiracy, possibly involving extraterrestrial beings. This narrative not only reinforces distrust in vaccines but also expands the idea that hidden or governmental forces are manipulating global events to control the population, justifying preparation for an apocalypse. Additionally, monetization can be found through the sale of courses on quantum magnetism, for example, where faith is instrumentalized to generate revenue for those who exploit the co-opted individuals.

**Figure 13.** Polarization and national politics themes

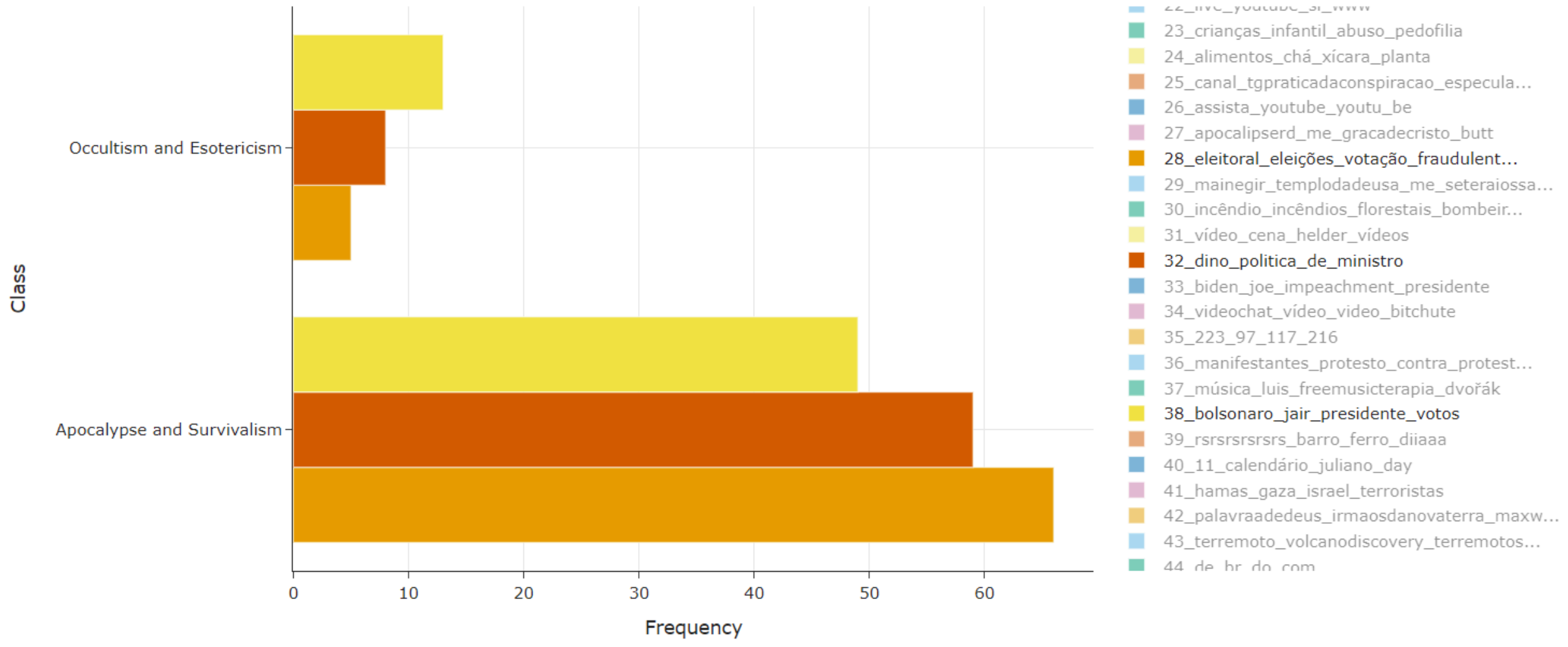

Source: Own elaboration (2024).

In Figure 13, survivalism themes overlap with discussions on polarization and national politics. Topics such as “Bolsonaro”, “impeachment”, and “election fraud” indicate that these communities perceive political instability as a sign of larger crises that justify preparation for an imminent collapse. The association between survivalism and national politics reflects a view that political events are indicators of an apocalyptic future, reinforcing the need to be prepared to survive political and social crises.

**Figure 14.** Supposedly planned catastrophes themes

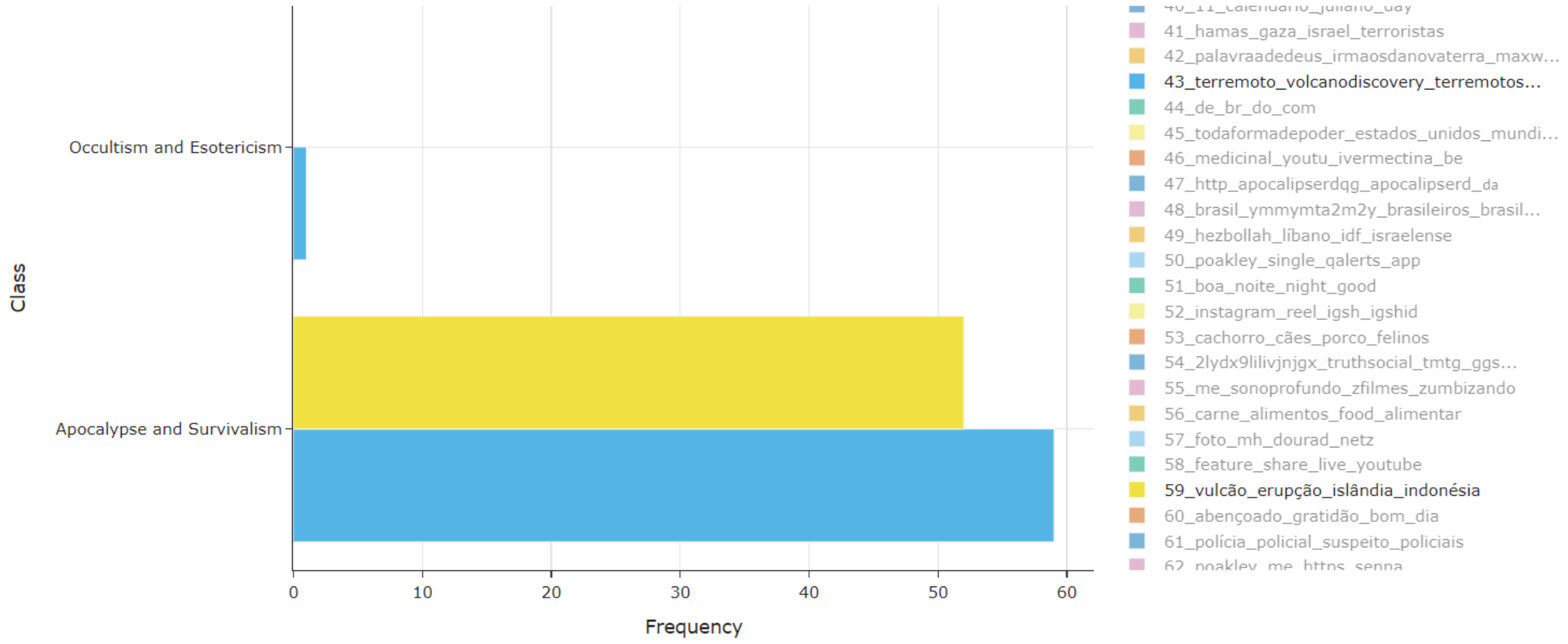

Source: Own elaboration (2024).

Figure 14 reveals how survivalism discussions intertwine with the belief in supposedly planned catastrophes. Topics such as “volcanoes”, “earthquakes”, and “tsunamis” are highlighted, suggesting that these communities believe natural disasters are orchestrated as part of a global agenda. The overlap with survivalism indicates that these beliefs are used to justify intensive preparation to survive these catastrophes, which are seen not as natural events but as deliberate actions to reduce or control the world's population.

**Figure 15.** Anti-Woke with narratives of supposed LGBT sexualization themes

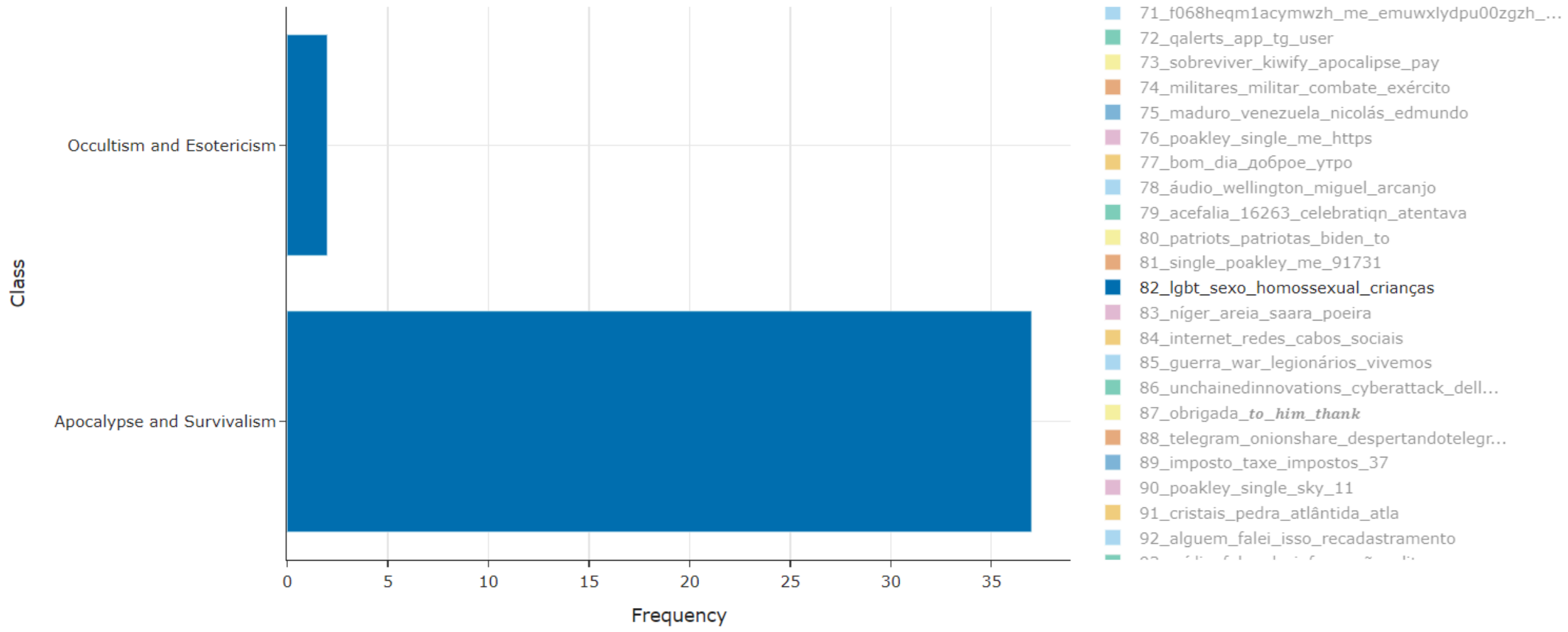

Source: Own elaboration (2024).

Figure 15 highlights the centrality of anti-Woke themes in survivalism discussions, focusing on narratives of supposed LGBT sexualization. "LGBT" and "sexualization" are addressed as part of a critique that unfoundedly insinuates a connection between LGBT individuals and child abuse. The overlap with survivalism suggests that these communities see these changes as part of a moral decline that justifies preparation for a cultural apocalypse, where survival depends on resistance to influences perceived as corrosive.

## 4. Reflections and future works

To answer the research question, **"how are Brazilian conspiracy theory communities on apocalypse, survivalism, occultism and esotericism topics characterized and articulated on Telegram?"**, this study adopted mirrored techniques in a series of seven publications aimed at characterizing and describing the phenomenon of conspiracy theories on Telegram, focusing on Brazil as a case study. After months of investigation, it was possible to extract a total of 72 Brazilian conspiracy theory communities on Telegram focused on apocalypse, survivalism, occultism and esotericism topics, amounting to 4,371,125 pieces of content published between May 2016 (initial publications) and August 2024 (when this study was conducted), with 192,138 users combined across the communities.

Four main approaches were adopted: **(i)** Network, which involved the creation of an algorithm to map connections between communities through invitations circulated among groups and channels; **(ii)** Time series, which used libraries like "Pandas" (McKinney, 2010) and "Plotly" (Plotly Technologies Inc., 2015) to analyze the evolution of publications and engagements over time; **(iii)** Content analysis, where textual analysis techniques were applied to identify patterns and word frequencies in the communities over the semesters; and **(iv)** Thematic agenda overlap, which utilized the BERTopic model (Grootendorst, 2020) to group and interpret large volumes of text, generating coherent topics from the analyzed publications. The main reflections are detailed below, followed by suggestions for future works.

### 4.1. Main reflections

**Occultism and esotericism communities serve as gateways to extreme survival and apocalypse theories, connecting to other conspiratorial themes:** Occultism and esotericism communities stand out as initial access points for survivalism and apocalypse narratives, recording 2,527,833 interactions. The interconnectivity of these themes highlights how the search for mystical answers fuels concerns about catastrophic scenarios, serving as a bridge to other conspiracy theories such as the New World Order (NWO);

**Conspiracies about the New World Order are amplified by apocalyptic discussions, significantly impacting health disinformation:** With 3,488,686 interactions, NWO-centered communities have a strong correlation with apocalyptic themes, reinforcing the idea of global control. This narrative is amplified during crises like the COVID-19 pandemic, linking apocalyptic theories to vaccine distrust and public health concerns, with over 1,345,060 interactions related to these issues;

**Survivalism narratives experience a 1,500% growth during the pandemic, evidencing their connection to other conspiracy theories:** The exponential growth of mentions of survivalism during the pandemic, from 1,000 to 35,000 between 2020 and 2021, reflects the impact of global crises on the proliferation of apocalyptic theories. These groups become catalysts for the acceptance of a worldview where social collapse is imminent;

**Occultism and esotericism are the largest sources of invitations to off-label drug communities, revealing a dangerous intersection between mysticism and scientific disinformation:** Occultism and esotericism, with 7,367 links directed to off-label drug communities, reinforce the dissemination of alternative and dangerous practices. This interconnection underscores the alliance between esoteric beliefs and the rejection of conventional science, creating a resistant cycle of disinformation. Additionally, monetization can be observed through the sale of courses on quantum magnetism, where faith is instrumentalized to generate revenue for those who exploit the co-opted individuals;

**Apocalypse and survivalism communities function as central hubs, connecting multiple conspiracy narratives and strengthening disinformation cycles:** Apocalypse and survivalism communities, with over 4,371,125 interactions, operate as central hubs within the disinformation network. These communities not only connect different conspiracy theories but also amplify them, creating a continuous cycle of mutual reinforcement;

**The overlap of esoteric and survival themes reveals a cohesive ideological bubble, resistant to scientific information:** The fusion of esoteric themes and survival narratives creates an ideological bubble that is difficult to penetrate. With 3,026,065 interactions linked

to occultism and esotericism, these groups perpetuate distrust in science, promoting an alternative worldview that feeds off itself;

**The false narrative associating vaccines with global control persists as one of the most discussed topics in apocalyptic and esoteric communities:** Even after repeated debunking, the idea that vaccines are part of a global control plan continues to be widely discussed, especially in communities associated with NWO and survivalism theories. This narrative serves to reinforce distrust in medicine;

**The interconnectivity between occultism, NWO, and apocalypse communities reflects a pattern of mutual reinforcement that amplifies disinformation:** Analyzing the connections between these communities reveals how seemingly distinct theories reinforce each other. With more than 5,421 invitation links between them, this interconnectivity creates an environment where different forms of disinformation become interdependent;

**Apocalypse discussions often serve as entry points for the acceptance of other conspiracy theories, expanding the reach of these beliefs:** Apocalypse and survivalism communities act as gateways to theories about NWO, globalism, and other conspiracies. This thematic expansion helps disseminate these beliefs to a broader audience, making it more challenging to dismantle these narratives;

**The COVID-19 pandemic catalyzed the growth of esoteric and survival communities, reflecting a reaction to global uncertainties:** The increase in interactions in esoteric and survival communities during the pandemic, with peaks of 1,200% for occultism and esotericism, suggests that global crises encourage the search for alternative explanations, often linked to conspiracy theories.

### 4.2. Future works

Based on the key findings of this study, several directions can be suggested for future research. Firstly, it is essential to explore in greater depth how apocalypse and survival themes connect with other conspiratorial narratives in the Brazilian context, especially considering the significant growth of these discussions during the pandemic. Future investigations could examine how these communities use global events, such as health crises and natural disasters, to reinforce their beliefs and attract new members. Additionally, the impact of these narratives on specific populations, such as young people and those in rural areas, deserves attention, considering the vulnerability of these demographics to disinformation.

Another relevant point is the need to map the interactions between occultism and esotericism communities with those promoting off-label drugs and alternative health practices. Studying the psychology behind the attraction to these esoteric beliefs could offer insights into how these practices gain traction within ideological bubbles. Future studies could focus on developing and testing interventions aimed at dismantling these beliefs, particularly in contexts where conventional science is strongly rejected. It is also important to investigate the persistence of apocalyptic narratives and how they are reintroduced during times of crisis.

Research could focus on identifying the mechanisms that allow these narratives to continue proliferating, even after being widely debunked. Understanding the resilience of these beliefs could be crucial for developing more effective strategies to combat disinformation.

Furthermore, the interconnectivity between communities discussing survival and occultism with other conspiracy theories, such as the New World Order and globalism, should be better understood. Future studies could explore how these networks form, sustain themselves, and interact, offering a detailed view of the internal dynamics of these communities. Understanding these interactions could help identify effective intervention points to reduce the spread of disinformation. Finally, the development of tools that enable real-time monitoring and identification of disinformation hubs emerging within these communities is a promising area for research. In times of global crises, such as pandemics or natural disasters, these tools could be used to quickly identify emerging narratives and implement interventions before they gain traction and cause significant harm.

## 5. References

Grootendorst, M. (2020). ***BERTopic:*** *Leveraging BERT and c-TF-IDF to create easily interpretable topics*. GitHub. https://github.com/MaartenGr/BERTopic

Loper, E., & Bird, S. (2002). ***NLTK:*** *The Natural Language Toolkit*. In *Proceedings of the ACL-02 Workshop on Effective Tools and Methodologies for Teaching Natural Language Processing and Computational Linguistics* - Volume 1 (pp. 63–70). Association for Computational Linguistics. https://doi.org/10.3115/1118108.1118117

McKinney, W. (2010). **Data structures for statistical computing in Python.** In *Proceedings of the 9th Python in Science Conference* (pp. 51–56). https://doi.org/10.25080/Majora-92bf1922-00a

Mueller, A. (2020). ***WordCloud.*** GitHub. https://github.com/amueller/word_cloud

Plotly Technologies Inc. (2015). ***Collaborative data science.*** Plotly Technologies Inc. https://plotly.com

Rocha, I., Silva, E. C. M., & Mielli, R. V. (2024). ***Catalytic conspiracism:*** *Exploring persistent homologies time series in the dissemination of disinformation in conspiracy theory communities on Telegram*. 14º Encontro da Associação Brasileira de Ciência Política. UFBA, Salvador, BA. https://www.abcp2024.sinteseeventos.com.br/trabalho/view?ID_TRABALHO=687

Silva, E. C. M. (2023, fevereiro). ***Web scraping Telegram posts and content.*** GitHub. https://github.com/ergoncugler/web-scraping-telegram/

Silva, E. C. M., & Sátiro, R. M. (2024). ***Conspiratorial convergence:*** *Comparing thematic agendas among conspiracy theory communities on Telegram using topic modeling*. 14º Encontro da Associação Brasileira de Ciência Política. UFBA, Salvador, BA. https://www.abcp2024.sinteseeventos.com.br/trabalho/view?ID_TRABALHO=903

[ english version / português abaixo ]

## 6. Author biography

**Ergon Cugler de Moraes Silva** has a Master's degree in Public Administration and Government (FGV), Postgraduate MBA in Data Science & Analytics (USP) and Bachelor's degree in Public Policy Management (USP). He is associated with the Bureaucracy Studies Center (NEB FGV), collaborates with the Interdisciplinary Observatory of Public Policies (OIPP USP), with the Study Group on Technology and Innovations in Public Management (GETIP USP) with the Monitor of Political Debate in the Digital Environment (Monitor USP) and with the Working Group on Strategy, Data and Sovereignty of the Study and Research Group on International Security of the Institute of International Relations of the University of Brasília (GEPSI UnB). He is also a researcher at the Brazilian Institute of Information in Science and Technology (IBICT), where he works for the Federal Government on strategies against disinformation. Brasília, Federal District, Brazil. Web site: https://ergoncugler.com/.

[ versão português / english above ]

# **Comunidades de apocalipse, sobrevivencialismo, ocultismo e esoterismo no Telegram brasileiro:** quando a fé é instrumentalizada para vender cursos quânticos e abrir portas para conspiracionismos nocivos

*Ergon Cugler de Moraes Silva*

Instituto Brasileiro de Informação
em Ciência e Tecnologia (IBICT)
Brasília, Distrito Federal, Brasil

contato@ergoncugler.com
www.ergoncugler.com

## Resumo

As comunidades brasileiras no Telegram têm se voltado cada vez mais para teorias de apocalipse e sobrevivencialismo, especialmente em momentos de crise como a Pandemia da COVID-19, onde as narrativas de ocultismo e esoterismo encontram terreno fértil. Dessa forma, esse estudo busca responder à pergunta de pesquisa: **como são caracterizadas e articuladas as comunidades de teorias da conspiração brasileiras sobre temáticas de apocalipse, sobrevivencialismo, ocultismo e esoterismo no Telegram?** Vale ressaltar que este estudo faz parte de uma série de um total de sete estudos que possuem como objetivo principal compreender e caracterizar as comunidades brasileiras de teorias da conspiração no Telegram. Esta série de sete estudos está disponibilizada abertamente e originalmente no arXiv da Cornell University, aplicando um método espelhado nos sete estudos, mudando apenas o objeto temático de análise e provendo uma replicabilidade de investigação, inclusive com códigos próprios e autorais elaborados, somando-se à cultura de software livre e de código aberto. No que diz respeito aos principais achados deste estudo, observa-se: Comunidades de ocultismo e esoterismo funcionam como portas de entrada para teorias de apocalipse; Conspirações sobre a Nova Ordem Mundial são amplificadas por discussões apocalípticas; Narrativas de sobrevivencialismo cresceram significativamente durante a Pandemia; Ocultismo e esoterismo são fontes de convites para comunidades de medicamentos *off label*, reforçando a desinformação científica; Discussões sobre apocalipse servem como ponto de partida para outras teorias conspiratórias, ampliando o alcance das mesmas.

### Principais descobertas

- ➔ Comunidades de ocultismo e esoterismo funcionam como portas de entrada para teorias de sobrevivência extrema e apocalipse, com 2.527.833 interações conectando esses temas a outras conspirações, evidenciando como a busca por respostas místicas alimenta preocupações catastróficas e reforça narrativas como a Nova Ordem Mundial;

- ➔ Conspirações sobre a Nova Ordem Mundial são amplificadas por discussões apocalípticas, com 3.488.686 interações reforçando a ideia de controle global, especialmente durante crises como a Pandemia da COVID-19, onde essas narrativas se conectam a desinformações sobre vacinas e saúde pública, somando 1.345.060 interações;

- ➔ Narrativas de sobrevivencialismo cresceram 1.500% durante a Pandemia, refletindo como crises impulsionam teorias apocalípticas, com menções subindo de 1.000 para 35.000 entre 2020 e 2021, tornando esses grupos catalisadores para a aceitação de um colapso iminente;
- ➔ Ocultismo e esoterismo são as maiores fontes de convites para comunidades de medicamentos *off label* e produtos químicos ilícitos que resultam em negacionismo vacinal, com 7.367 links, mostrando uma interseção perigosa entre misticismo e desinformação científica, que reforça a disseminação de práticas alternativas e rejeição da ciência convencional. Além disso, é possível encontrar a monetização com venda de cursos de magnetismo quântico, por exemplo;
- ➔ Comunidades de apocalipse e sobrevivencialismo operam como *hubs* centrais na rede de desinformação, com 4.371.125 interações, conectando e amplificando diferentes teorias conspiratórias, criando um ciclo contínuo de reforço mútuo que fortalece essas narrativas;
- ➔ A sobreposição de temáticas esotéricas e de sobrevivência cria uma bolha ideológica coesa, resistente à informação científica, com 3.026.065 interações ligadas a ocultismo e esoterismo, perpetuando desconfiança na ciência e promovendo uma visão de mundo alternativa;
- ➔ A narrativa falsa que associa vacinas ao controle global persiste como uma das mais discutidas em comunidades apocalípticas e esotéricas, reforçando a desconfiança na medicina mesmo após desmentidos, especialmente em grupos de teorias da NOM e sobrevivencialismo;
- ➔ A interconectividade entre comunidades de ocultismo, NOM e apocalipse revela um padrão de reforço mútuo que amplifica a desinformação, com 5.421 links de convites entre elas, criando um ambiente onde diferentes formas de desinformação se tornam interdependentes;
- ➔ Discussões sobre apocalipse frequentemente servem como ponto de partida para a aceitação de outras teorias conspiratórias, com comunidades de sobrevivencialismo atuando como portas de entrada para teorias sobre NOM, globalismo e outras conspirações, ampliando o alcance.
- ➔ A Pandemia da COVID-19 catalisou o crescimento de comunidades esotéricas e de sobrevivência, com picos de 1.200% nas interações em ocultismo e esoterismo — embasando desinformação vacinal e promovendo curandeirismos —, refletindo a reação às incertezas globais e incentivando a busca por explicações alternativas ligadas a teorias conspiratórias.

## 1. Introdução

Após percorrer milhares de comunidades brasileiras de teorias da conspiração no Telegram, extrair dezenas de milhões de conteúdos dessas comunidades, elaborados e/ou compartilhados por milhões de usuários que as compõem, este estudo tem o objetivo de compor uma série de um total de sete estudos que tratam sobre o fenômeno das teorias da conspiração no Telegram, adotando o Brasil como estudo de caso. Com as abordagens de identificação implementadas, foi possível alcançar um total de 72 comunidades de teorias da conspiração brasileiras no Telegram sobre temáticas de apocalipse, sobrevivencialismo, ocultismo e esoterismo, estas somando 4.371.125 de conteúdos publicados entre maio de 2016 (primeiras publicações) até agosto de 2024 (realização deste estudo), com 192.138 usuários somados dentre as comunidades. Dessa forma, este estudo tem como objetivo compreender e caracterizar as comunidades sobre temáticas de apocalipse, sobrevivencialismo, ocultismo e esoterismo presentes nessa rede brasileira de teorias da conspiração identificada no Telegram.

Para tal, será aplicado um método espelhado em todos os sete estudos, mudando apenas o objeto temático de análise e provendo uma replicabilidade de investigação. Assim, abordaremos técnicas para observar as conexões, séries temporais, conteúdos e sobreposições temáticas das comunidades de teorias da conspiração. Além desse estudo, é possível encontrar os seis demais disponibilizados abertamente e originalmente no arXiv da Cornell University. Essa série contou com a atenção redobrada para garantir a integridade dos dados e o respeito à privacidade dos usuários, conforme a legislação brasileira prevê (Lei nº 13.709/2018).

Portanto questiona-se: **como são caracterizadas e articuladas as comunidades de teorias da conspiração brasileiras sobre temáticas de apocalipse, sobrevivencialismo, ocultismo e esoterismo no Telegram?**

## 2. Materiais e métodos

A metodologia deste estudo está organizada em três subseções, sendo: **2.1. Extração de dados**, que descreve o processo e as ferramentas utilizadas para coletar as informações das comunidades no Telegram; **2.2. Tratamento de dados**, onde são abordados os critérios e métodos aplicados para classificar e anonimizar os dados coletados; e **2.3. Abordagens para análise de dados**, que detalha as técnicas utilizadas para investigar as conexões, séries temporais, conteúdos e sobreposições temáticas das comunidades de teorias da conspiração.

### 2.1. Extração de dados

Este projeto teve início em fevereiro de 2023, com a publicação da primeira versão do TelegramScrap (Silva, 2023), uma ferramenta própria e autoral, de software livre e de código aberto, que faz uso da Application Programming Interface (API) da plataforma Telegram por meio da biblioteca Telethon e organiza ciclos de extração de dados de grupos e canais abertos no Telegram. Ao longo dos meses, a base de dados pôde ser ampliada e qualificada fazendo uso de quatro abordagens de identificação de comunidades de teorias da conspiração:

**(i) Uso de palavras chave:** no início do projeto, foram elencadas palavras-chave para identificação diretamente no buscador de grupos e canais brasileiros no Telegram, tais como “apocalipse”, “sobrevivencialismo”, “mudanças climáticas”, “terra plana”, “teoria da conspiração”, “globalismo”, “nova ordem mundial”, “ocultismo”, “esoterismo”, “curas alternativas”, “qAnon”, “reptilianos”, “revisionismo”, “alienígenas”, dentre outras. Essa primeira abordagem forneceu algumas comunidades cujos títulos e/ou descrições dos grupos e canais contassem com os termos explícitos relacionados a teorias da conspiração. Contudo, com o tempo foi possível identificar outras diversas comunidades cujas palavras-chave elencadas não davam conta de abarcar, algumas inclusive propositalmente com caracteres trocados para dificultar a busca de quem a quisesse encontrar na rede;

**(ii) Mecanismo de recomendação de canais do Telegram:** com o tempo, foi identificado que canais do Telegram (exceto grupos) contam com uma aba de recomendação chamada de “canais similares”, onde o próprio Telegram sugere dez canais que tenham alguma similaridade com o canal que se está observando. A partir desse mecanismo de recomendação do próprio Telegram, foi possível encontrar mais comunidades de teorias da conspiração brasileiras, com estas sendo recomendadas pela própria plataforma;

**(iii) Abordagem de bola de neve para identificação de convites:** após algumas comunidades iniciais serem acumuladas para a extração, foi elaborado um algoritmo próprio autoral de identificação de urls que contivessem "t.me/", sendo o prefixo de qualquer convite para grupos e canais do Telegram. Acumulando uma frequência de centenas de milhares de links que atendessem a esse critério, foram elencados os endereços únicos e contabilizadas as suas repetições. Dessa forma, foi possível fazer uma investigação de novos grupos e canais brasileiros mencionados nas próprias mensagens dos já investigados, ampliando a rede. Esse processo foi sendo repetido periodicamente, buscando identificar novas comunidades que tivessem identificação com as temáticas de teorias da conspiração no Telegram;

**(iv) Ampliação para tweets publicados no X que mencionassem convites:** com o objetivo de diversificar ainda mais a fonte de comunidades de teorias da conspiração brasileiras no Telegram, foi elaborada uma query de busca própria que pudesse identificar as palavras-chave de temáticas de teorias da conspiração, porém usando como fonte tweets publicados no X (antigo Twitter) e que, além de conter alguma das palavras-chave, contivesse também o "t.me/", sendo o prefixo de qualquer convite para grupos e canais do Telegram, "https://x.com/search?q=lang%3Apt%20%22t.me%2F%22%20TERMO-DE-BUSCA".

Com as abordagens de identificação de comunidades de teorias da conspiração implementadas ao longo de meses de investigação e aprimoramento de método, foi possível construir uma base de dados do projeto com um total de 855 comunidades de teorias da conspiração brasileiras no Telegram (considerando as demais temáticas também não incluídas nesse estudo), estas somando 27.227.525 de conteúdos publicados entre maio de 2016 (primeiras publicações) até agosto de 2024 (realização deste estudo), com 2.290.621 usuários somados dentre as comunidades brasileiras. Há de se considerar que este volume de usuários conta com dois elementos, o primeiro é que trata-se de uma variável, pois usuários podem entrar e sair diariamente, portanto este valor representa o registrado na data de extração de publicações da comunidade; além disso, é possível que um mesmo usuário esteja em mais de um grupo e, portanto, é contabilizado mais de uma vez. Nesse sentido, o volume ainda sinaliza ser expressivo, mas pode ser levemente menor quando considerado o volume de cidadãos deduplicados dentro dessas comunidades brasileiras de teorias da conspiração.

### 2.2. Tratamento de dados

Com todos os grupos e canais brasileiros de teorias da conspiração no Telegram extraídos, foi realizada uma classificação manual considerando o título e a descrição da comunidade. Caso houvesse menção explícita no título ou na descrição da comunidade a alguma temática, esta foi classificada entre: (i) "Anticiência"; (ii) "Anti-Woke e Gênero"; (iii) "Antivax"; (iv) "Apocalipse e Sobrevivencialismo"; (v) "Mudanças Climáticas"; (vi) Terra Plana; (vii) "Globalismo"; (viii) "Nova Ordem Mundial"; (ix) "Ocultismo e Esoterismo"; (x) "Off Label e Charlatanismo"; (xi) "QAnon"; (xii) "Reptilianos e Criaturas"; (xiii)

“Revisionismo e Discurso de Ódio”; (xiv) “OVNI e Universo”. Caso não houvesse nenhuma menção explícita relacionada às temáticas no título ou na descrição da comunidade, esta foi classificada como (xv) “Conspiração Geral”. No Quadro a seguir, podemos observar as métricas relacionadas à classificação dessas comunidades de teorias da conspiração no Brasil.

**Tabela 01.** Comunidades de teorias da conspiração no Brasil (métricas até agosto de 2024)

| **Categorias** | **Grupos** | **Usuários** | **Publicações** | **Comentários** | **Total** |
|---|---|---|---|---|---|
| **Anticiência** | 22 | 58.138 | 187.585 | 784.331 | 971.916 |
| **Anti-*Woke* e Gênero** | 43 | 154.391 | 276.018 | 1.017.412 | 1.293.430 |
| **Antivacinas (*Antivax*)** | 111 | 239.309 | 1.778.587 | 1.965.381 | 3.743.968 |
| **Apocalipse e Sobrevivência** | 33 | 109.266 | 915.584 | 429.476 | 1.345.060 |
| **Mudanças Climáticas** | 14 | 20.114 | 269.203 | 46.819 | 316.022 |
| **Terraplanismo** | 33 | 38.563 | 354.200 | 1.025.039 | 1.379.239 |
| **Conspirações Gerais** | 127 | 498.190 | 2.671.440 | 3.498.492 | 6.169.932 |
| **Globalismo** | 41 | 326.596 | 768.176 | 537.087 | 1.305.263 |
| **Nova Ordem Mundial (NOM)** | 148 | 329.304 | 2.411.003 | 1.077.683 | 3.488.686 |
| **Ocultismo e Esoterismo** | 39 | 82.872 | 927.708 | 2.098.357 | 3.026.065 |
| **Medicamentos *off label*** | 84 | 201.342 | 929.156 | 733.638 | 1.662.794 |
| **QAnon** | 28 | 62.346 | 531.678 | 219.742 | 751.420 |
| **Reptilianos e Criaturas** | 19 | 82.290 | 96.262 | 62.342 | 158.604 |
| **Revisionismo e Ódio** | 66 | 34.380 | 204.453 | 142.266 | 346.719 |
| **OVNI e Universo** | 47 | 58.912 | 862.358 | 406.049 | 1.268.407 |
| **Total** | **855** | **2.296.013** | **13.183.411** | **14.044.114** | **27.227.525** |

Fonte: Elaboração própria (2024).

Com esse volume de dados extraídos, foi possível segmentar para apresentar neste estudo apenas comunidades e conteúdos referentes às temáticas de apocalipse, sobrevivencialismo, ocultismo e esoterismo. Em paralelo, as demais temáticas de comunidades brasileiras de teorias da conspiração também contaram com estudos elaborados para a caracterização da extensão e da dinâmica da rede, estes sendo disponibilizados abertamente e originalmente no arXiv da Cornell University.

Além disso, cabe citar que apenas foram extraídas comunidades abertas, isto é, não apenas identificáveis publicamente, mas também sem necessidade de solicitação para acessar ao conteúdo, estando aberto para todo e qualquer usuário com alguma conta do Telegram sem que este necessite ingressar no grupo ou canal. Além disso, em respeito à legislação brasileira e especialmente da Lei Geral de Proteção de Dados Pessoais (LGPD), ou Lei nº 13.709/2018, que trata do controle da privacidade e do uso/tratamento de dados pessoais, todos os dados extraídos foram anonimizados para a realização de análises e investigações. Dessa forma, nem mesmo a identificação das comunidades é possível por meio deste estudo, estendendo aqui a privacidade do usuário ao anonimizar os seus dados para além da própria comunidade à qual ele se submeteu ao estar em um grupo ou canal público e aberto no Telegram.

### 2.3. Abordagens para análise de dados

Totalizando 72 comunidades selecionadas nas temáticas de apocalipse, sobrevivencialismo, ocultismo e esoterismo, contendo 4.371.125 publicações e 192.138 usuários somados, quatro abordagens serão utilizadas para investigar as comunidades de teorias da conspiração selecionadas para o escopo do estudo. Tais métricas são detalhadas no Quadro a seguir:

**Tabela 02.** Comunidades selecionadas para análise (métricas até agosto de 2024)

| Categorias | Grupos | Usuários | Publicações | Comentários | Total |
|---|---|---|---|---|---|
| **Apocalipse e Sobrevivência** | 33 | 109.266 | 915.584 | 429.476 | 1.345.060 |
| **Ocultismo e Esoterismo** | 39 | 82.872 | 927.708 | 2.098.357 | 3.026.065 |
| **Total** | **72** | **192.138** | **1.843.292** | **2.527.833** | **4.371.125** |

Fonte: Elaboração própria (2024).

**(i) Rede:** com a elaboração de um algoritmo próprio para a identificação de mensagens que contenham o termo de “t.me/” (de convite para entrarem em outras comunidades), propomos apresentar volumes e conexões observadas sobre como **(a)** as comunidades da temática de apocalipse, sobrevivencialismo, ocultismo e esoterismo circulam convites para que os seus usuários conheçam mais grupos e canais da mesma temática, reforçando os sistemas de crença que comungam; e como **(b)** essas mesmas comunidades circulam convites para que os seus usuários conheçam grupos e canais que tratem de outras teorias da conspiração, distintas de seu propósito explícito. Esta abordagem é interessante para observar se essas comunidades utilizam a si próprias como fonte de legitimação e referência e/ou se embasam-se em demais temáticas de teorias da conspiração, inclusive abrindo portas para que seus usuários conheçam outras conspirações. Além disso, cabe citar o estudo de Rocha *et al.* (2024) em que uma abordagem de identificação de rede também foi aplicada em comunidades do Telegram, porém observando conteúdos similares a partir de um ID gerado para cada mensagem única e suas similares;

**(ii) Séries temporais:** utiliza-se da biblioteca “Pandas” (McKinney, 2010) para organizar os data frames de investigação, observando **(a)** o volume de publicações ao longo dos meses; e **(b)** o volume de engajamento ao longo dos meses, considerando metadados de visualizações, reações e comentários coletados na extração; Além da volumetria, a biblioteca “Plotly” (Plotly Technologies Inc., 2015) viabilizou a representação gráfica dessa variação;

**(iii) Análise de conteúdo:** além da análise geral de palavras com identificação das frequências, são aplicadas séries temporais na variação das palavras mais frequentes ao longo dos semestres — observando entre maio de 2016 (primeiras publicações) até agosto de 2024 (realização deste estudo). E com as bibliotecas “Pandas” (McKinney, 2010) e “WordCloud” (Mueller, 2020), os resultados são apresentados tanto volumetricamente quanto graficamente;

**(iv) Sobreposição de agenda temática:** seguindo a abordagem proposta por Silva & Sátiro (2024) para identificação de sobreposição de agenda temática em comunidades do Telegram, utilizamos o modelo “BERTopic” (Grootendorst, 2020). O BERTopic é um algoritmo de modelagem de tópicos que facilita a geração de representações temáticas a partir de grandes quantidades de textos. Primeiramente, o algoritmo extrai embeddings dos documentos usando modelos transformadores de sentenças, como o “all-MiniLM-L6-v2”. Em seguida, essas embeddings têm sua dimensionalidade reduzida por técnicas como “UMAP”, facilitando o processo de agrupamento. A clusterização é realizada usando “HDBSCAN”, uma técnica baseada em densidade que identifica clusters de diferentes formas e tamanhos, além de detectar outliers. Posteriormente, os documentos são tokenizados e representados em uma estrutura de bag-of-words, que é normalizada (L1) para considerar as diferenças de tamanho entre os clusters. A representação dos tópicos é refinada usando uma versão modificada do “TF-IDF”, chamada “Class-TF-IDF”, que considera a importância das palavras dentro de cada cluster (Grootendorst, 2020). Cabe destacar que, antes de aplicar o BERTopic, realizamos a limpeza da base removendo “stopwords” em português, por meio da biblioteca “NLTK” (Loper & Bird, 2002). Para a modelagem de tópicos, utilizamos o backend “loky” para otimizar o desempenho durante o ajuste e a transformação dos dados.

Em síntese, a metodologia aplicada compreendeu desde a extração de dados com a ferramenta própria autoral TelegramScrap (Silva, 2023), até o tratamento e a análise de dados coletados, utilizando diversas abordagens para identificar e classificar comunidades de teorias da conspiração brasileiras no Telegram. Cada uma das etapas foi cuidadosamente implementada para garantir a integridade dos dados e o respeito à privacidade dos usuários, conforme a legislação brasileira prevê. A seguir, serão apresentados os resultados desses dados, com o intuito de revelar as dinâmicas e as características das comunidades estudadas.

## 3. Resultados

A seguir, os resultados são detalhados na ordem prevista na metodologia, iniciando com a caracterização da rede e suas fontes de legitimação e referência, avançando para as séries temporais, incorporando a análise de conteúdo e concluindo com a identificação de sobreposição de agenda temática dentre as comunidades de teorias da conspiração.

### 3.1. Rede

A análise das redes que envolvem apocalipse, sobrevivencialismo, ocultismo e esoterismo revela uma estrutura intricada e interconectada de comunidades que promovem e perpetuam narrativas de medo, desconfiança e preparação para catástrofes iminentes. Na primeira figura, a rede interna entre essas temáticas mostra como as comunidades de sobrevivencialismo e apocalipse se entrelaçam com aquelas voltadas para o ocultismo e esoterismo. A presença de dois grandes núcleos na rede evidencia a existência de pontos de centralização onde essas narrativas se encontram e se reforçam mutuamente. A interdependência dessas comunidades sugere que o medo de eventos cataclísmicos e a busca por respostas esotéricas andam lado a lado, criando um ambiente em que a incerteza é amplificada e a adesão a essas crenças é intensificada.

A segunda figura foca nas comunidades que servem como portas de entrada para narrativas apocalípticas e de sobrevivencialismo. A forte conexão dessas comunidades com temas como ocultismo e conspirações gerais indica que o sobrevivencialismo não só atrai novos membros interessados em cenários de fim do mundo, mas também os introduz a uma rede mais ampla de teorias conspiratórias. Essas comunidades atuam como *hubs* centrais, capturando a atenção de indivíduos que, a partir de uma preocupação inicial com a sobrevivência, acabam sendo expostos a um leque amplo de ideias conspiratórias e esotéricas.

Na terceira figura, a rede de comunidades que operam como portas de saída demonstra como as discussões sobre apocalipse e sobrevivencialismo frequentemente levam os indivíduos a explorar outras áreas de desinformação, como a Nova Ordem Mundial e o ocultismo. A centralidade das comunidades de ocultismo na rede sugere que essas discussões sobre sobrevivência extrema muitas vezes servem como um ponto de partida para a aceitação de narrativas sobre ordens secretas e conspirações globais. Isso reflete como as preocupações com eventos cataclísmicos podem ser manipuladas para introduzir e perpetuar outras teorias conspiratórias, expandindo a amplitude e a complexidade das crenças dos seguidores.

O gráfico de fluxo de links de convites entre comunidades de apocalipse e sobrevivência destaca a forte conexão dessas narrativas com a Nova Ordem Mundial (NOM) e Antivacinas, sugerindo que essas temáticas não apenas atraem, mas também amplificam a disseminação de desinformação relacionada a conspirações globais e saúde pública. O Apocalipse e Sobrevivência funcionam como catalisadores para a aceitação de uma visão de mundo onde o colapso social é inevitável, alimentando um ciclo de medo e desconfiança que predispõe os indivíduos a abraçar outras formas de desinformação.

Por fim, o gráfico de fluxo de links entre comunidades de ocultismo e esoterismo revela como essas temáticas estão profundamente interligadas com outras narrativas conspiratórias, como Conspirações Gerais e OVNI e Universo. O ocultismo, frequentemente associado a uma rejeição da ciência moderna, também emite convites significativos para Medicamentos *off label*, indicando uma tendência de unir o misticismo com a desconfiança na medicina convencional. Essa interconexão reflete um padrão de reforço mútuo entre crenças esotéricas e desinformação científica, criando um círculo vicioso de desconfiança institucional que perpetua uma visão de mundo onde o conhecimento alternativo é preferido em detrimento da autoridade científica estabelecida.

[ versão português / english above ]

**Figura 01.** Rede interna entre apocalipse, sobrevivencialismo, ocultismo e esoterismo

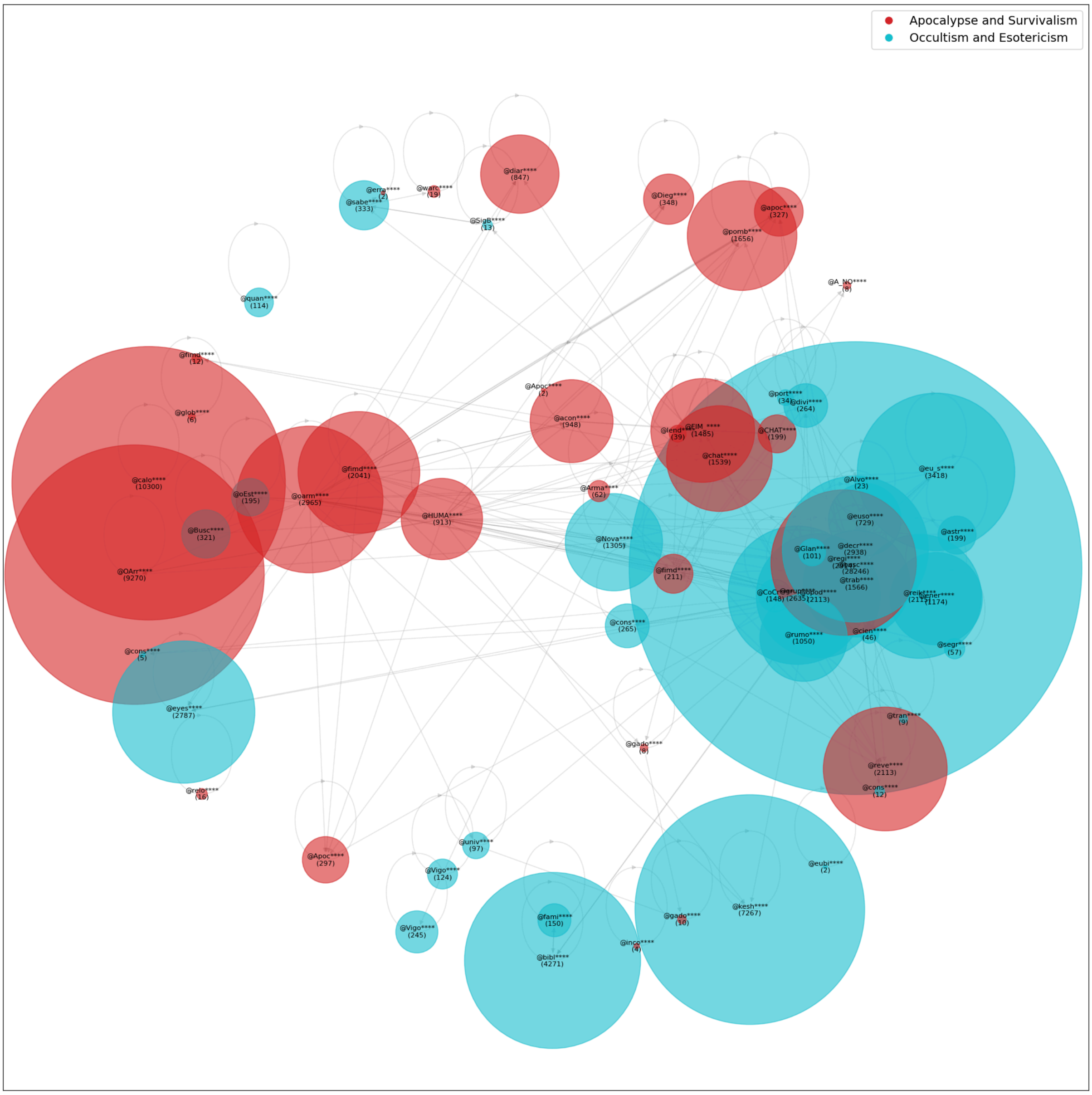

Fonte: Elaboração própria (2024).

A figura expõe uma rede interna que conecta comunidades focadas em apocalipse, sobrevivencialismo, ocultismo e esoterismo. A rede, caracterizada por dois grandes núcleos, reflete uma interligação entre a preparação para eventos catastróficos e crenças ocultistas. Esses temas, embora distintos, se cruzam frequentemente, criando um ciclo onde o medo do apocalipse é alimentado por narrativas esotéricas e vice-versa. Os grandes nós no gráfico indicam comunidades que atuam como centros de influência, onde ideias sobre a sobrevivência em um mundo pós-apocalíptico são reforçadas por práticas ocultistas. A existência de várias comunidades periféricas sugere que essas ideias estão se espalhando para além do núcleo central, mas ainda fortemente influenciadas pelos principais disseminadores dessas crenças. A interdependência entre sobrevivencialismo e ocultismo cria um ambiente onde a incerteza e o medo são continuamente exacerbados, alimentando a necessidade de pertencer a essas comunidades para encontrar soluções e respostas.

**Figura 02.** Rede de comunidades que abrem portas para a temática (porta de entrada)

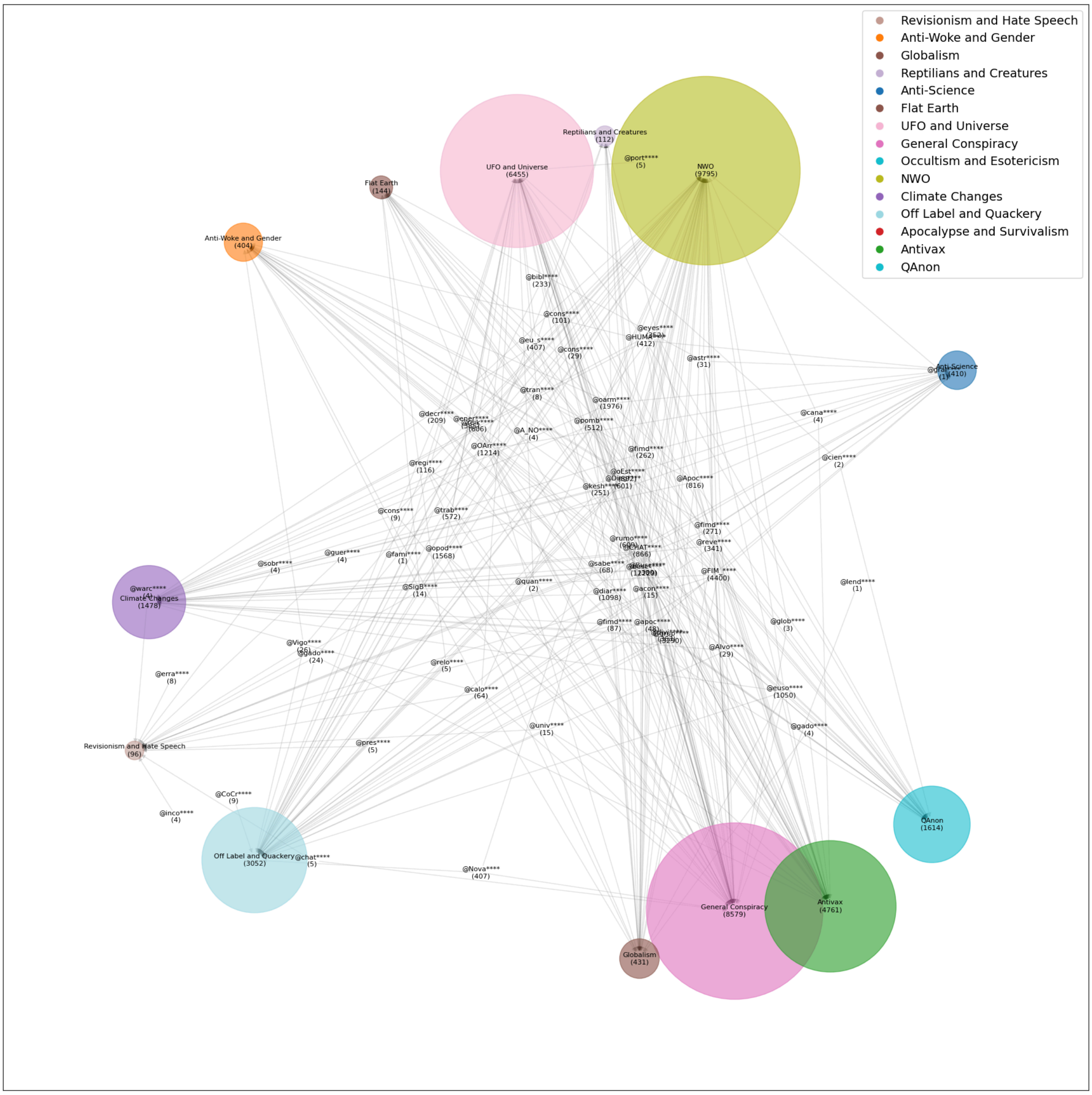

Fonte: Elaboração própria (2024).

Neste gráfico, a figura destaca as comunidades que servem como porta de entrada para narrativas apocalípticas e de sobrevivencialismo. As grandes esferas vermelhas e azuis representam comunidades centrais que atraem indivíduos interessados em cenários de fim do mundo e práticas de sobrevivência. A interconexão com outras comunidades, como ocultismo e teorias de conspiração gerais, sugere que essas ideias se alimentam mutuamente. Assim, o sobrevivencialismo não apenas atrai novos membros com interesses em apocalipse, mas também pode servir como um ponto de entrada para outras teorias, como a crença em poderes ocultos e a conspiração global. A estrutura dessa rede reflete a forma como narrativas de medo e preparação extrema podem expandir o horizonte de conspirações dos indivíduos, conectando-os a um conjunto mais amplo de teorias.

**Figura 03.** Rede de comunidades cuja temática abre portas (porta de saída)

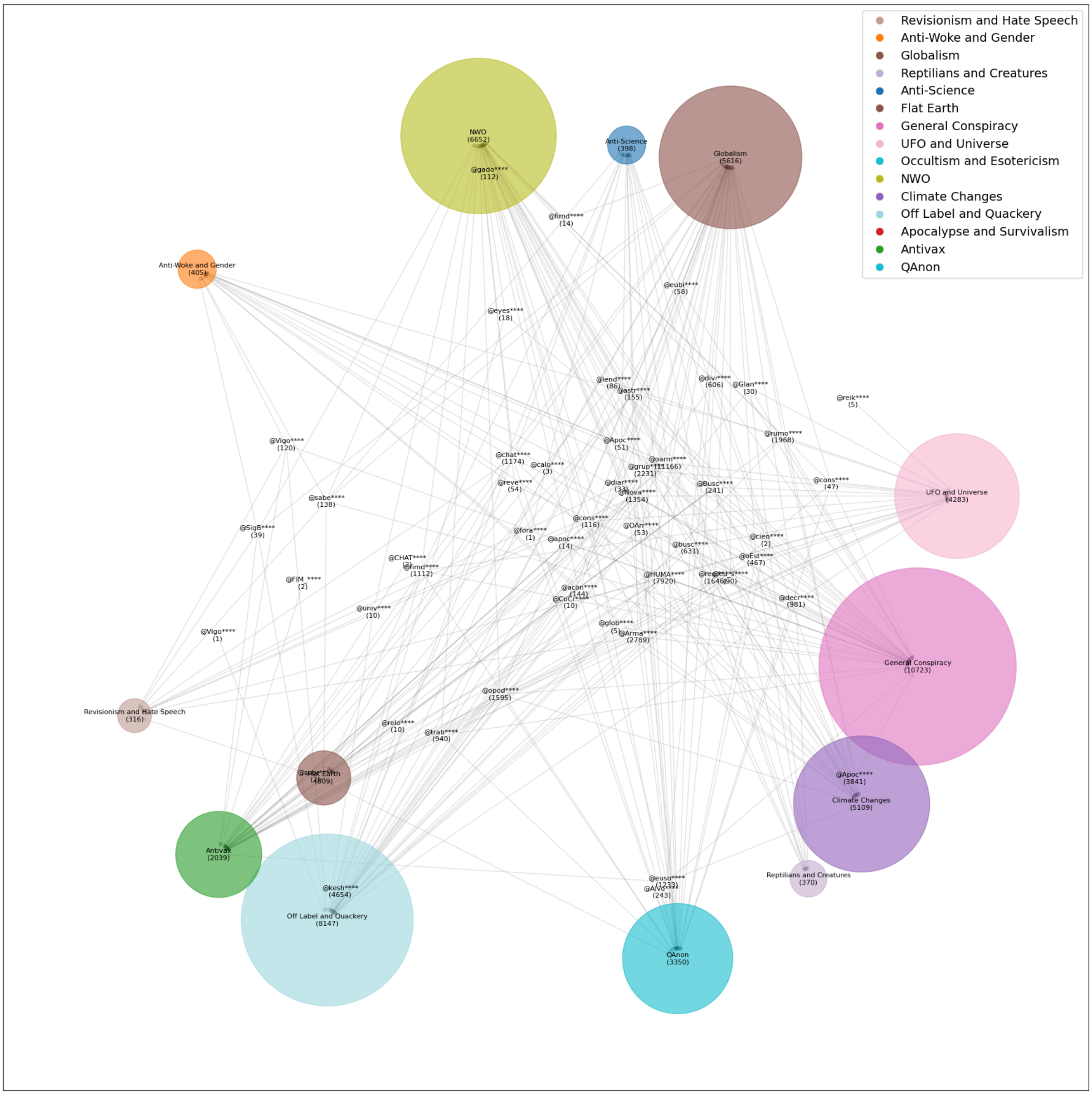

Fonte: Elaboração própria (2024).

Este gráfico explora as conexões entre comunidades focadas no apocalipse, sobrevivencialismo, ocultismo e esoterismo, revelando como esses temas interagem dentro do ecossistema conspiratório. As grandes bolhas de Ocultismo e Nova Ordem Mundial destacam-se como pontos centrais, sugerindo que discussões sobre sobrevivencialismo e apocalipse frequentemente conduzem os indivíduos para temas mais amplos e interconectados, como conspirações globais. A figura ilustra a transição fluida de uma preocupação com eventos cataclísmicos e sobrevivência para uma aceitação de narrativas sobre ordens secretas. Isso reflete como as preocupações com o futuro e a sobrevivência podem ser manipuladas para introduzir e perpetuar outras teorias conspiratórias, expandindo assim o alcance e a complexidade das crenças conspiratórias dos seguidores.

**Figura 04.** Fluxo de links de convites entre comunidades de apocalipse e sobrevivência

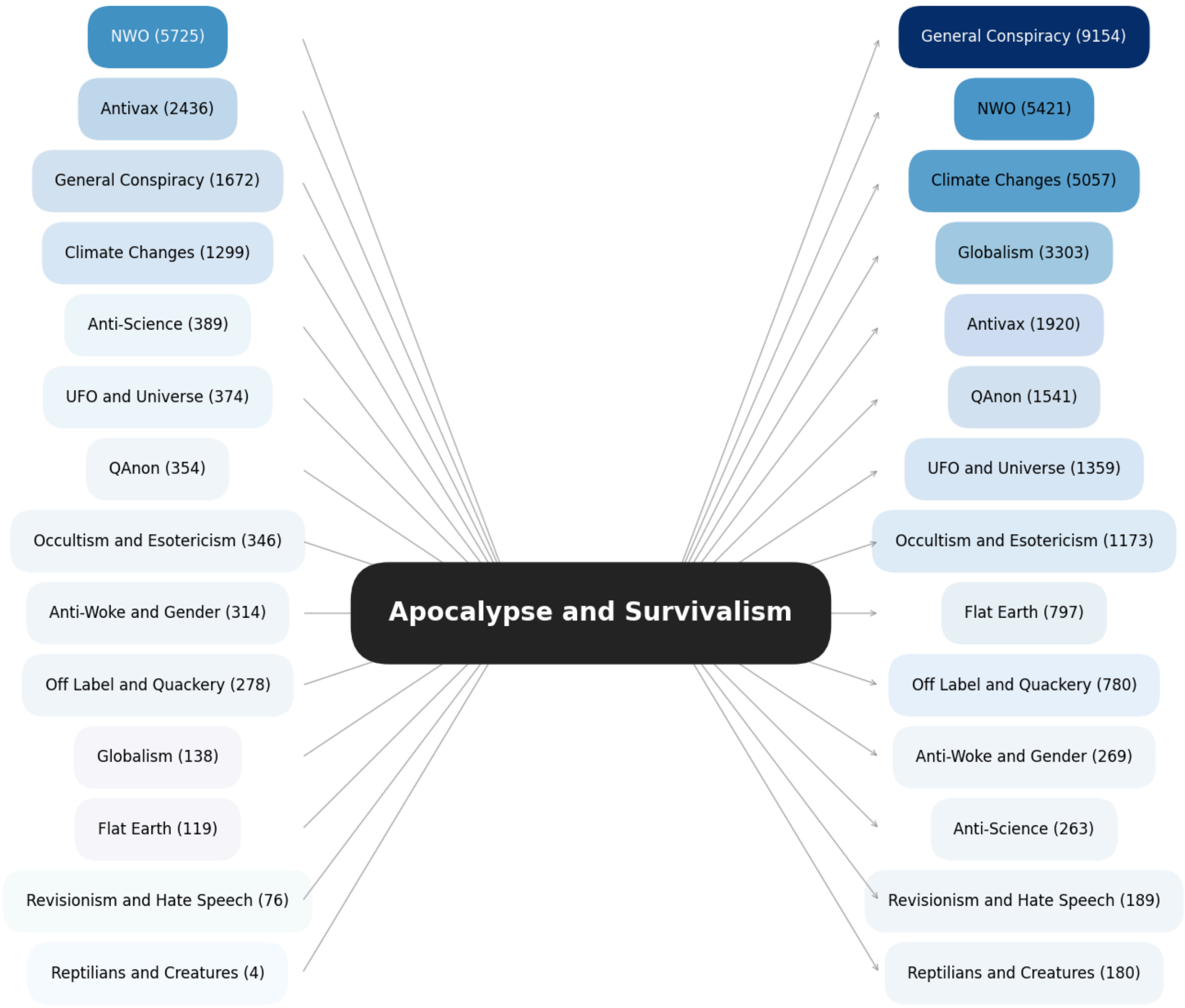

Fonte: Elaboração própria (2024).

O gráfico de Apocalipse e Sobrevivência apresenta uma forte conexão com a NOM (5.725 links) e Antivacinas (2.436 links), sugerindo que essas narrativas de sobrevivência extrema e preparação para catástrofes são frequentemente associadas a uma visão de mundo onde o colapso social é iminente e inevitável. Esses dados indicam que a narrativa apocalíptica serve como um meio eficaz de engajar os indivíduos em um estado constante de medo e alerta, que os predispõe a aceitar e propagar outras formas de desinformação. Ao analisar os convites emitidos para Conspirações Gerais (9.154 links) e NOM (5.421 links), é possível observar como essas comunidades funcionam como amplificadores de uma visão de mundo onde as crises globais são vistas não como eventos isolados, mas como parte de um plano maior de dominação e controle. O Apocalipse e Sobrevivência, portanto, serve como um catalisador para a aceitação de narrativas extremas, promovendo uma mentalidade de "nós contra eles" que é fundamental para a perpetuação e disseminação de teorias conspiratórias.

**Figura 05.** Fluxo de links de convites entre comunidades de ocultismo e esoterismo

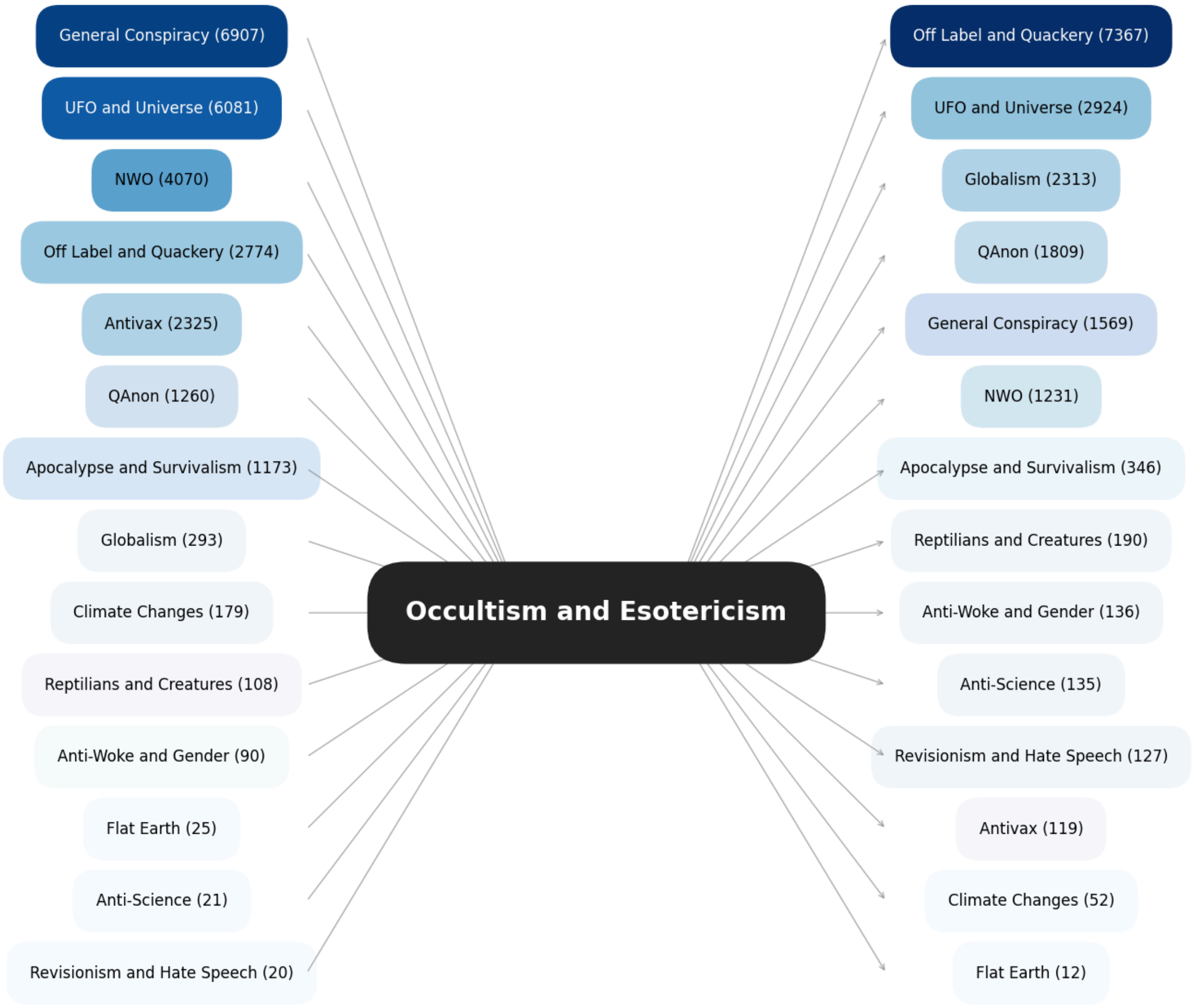

Fonte: Elaboração própria (2024).

O gráfico de Ocultismo e Esoterismo revela que essas comunidades estão profundamente entrelaçadas com Conspirações Gerais (6.907 links) e OVNI e Universo (6.081 links), o que sugere que o interesse pelo oculto e esotérico é frequentemente um ponto de entrada para narrativas conspiratórias mais amplas. Essa interconexão não é apenas uma sobreposição de interesses, mas sim uma fusão de ideologias que se alimentam mutuamente, criando um ambiente onde a crença em poderes ocultos e conspirações globais se reforçam e legitimam. Por outro lado, o fato de Ocultismo e Esoterismo emitir convites significativos para Medicamentos *off label* (7.367 links) e OVNI e Universo (2.924 links) indica uma tendência em expandir as fronteiras da crença, unindo o misticismo com a desconfiança em relação à medicina convencional e à ciência moderna. Essa convergência reflete um padrão onde o esoterismo e a desinformação científica caminham juntos, formando uma aliança que se baseia na rejeição da autoridade e na busca por verdades alternativas. Assim, Ocultismo e Esoterismo não é apenas uma temática isolada, mas um núcleo que conecta e expande diferentes formas de desinformação, criando um círculo vicioso de crenças que perpetuam a desconfiança institucional.

### 3.2. Séries temporais

No próximo gráfico, analisaremos como as menções a Apocalipse e Sobrevivência, e Ocultismo e Esoterismo evoluíram nos últimos anos. Observa-se que, com o advento da Pandemia da COVID-19 e as crises globais subsequentes, houve um crescimento exponencial das menções a essas temáticas, especialmente entre 2020 e 2021. Este período foi marcado por um aumento das incertezas globais, o que pode ter contribuído para a proliferação de teorias apocalípticas e a busca por explicações místicas. Embora ambas as temáticas apresentem uma leve queda após os picos, os níveis de interesse permanecem elevados, sugerindo que, em tempos de crise, as narrativas apocalípticas e esotéricas seguem a crescer.

**Figura 06.** Gráfico de linhas do período

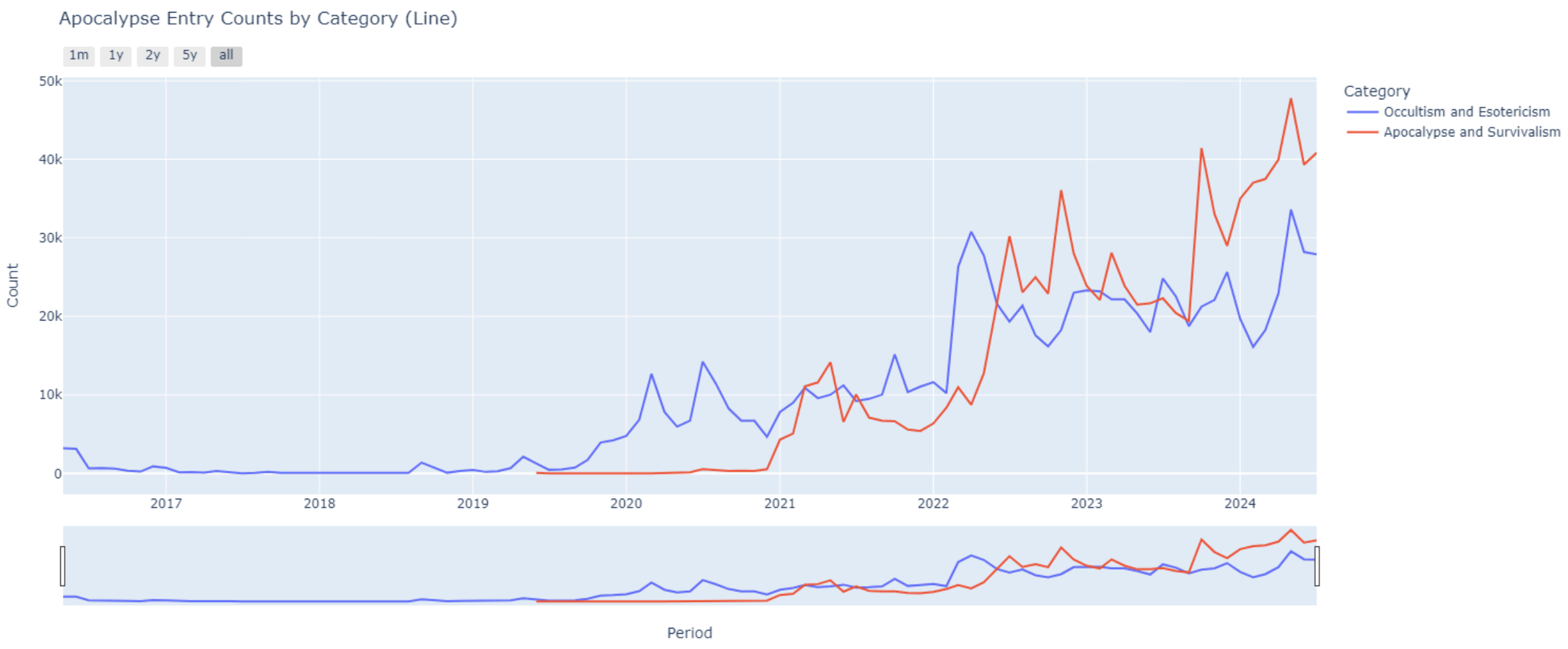

Fonte: Elaboração própria (2024).

Para Apocalipse e Sobrevivência, as menções cresceram 1.500% entre 2020 e o pico em abril de 2021, de aproximadamente 1.000 para 35.000 menções. Este aumento significativo está correlacionado com o aumento de incertezas globais, como novas variantes da COVID-19 e crises econômicas globais, que incentivaram discussões sobre cenários apocalípticos. Ocultismo e Esoterismo teve um aumento de 1.200% no mesmo período, subindo de cerca de 2.000 para 28.000 menções em novembro de 2021. Isso mostra como a busca por explicações alternativas e místicas também cresceu durante períodos de incerteza, alimentando teorias apocalípticas. Após os picos, ambos os temas mostram uma queda de aproximadamente 25%, indicando uma leve desaceleração, mas ainda com níveis de interesse elevados em comparação ao período pré-Pandemia.

**Figura 07.** Gráfico de área absoluta do período

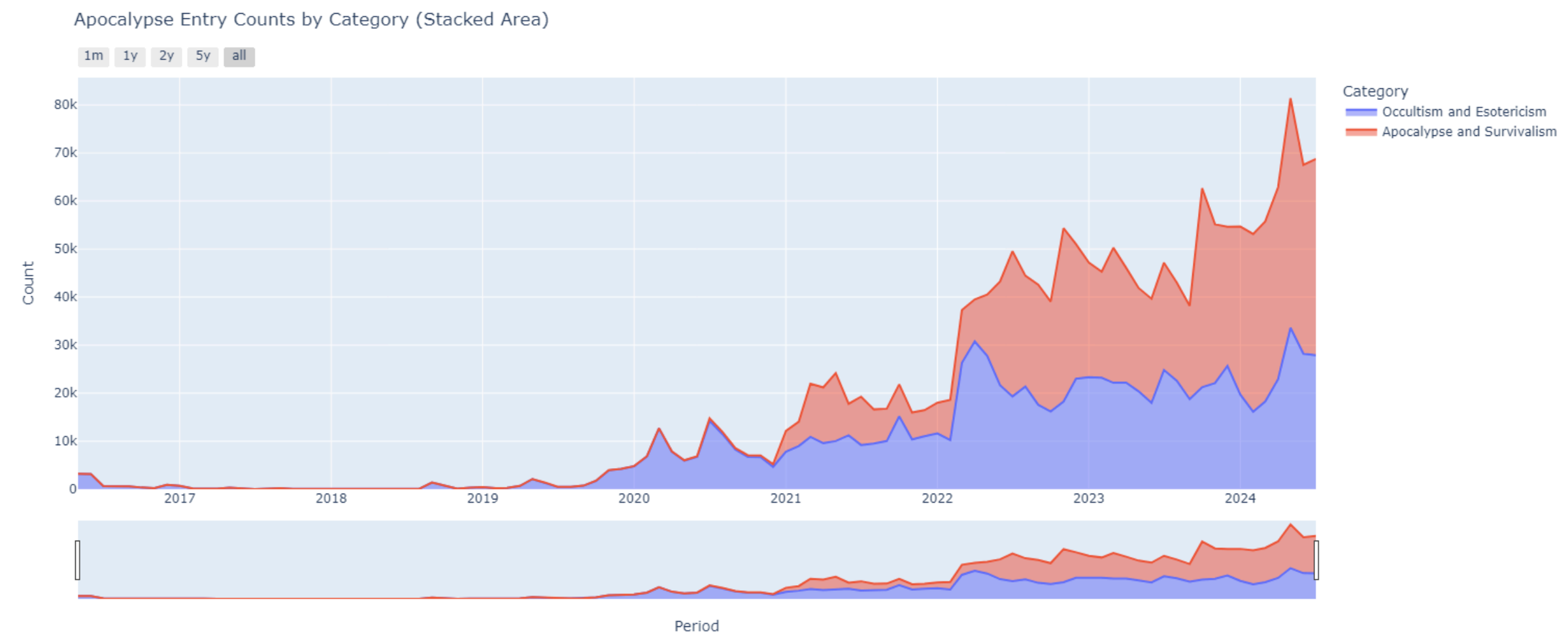

Fonte: Elaboração própria (2024).

O gráfico de área absoluta destaca o aumento contínuo das discussões em torno de Ocultismo e Esoterismo e Apocalipse e Sobrevivência. A partir de 2020, as menções a Apocalipse e Sobrevivência crescem substancialmente, com um pico em 2022, atingindo mais de 50 mil entradas. Isso reflete o aumento do medo e da incerteza provocados pela Pandemia da COVID-19, que alimentou narrativas apocalípticas e teorias sobre a necessidade de preparação para catástrofes. A Pandemia, assim como as crises econômicas e ambientais globais, contribuiu para a proliferação dessas teorias. O crescimento simultâneo de Ocultismo e Esoterismo também é significativo, sugerindo que em tempos de crise, as pessoas buscam explicações sobrenaturais ou alternativas, o que muitas vezes se conecta com narrativas de sobrevivência apocalíptica. Esses dois temas parecem se reforçar mutuamente, com adeptos de teorias esotéricas sendo atraídos por discursos apocalípticos e vice-versa.

**Figura 08.** Gráfico de área relativa do período

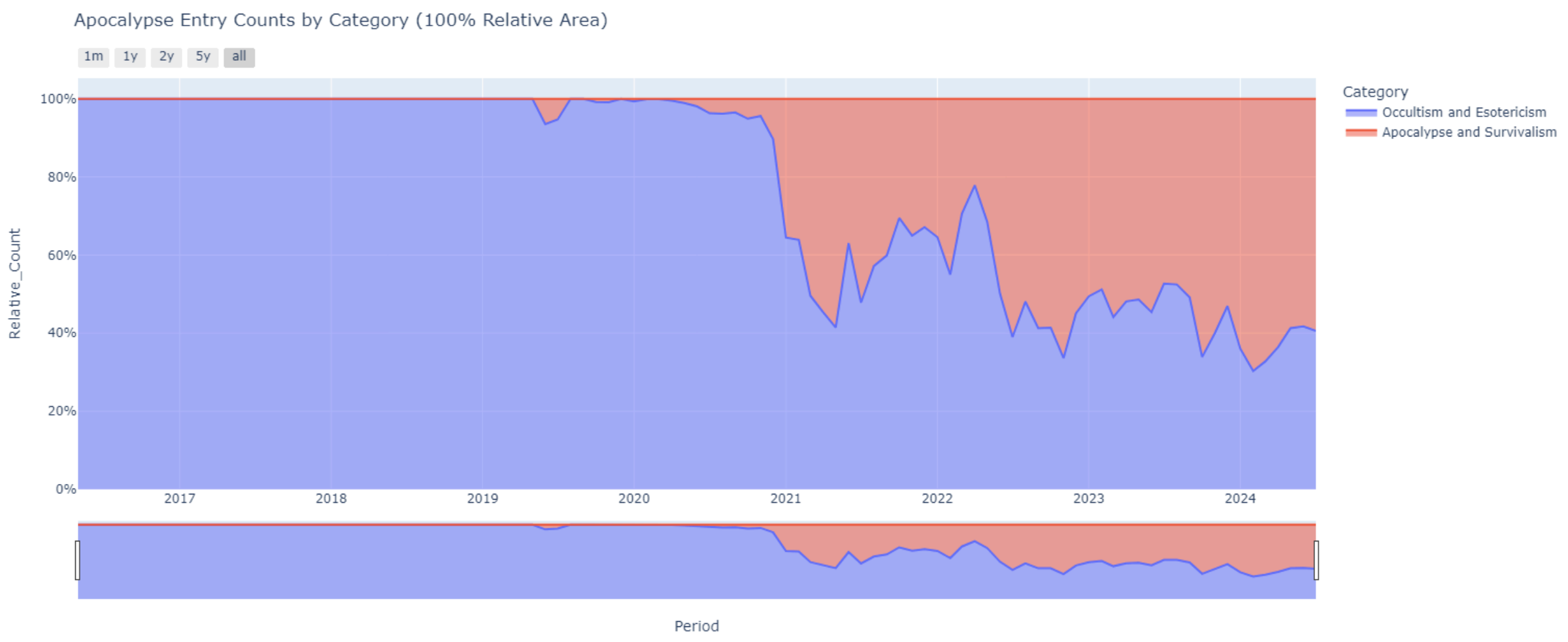

Fonte: Elaboração própria (2024).

O gráfico de área relativa revela que, embora Ocultismo e Esoterismo tenha sido mais dominante antes de 2020, Apocalipse e Sobrevivência passou a dominar as discussões a partir de 2021. Isso pode ser interpretado como um reflexo direto da Pandemia e da percepção de que eventos catastróficos globais exigem respostas extremas. O gráfico também indica que, embora Ocultismo e Esoterismo mantenha uma presença significativa, ele é cada vez mais absorvido dentro do contexto apocalíptico, sugerindo uma convergência de narrativas. O deslocamento no foco sugere que, em tempos de crise, as preocupações com sobrevivência e preparação para desastres se tornam mais proeminentes, enquanto o esoterismo e o ocultismo servem como bases explicativas para esses medos e preparações.

### 3.3. Análise de conteúdo

A análise de conteúdo das comunidades de apocalipse, ocultismo e esoterismo, por meio de nuvens de palavras, revela como certos conceitos e termos se destacam e se transformam ao longo do tempo dentro dessas discussões. As palavras mais frequentes, como “vida”, “luz”, “Deus” e “mundo”, indicam uma forte interseção entre temas espirituais e esotéricos, muitas vezes ligados a percepções de transformação pessoal e global. A presença de termos como “energia”, “alma” e “consciência” sugere uma busca por entendimento além da realidade material, refletindo uma tentativa de conectar eventos apocalípticos a um propósito maior ou a uma evolução espiritual. Além disso, o foco em palavras como “agora” e “verdade” demonstra a urgência percebida por esses grupos em relação a essas mudanças, apontando para uma narrativa que mistura o iminente com o eterno, o físico com o metafísico. Essa análise permite compreender não apenas os tópicos centrais dessas comunidades, mas também como essas ideias são disseminadas e reformuladas em resposta a eventos contemporâneos e mudanças culturais.

**Figura 09.** Nuvem de palavras consolidadas de apocalipse, ocultismo e esoterismo

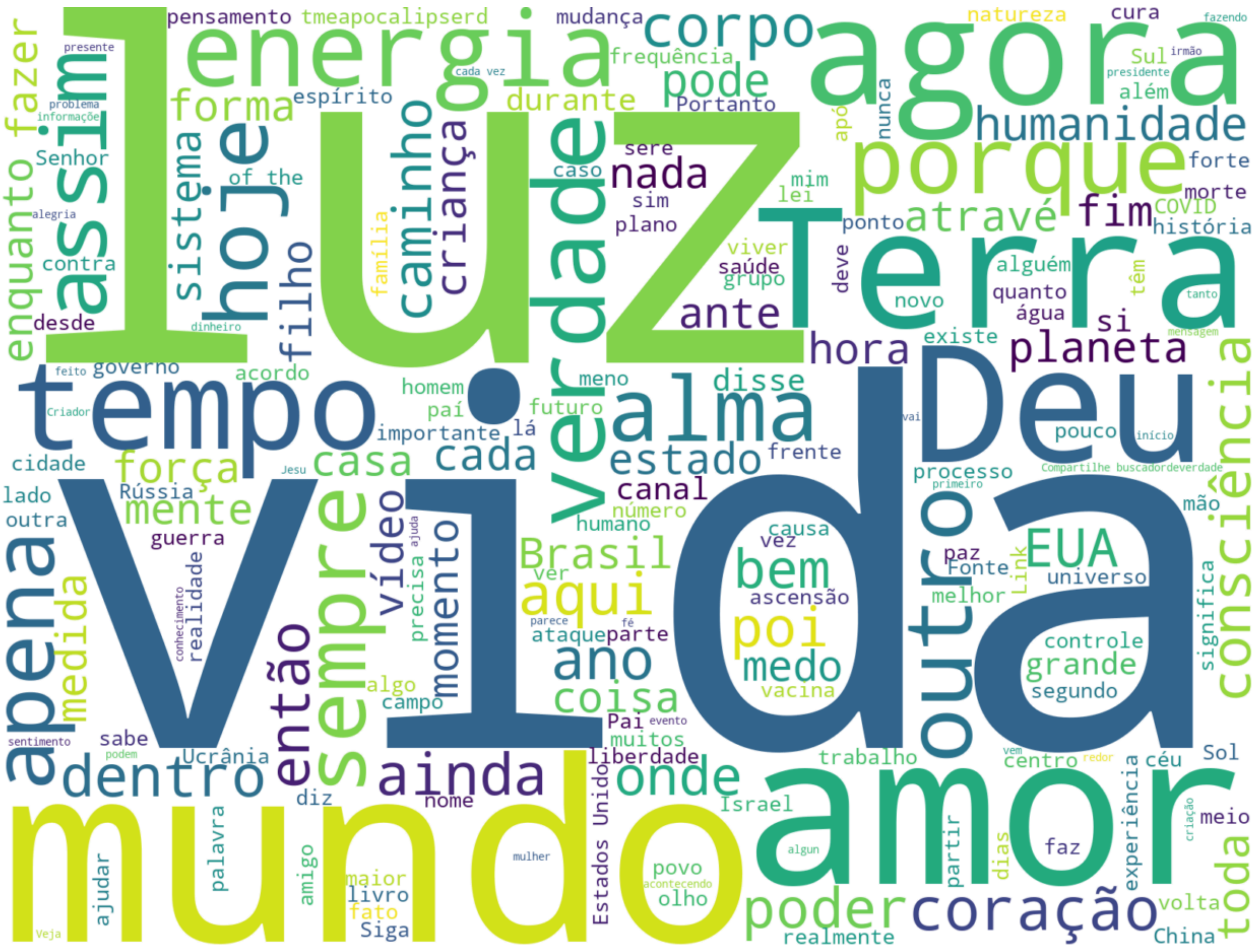

Fonte: Elaboração própria (2024).

A nuvem de palavras consolidada das discussões sobre apocalipse, ocultismo e esoterismo revela a centralidade de termos profundamente carregados de significado espiritual e existencial, como “vida”, “luz”, “Deus” e “mundo”. O destaque para a palavra “vida” sugere uma preocupação fundamental com a existência e o propósito, temas que permeiam as discussões dentro dessas comunidades. “Luz” e “energia” indicam uma ênfase em conceitos esotéricos de iluminação e força vital, frequentemente associados à ideia de transformação ou ascensão espiritual. A recorrência de “Deus” e “alma” aponta para uma tentativa de conectar eventos apocalípticos com uma narrativa de redenção ou julgamento divino, enquanto “mundo” e “verdade” refletem uma visão dicotômica entre o conhecido e o desconhecido, o visível e o invisível. A palavra “agora” sugere uma urgência nas discussões, como se esses eventos fossem iminentes e exigissem uma resposta imediata. Por outro lado, “medo” e “poder” indicam as emoções e forças em jogo nessas narrativas, onde o apocalipse é visto tanto como um tempo de destruição quanto de potencial renascimento.

[ versão português / english above ]

**Quadro 01.** Nuvem de palavras em série temporal de apocalipse e sobrevivência

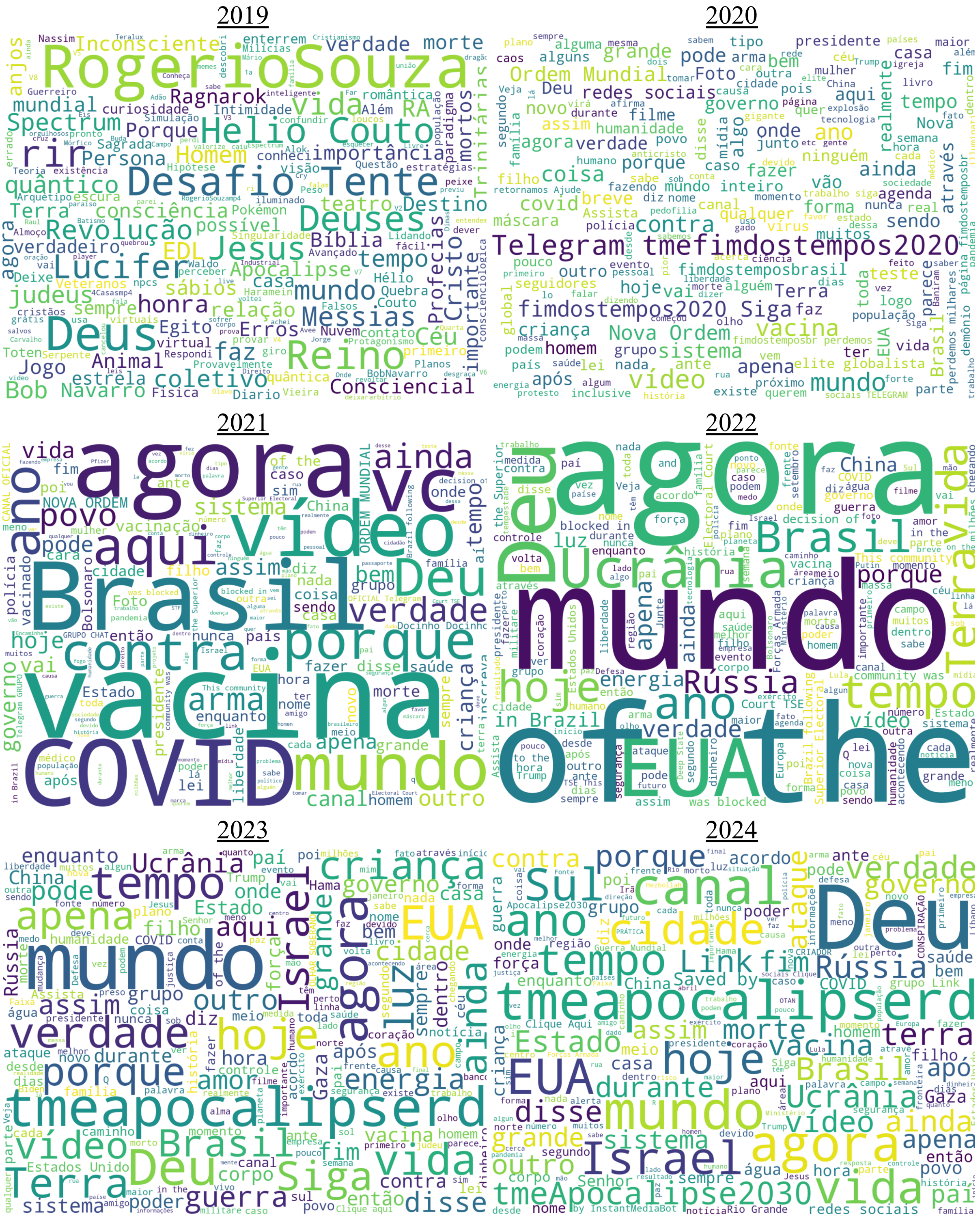

Fonte: Elaboração própria (2024).

O quadro de nuvens de palavras em série temporal sobre apocalipse e sobrevivência revela como os temas e preocupações dessas comunidades evoluíram de 2019 a 2024. Em 2019, termos como “Deus” e “Reino” sugerem uma forte conexão com narrativas religiosas de fim dos tempos, enquanto “vida” e “mundo” indicam a dualidade entre o presente e o que está por vir. Em 2020, com o advento da Pandemia da COVID-19, palavras como “fim” e “mundo” ganham destaque, refletindo o temor e a incerteza trazidos pela crise global. Em 2021 e 2022, “vacina” e “COVID” aparecem com frequência, indicando que a Pandemia se tornou um ponto central nas discussões apocalípticas, com a vacina sendo interpretada em alguns casos como parte de uma conspiração maior. Nos anos seguintes, o foco parece se expandir para incluir temas como “energia” e “sistema”, sugerindo uma crescente preocupação com questões de sobrevivência e autossuficiência em um mundo percebido como cada vez mais instável e perigoso. A continuidade de termos como “Deus” e “apocalipse” ao longo dos anos mostra como essas comunidades mantêm uma narrativa consistente de preparação para um fim iminente, adaptando-a às mudanças do contexto global.

**Quadro 02.** Nuvem de palavras em série temporal de ocultismo e esoterismo

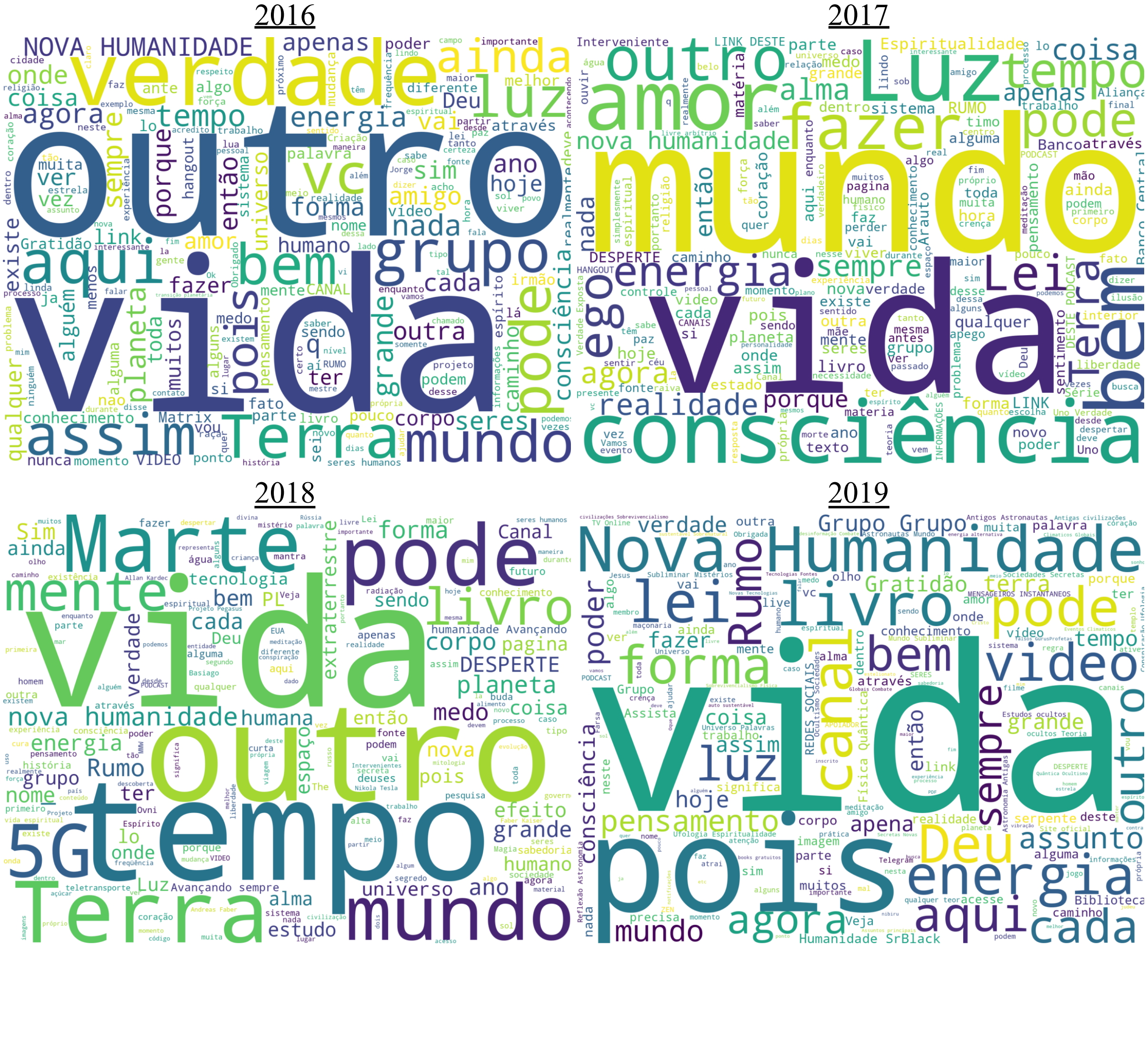

[ versão português / english above ]

2020

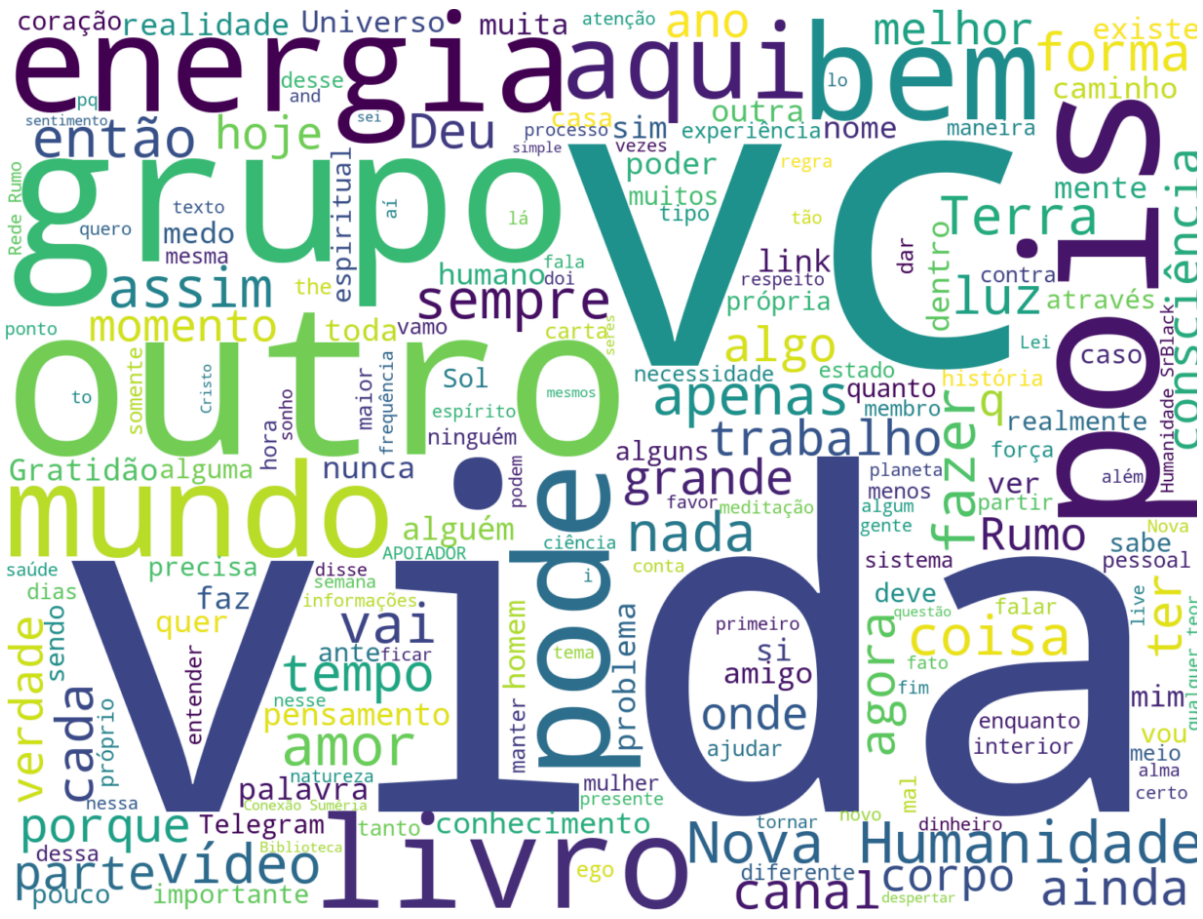

2021

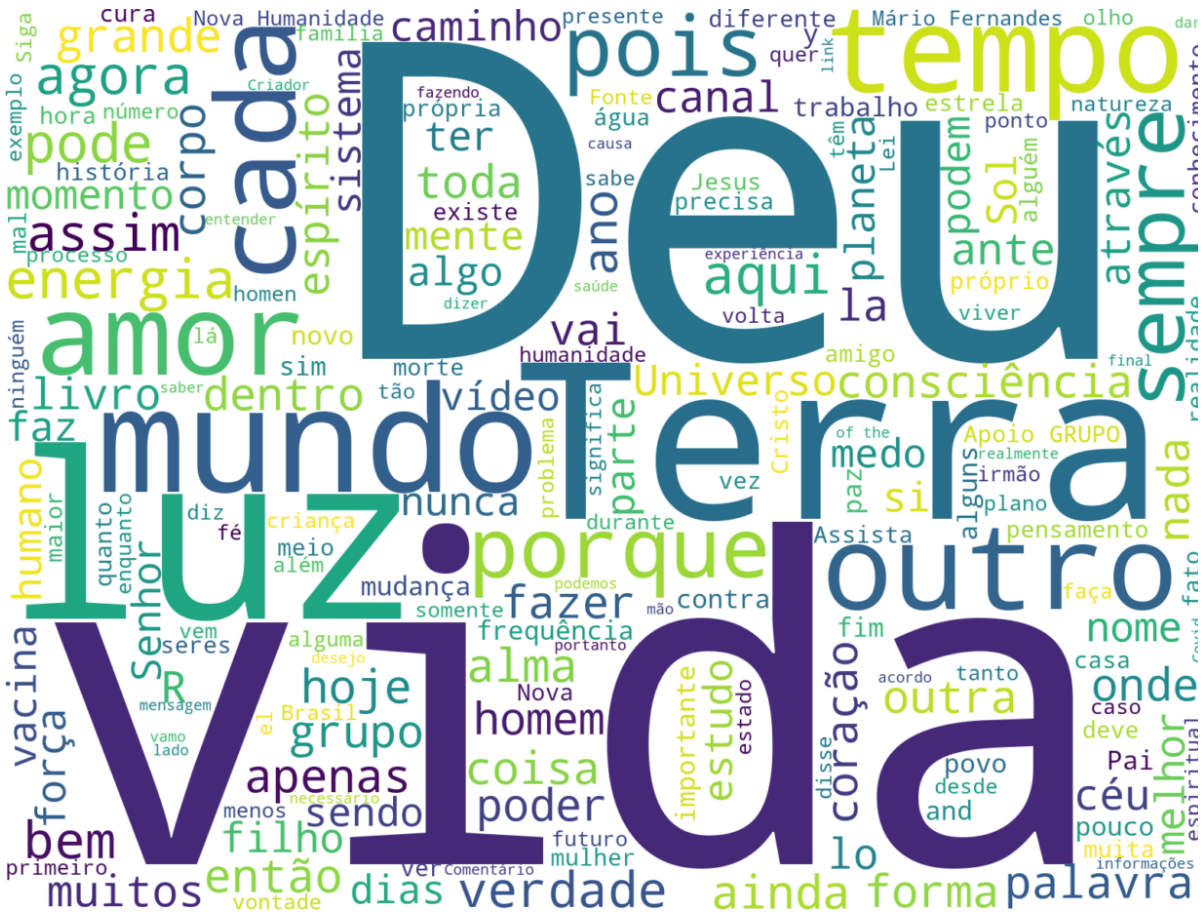

2022

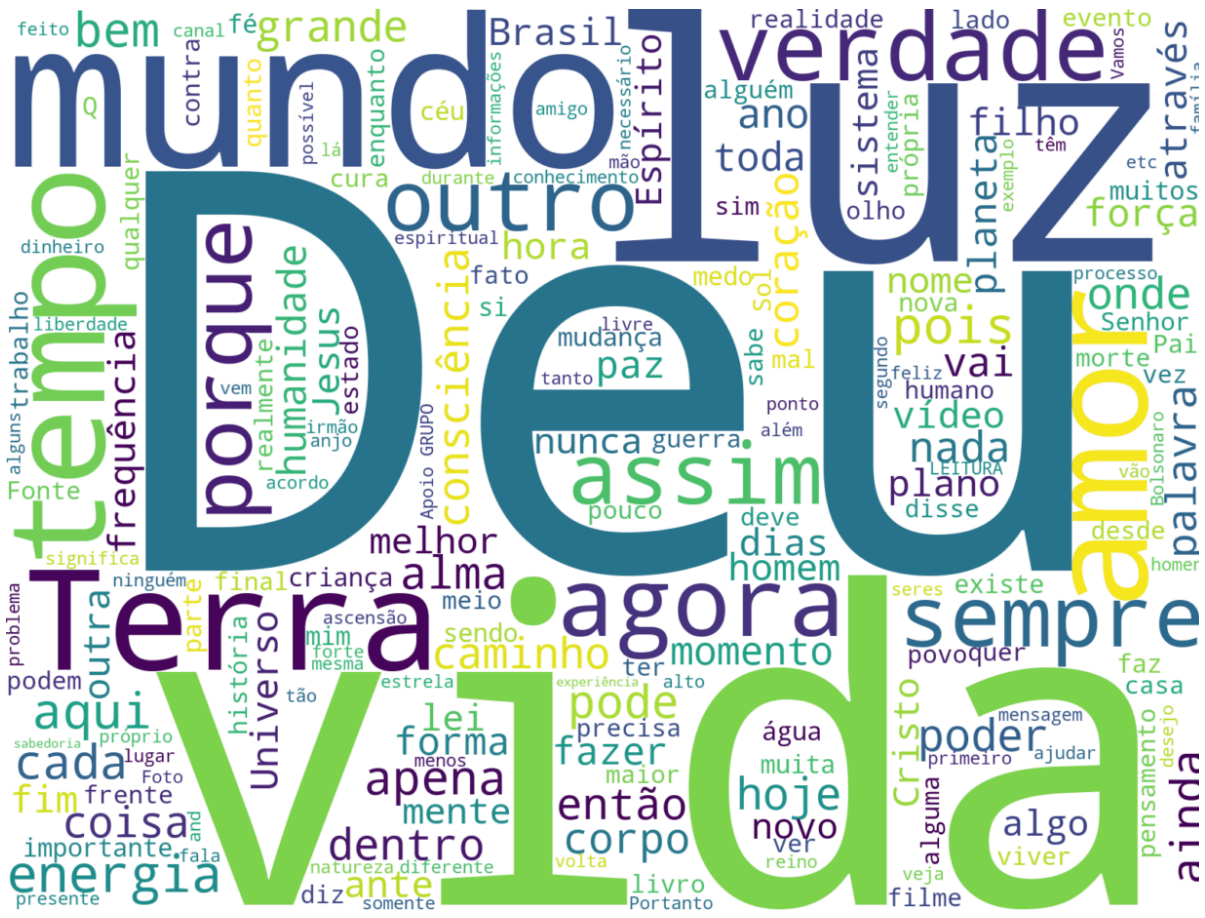

2023

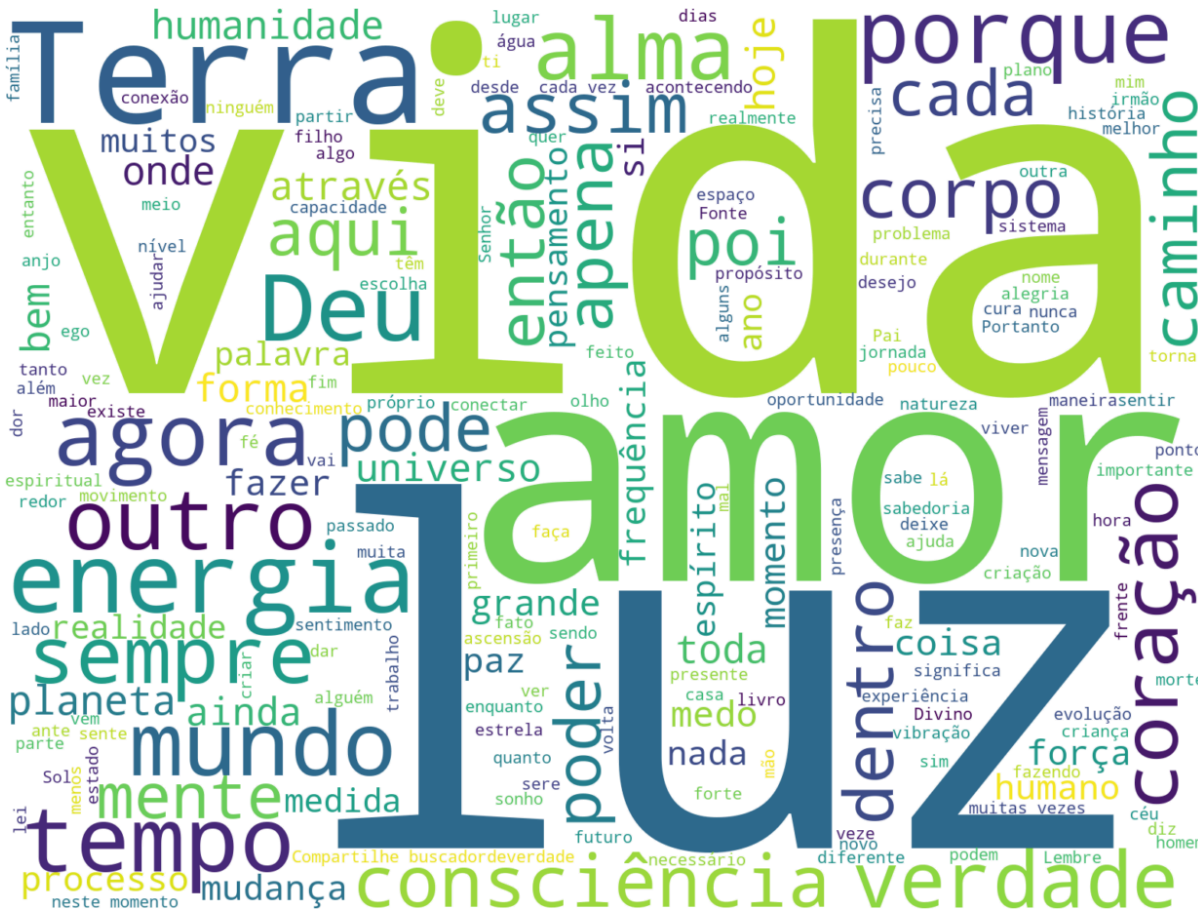

2024

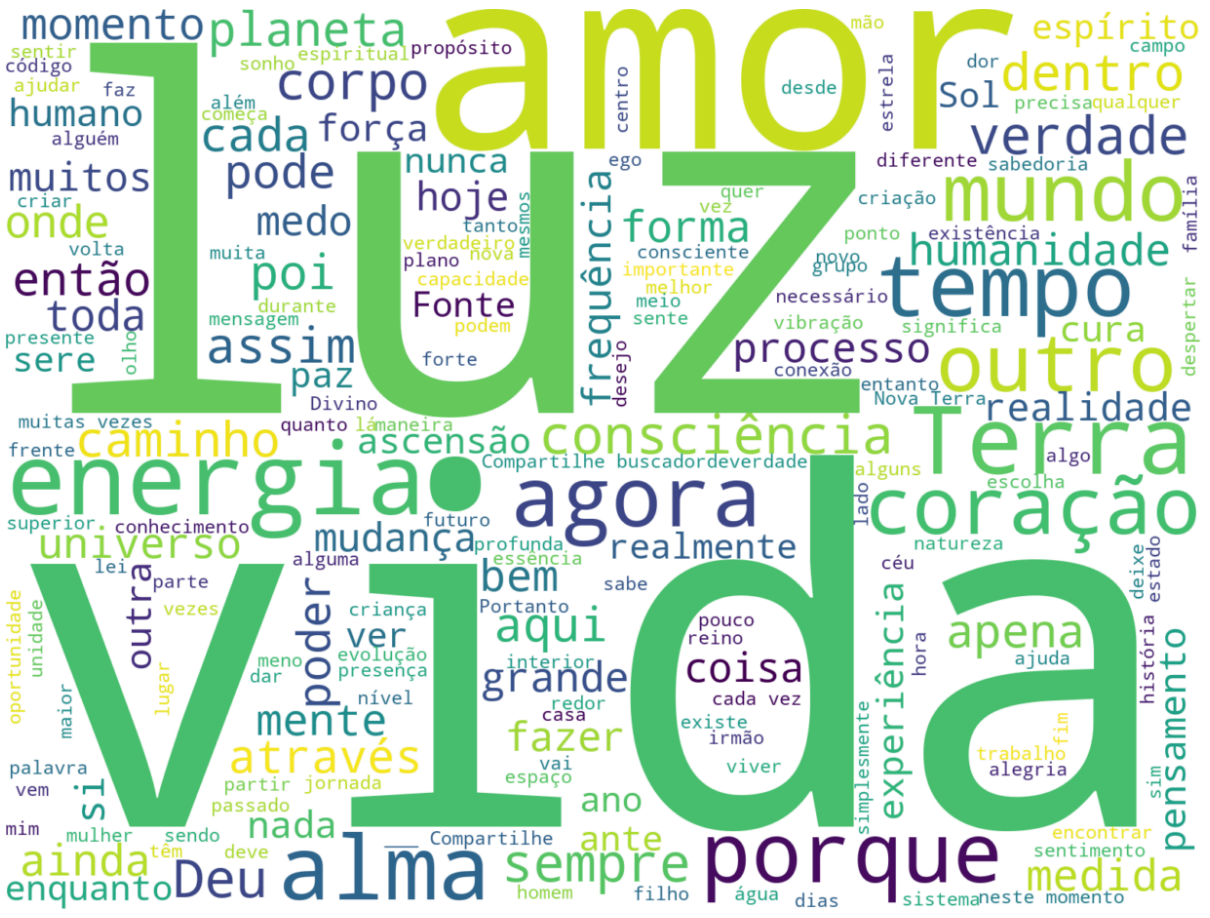

Fonte: Elaboração própria (2024).

O quadro de nuvens de palavras em série temporal sobre ocultismo e esoterismo destaca a persistência e evolução de temas espirituais e místicos entre 2016 e 2024. Em 2016, termos como “vida” e “mundo” são predominantes, sugerindo uma busca por significado e compreensão do universo, muitas vezes através de uma lente esotérica. Ao longo dos anos, palavras como “luz” e “energia” ganham importância, refletindo uma crescente ênfase em conceitos de iluminação e força vital. Em 2020, durante o pico da Pandemia, “Deus” e “verdade” aparecem com destaque, talvez indicando uma busca por respostas espirituais em meio ao caos global. Em 2021 e 2022, há uma diversificação dos temas, com o surgimento de termos como “consciência” e “terra”, sugerindo uma ampliação das discussões para incluir questões de consciência planetária e conexão com a natureza. Em 2023 e 2024, “amor” e “vida” continuam a ser centrais, apontando para uma narrativa que enfatiza a importância da transformação espiritual e do autoconhecimento como respostas às crises externas. A recorrência de termos ao longo do tempo sugere que, enquanto o contexto muda, as preocupações fundamentais dessas comunidades permanecem focadas na busca por entendimento e iluminação espiritual.

### 3.4. Sobreposição de agenda temática

As figuras a seguir exploram como comunidades de teorias da conspiração relacionadas ao Apocalipse e ao Sobrevivencialismo se conectam com outras narrativas, formando um discurso coeso que reforça as crenças dessas comunidades. Através da análise dos tópicos abordados, é possível observar a centralidade desses temas na disseminação de desinformação, onde crenças esotéricas, teorias de globalismo e anti-vacinação são integradas em uma narrativa de preparação para um futuro apocalíptico. Esta análise evidencia como a sobreposição de temas amplifica o alcance dessas comunidades, dificultando intervenções baseadas em evidências científicas.

**Figura 10.** Temáticas de energia e iluminação esotérica

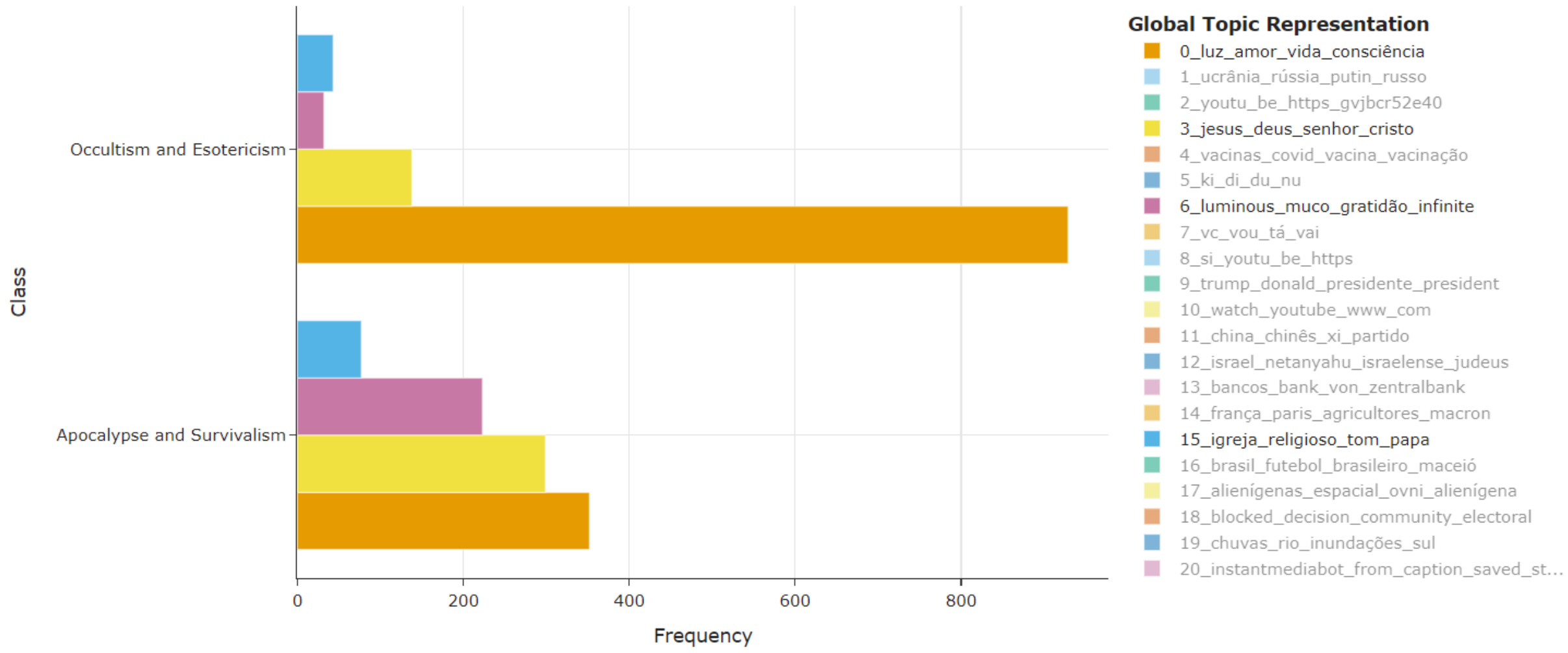

Fonte: Elaboração própria (2024).

A Figura 10 destaca a predominância de discussões sobre “energia” e “iluminação” dentro do contexto esotérico, em contraste com as discussões mais abruptas sobre apocalipse e sobrevivencialismo que vemos nas comunidades. Tópicos como “luz”, “amor” e “vida” são amplamente discutidos, sugerindo que as comunidades esotéricas integram esses conceitos como parte de uma preparação espiritual para eventos catastróficos. Essa sobreposição reflete uma busca por significados místicos e uma tentativa de encontrar orientação espiritual em meio à antecipação de crises apocalípticas.

**Figura 11.** Temáticas de globalismo e disputas geopolíticas

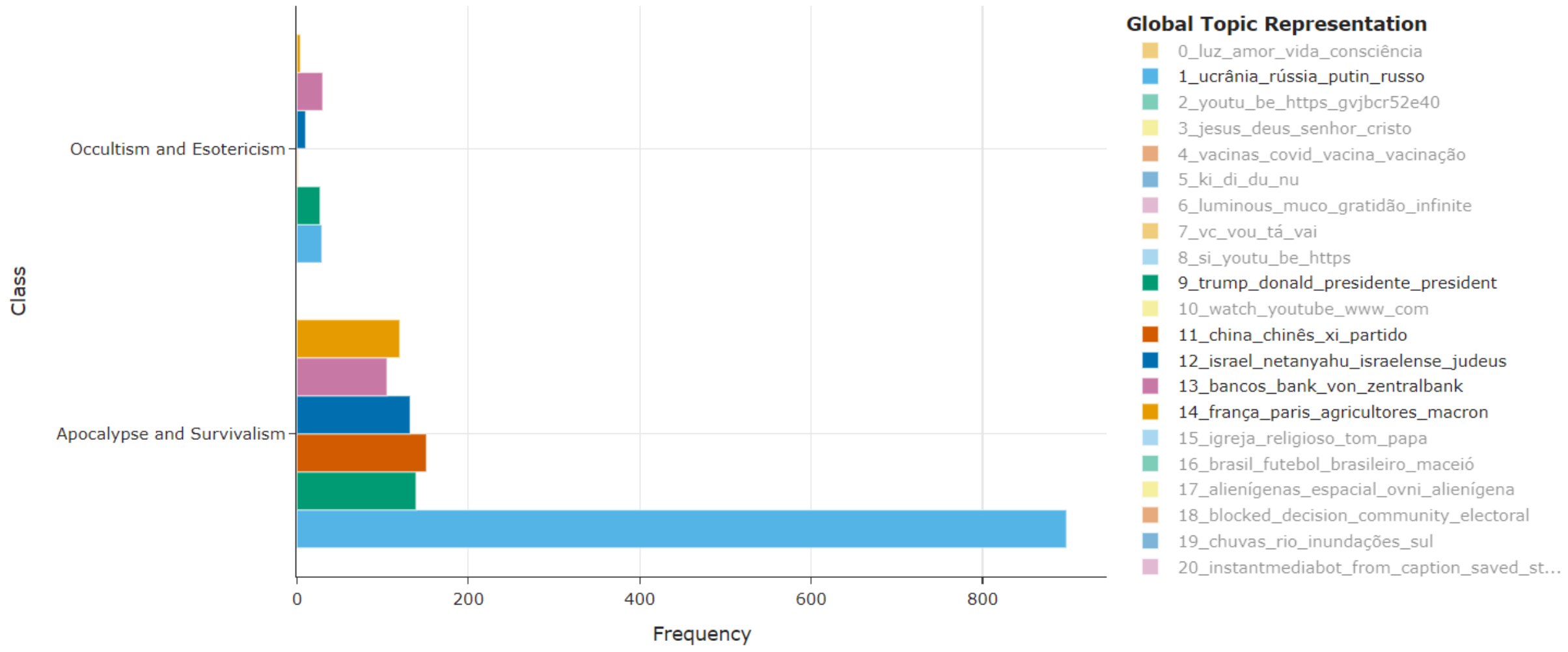

Fonte: Elaboração própria (2024).

Na Figura 11, as discussões de apocalipse e sobrevivencialismo dominam, com foco em temas de globalismo e disputas geopolíticas. Tópicos como “Rússia”, “Ucrânia” e “Putin” são amplamente abordados, sugerindo que essas comunidades utilizam eventos geopolíticos recentes para reforçar suas narrativas de um controle global iminente. A sobreposição com temas de sobrevivencialismo indica que essas teorias são usadas para justificar a preparação para um colapso global, onde as disputas internacionais são vistas como precursores de um apocalipse. Disso, nascem preocupações que reverberam em alertas cotidianos emitidos por comunidades que agregam, inclusive, cidadãos preocupados com a sua integridade.

**Figura 12.** Temáticas de negacionismo anti-vacinas, alienígenas e QAnon

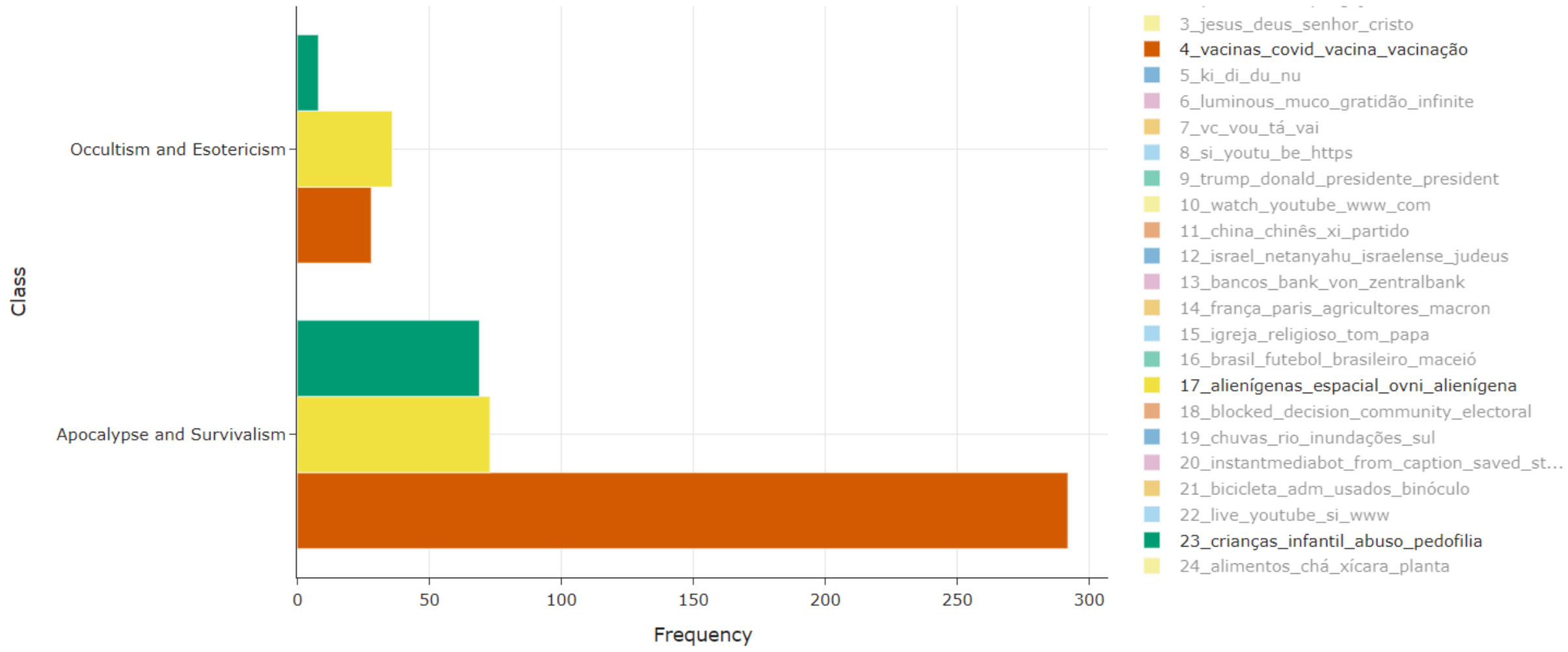

Fonte: Elaboração própria (2024).

A Figura 12 mostra uma forte interconexão entre o negacionismo anti-vacinas, teorias sobre alienígenas e QAnon. Tópicos como “vacinas”, “covid” e “alienígenas” sugerem que essas comunidades veem a vacinação e a Pandemia como partes de uma conspiração maior, possivelmente envolvendo seres extraterrestres. Essa narrativa não apenas reforça a desconfiança em relação às vacinas, mas também amplia a ideia de que forças ocultas ou governamentais estão manipulando eventos globais para controlar a população, justificando a preparação para um apocalipse. Além disso, é possível encontrar a monetização com venda de cursos de magnetismo quântico, por exemplo, onde a fé é instrumentalizada para gerar faturamento para quem se aproveita da cooptação.

**Figura 13.** Temáticas de polarização e política nacional

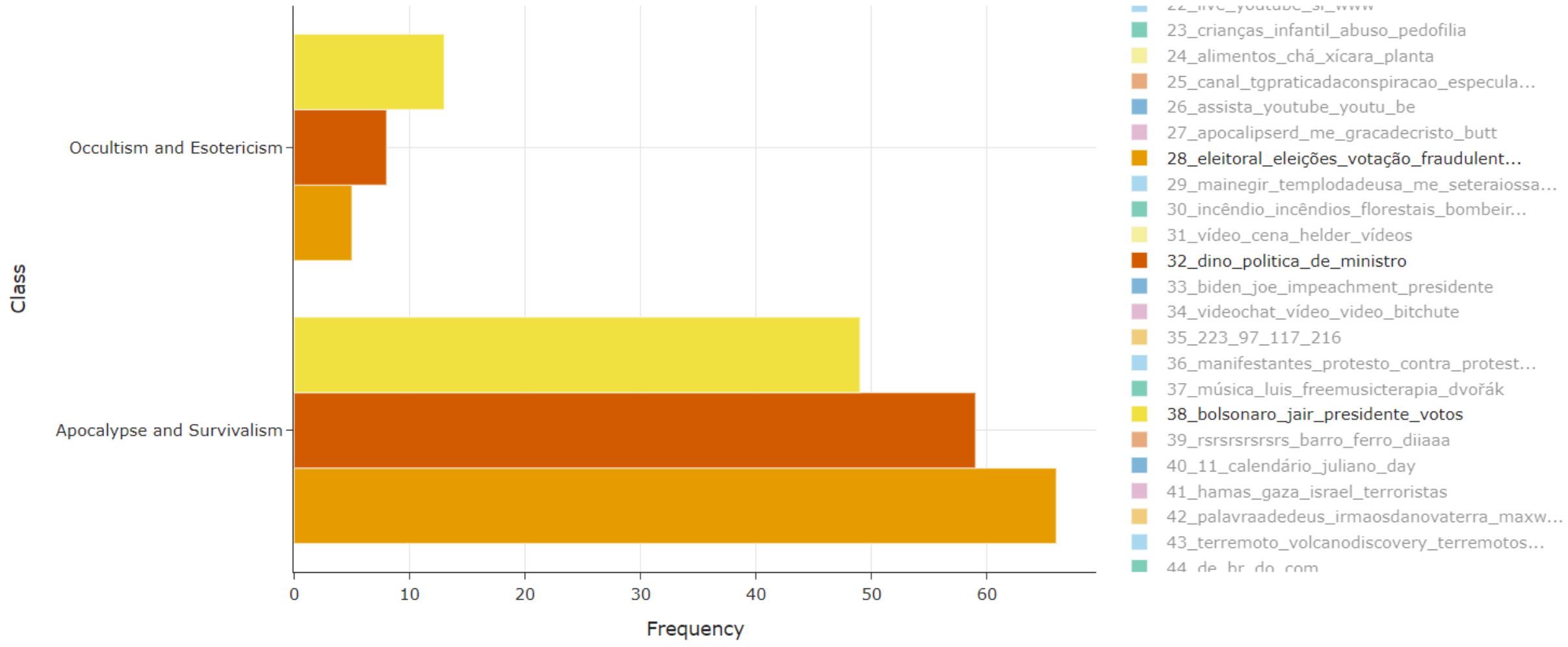

Fonte: Elaboração própria (2024).

Na Figura 13, as temáticas de sobrevivencialismo se sobrepõem a discussões sobre polarização e política nacional. Tópicos como “Bolsonaro”, “impeachment” e “fraude eleitoral” indicam que essas comunidades percebem a instabilidade política como um sinal de crises maiores que justificam a preparação para um colapso iminente. A associação entre sobrevivencialismo e política nacional reflete uma visão de que eventos políticos são indicadores de um futuro apocalíptico, reforçando a necessidade de estar preparado para sobreviver a crises políticas e sociais.

**Figura 14.** Temáticas de catástrofes supostamente planejadas

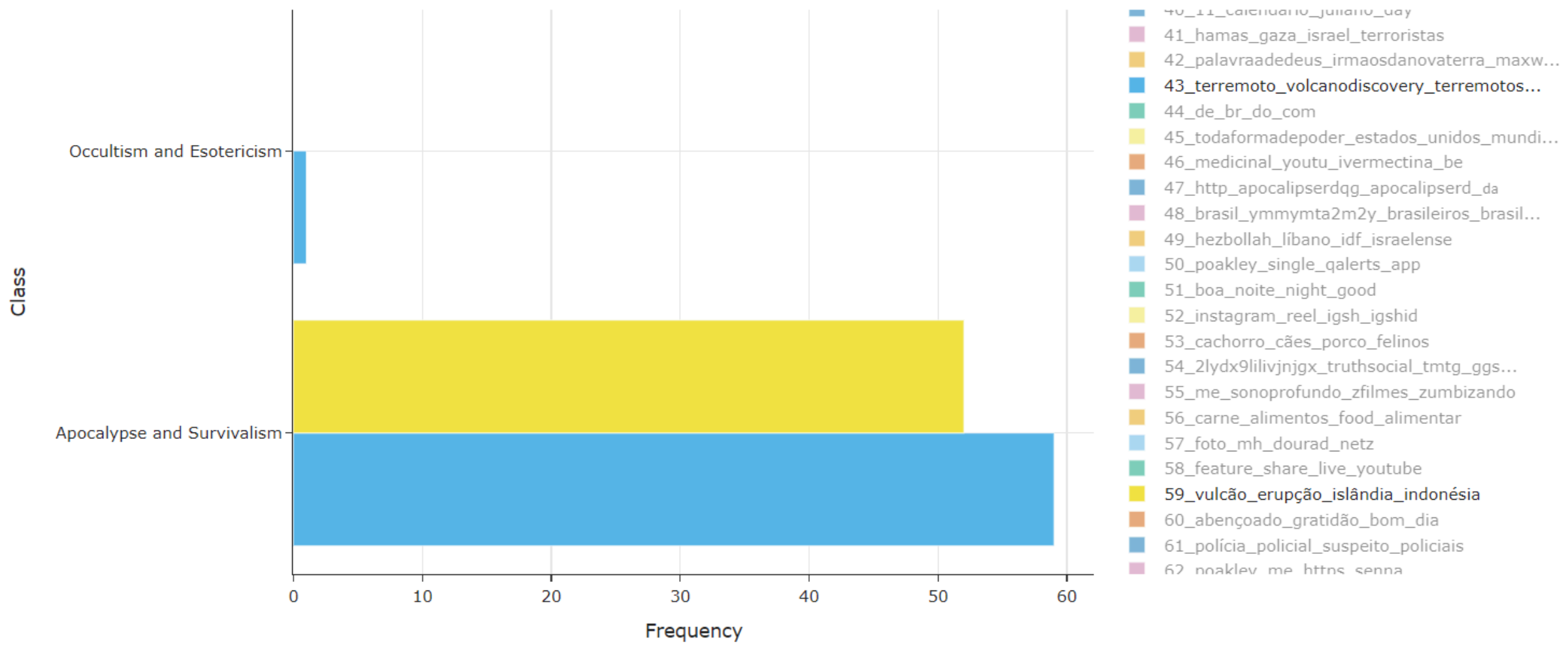

Fonte: Elaboração própria (2024).

A Figura 14 revela como as discussões de sobrevivencialismo se entrelaçam com a crença em catástrofes supostamente planejadas. Tópicos como "vulcões", "terremotos" e "tsunamis" são destacados, sugerindo que essas comunidades acreditam que desastres naturais são orquestrados como parte de uma agenda global. A sobreposição com sobrevivencialismo indica que essas crenças são usadas para justificar a preparação intensiva para sobreviver a essas catástrofes, que são vistas não como eventos naturais, mas como ações deliberadas para reduzir ou controlar a população mundial.

**Figura 15.** Temáticas de anti-*Woke* com narrativa se suposta sexualização LGBT

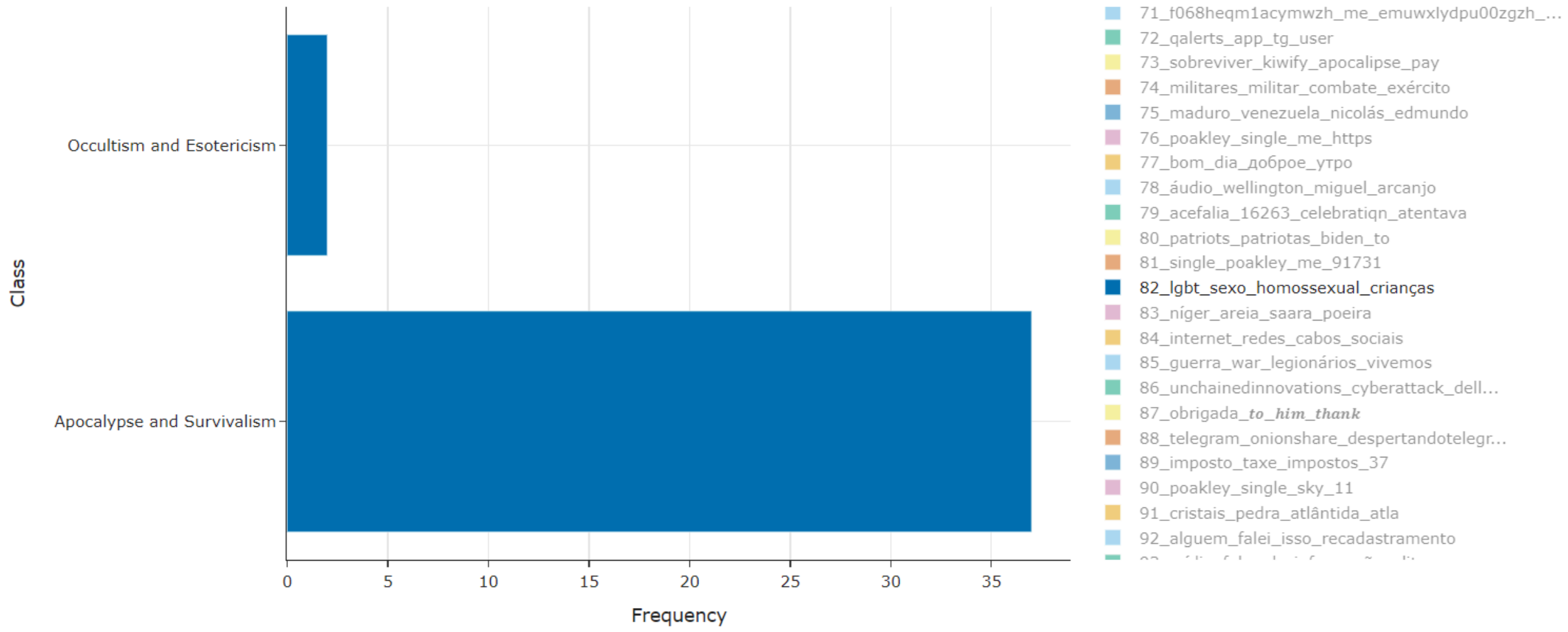

Fonte: Elaboração própria (2024).

A Figura 15 destaca a centralidade das temáticas de anti-*Woke* nas discussões de sobrevivencialismo, com foco em narrativas de suposta sexualização LGBT. Tópicos como "LGBT" e "sexualização" são abordados como parte de uma crítica inclusive que insinua de forma infundada essa relação de LGBTs com abusos de menores de idade. A sobreposição com sobrevivencialismo sugere que essas comunidades veem essas mudanças como parte de um declínio moral que justifica a preparação para um apocalipse cultural, onde a sobrevivência depende da resistência às influências percebidas como corrosivas.

## 4. Reflexões e trabalhos futuros

Para responder a pergunta de pesquisa "**como são caracterizadas e articuladas as comunidades de teorias da conspiração brasileiras sobre temáticas de apocalipse, sobrevivencialismo, ocultismo e esoterismo no Telegram?**", este estudo adotou técnicas espelhadas em uma série de sete publicações que buscam caracterizar e descrever o fenômeno das teorias da conspiração no Telegram, adotando o Brasil como estudo de caso. Após meses de investigação, foi possível extrair um total de 72 comunidades de teorias da conspiração brasileiras no Telegram sobre temáticas de apocalipse, sobrevivencialismo, ocultismo e esoterismo, estas somando 4.371.125 de conteúdos publicados entre maio de 2016 (primeiras publicações) até agosto de 2024 (realização deste estudo), com 192.138 usuários somados dentre as comunidades.

Foram adotadas quatro abordagens principais: **(i)** Rede, que envolveu a criação de um algoritmo para mapear as conexões entre as comunidades por meio de convites circulados entre grupos e canais; **(ii)** Séries temporais, que utilizou bibliotecas como "Pandas" (McKinney, 2010) e "Plotly" (Plotly Technologies Inc., 2015) para analisar a evolução das publicações e engajamentos ao longo do tempo; **(iii)** Análise de conteúdo, sendo aplicadas técnicas de análise textual para identificar padrões e frequências de palavras nas comunidades ao longo dos semestres; e **(iv)** Sobreposição de agenda temática, que utilizou o modelo BERTopic (Grootendorst, 2020) para agrupar e interpretar grandes volumes de textos, gerando tópicos coerentes a partir das publicações analisadas. A seguir, as principais reflexões são detalhadas, sendo seguidas por sugestões para trabalhos futuros.

### 4.1. Principais reflexões

**Comunidades de ocultismo e esoterismo funcionam como portas de entrada para teorias de sobrevivência extrema e apocalipse, conectando-se a outros temas conspiratórios:** As comunidades de ocultismo e esoterismo destacam-se como um ponto inicial de acesso para narrativas de sobrevivencialismo e apocalipse, registrando 2.527.833 interações. A interconectividade dessas temáticas evidencia como a busca por respostas místicas alimenta preocupações sobre cenários catastróficos, servindo de ponte para outras teorias conspiratórias como a Nova Ordem Mundial (NOM);

**Conspirações sobre a Nova Ordem Mundial são amplificadas por discussões sobre apocalipse, com significativo impacto na desinformação sobre saúde:** Com 3.488.686 interações, as comunidades centradas na NOM têm forte correlação com temas apocalípticos, reforçando a ideia de controle global. Essa narrativa é ampliada durante crises como a Pandemia da COVID-19, associando teorias sobre apocalipse a desconfianças em vacinas e saúde pública, com mais de 1.345.060 interações vinculadas a essas preocupações;

**Narrativas de sobrevivencialismo experimentam um crescimento de 1.500% durante a Pandemia, evidenciando sua conexão com outras teorias conspiratórias:** O crescimento exponencial das menções a sobrevivencialismo durante a Pandemia, de 1.000 para 35.000 entre 2020 e 2021, reflete o impacto das crises globais na proliferação de teorias apocalípticas. Esses grupos se tornam catalisadores para a aceitação de uma visão de mundo onde o colapso social é iminente;

**Ocultismo e esoterismo são as maiores fontes de convites para comunidades de medicamentos *off label*, mostrando uma intersecção perigosa entre misticismo e desinformação científica:** Ocultismo e esoterismo, com 7.367 links direcionados a comunidades de medicamentos *off label*, reforçam a disseminação de práticas alternativas e perigosas. Essa interconexão sublinha a aliança entre crenças esotéricas e a rejeição da ciência convencional, criando um ciclo de desinformação resistente. Além disso, é possível encontrar a monetização com venda de cursos de magnetismo quântico, por exemplo, onde a fé é instrumentalizada para gerar faturamento para quem se aproveita da cooptação;

**Comunidades de apocalipse e sobrevivência funcionam como *hubs* centrais, conectando múltiplas narrativas conspiratórias e fortalecendo ciclos de desinformação:** As comunidades de apocalipse e sobrevivencialismo, com mais de 4.371.125 interações, operam como *hubs* centrais dentro da rede de desinformação. Essas comunidades não apenas conectam diferentes teorias conspiratórias, mas também as amplificam, criando um ciclo contínuo de reforço mútuo;

**A sobreposição de temáticas esotéricas e de sobrevivência revela uma bolha ideológica coesa, resistente à informação científica:** A fusão entre temas esotéricos e narrativas de sobrevivência cria uma bolha ideológica difícil de penetrar. Com 3.026.065 interações ligadas a ocultismo e esoterismo, esses grupos perpetuam a desconfiança em relação à ciência, promovendo uma visão de mundo alternativa que se retroalimenta;

**A narrativa falsa que associa vacinas ao controle global persiste como uma das mais discutidas em comunidades apocalípticas e esotéricas:** Mesmo após repetidos desmentidos, a ideia de que vacinas fazem parte de um plano de controle global continua a ser amplamente discutida, especialmente em comunidades associadas a teorias da NOM e sobrevivencialismo. Essa narrativa serve para reforçar a desconfiança em relação à medicina;

**A interconectividade entre comunidades de ocultismo, NOM, e apocalipse reflete um padrão de reforço mútuo que amplifica a desinformação:** A análise das conexões entre essas comunidades revela como teorias aparentemente distintas se reforçam mutuamente. Com mais de 5.421 links de convites entre elas, essa interconectividade cria um ambiente onde diferentes formas de desinformação se tornam interdependentes;

**Discussões sobre apocalipse frequentemente servem como ponto de partida para a aceitação de outras teorias conspiratórias, ampliando o alcance dessas crenças:** As comunidades de apocalipse e sobrevivencialismo atuam como portas de entrada para teorias

sobre NOM, globalismo e outras conspirações. Essa expansão temática ajuda a disseminar essas crenças para um público mais amplo, dificultando a desmobilização dessas narrativas;

**A Pandemia da COVID-19 catalisou o crescimento de comunidades esotéricas e de sobrevivência, refletindo uma reação às incertezas globais:** O aumento das interações em comunidades esotéricas e de sobrevivência durante a Pandemia, com picos de 1.200% para ocultismo e esoterismo, sugere que as crises globais incentivam a busca por explicações alternativas, muitas vezes ligadas a teorias conspiratórias.

### 4.2. Trabalhos futuros

A partir dos principais achados deste estudo, várias direções podem ser sugeridas para futuras pesquisas. Primeiramente, é essencial explorar em mais profundidade como as temáticas de apocalipse e sobrevivência se conectam com outras narrativas conspiratórias no contexto brasileiro, especialmente considerando o crescimento significativo dessas discussões durante a Pandemia. Investigações futuras poderiam examinar como essas comunidades utilizam eventos globais, como crises de saúde e desastres naturais, para reforçar suas crenças e atrair novos membros. Além disso, o impacto dessas narrativas em populações específicas, como jovens e pessoas de áreas rurais, merece atenção, considerando a vulnerabilidade dessas demografias à desinformação.

Outro ponto relevante é a necessidade de mapear as interações entre comunidades de ocultismo e esoterismo com aquelas que promovem medicamentos *off label* e práticas de saúde alternativas. Estudar a psicologia por trás da atração por essas crenças esotéricas pode oferecer insights sobre como essas práticas ganham tração dentro de bolhas ideológicas. Futuros estudos poderiam se concentrar em desenvolver e testar intervenções que visem desmantelar essas crenças, especialmente em contextos onde a ciência convencional é fortemente rejeitada. Também é importante investigar a persistência de narrativas apocalípticas e a maneira como elas são reintroduzidas em momentos de crise. A pesquisa pode focar em identificar os mecanismos que permitem que essas narrativas continuem a proliferar, mesmo após serem amplamente desmentidas. Entender a resiliência dessas crenças pode ser crucial para desenvolver estratégias mais eficazes de combate à desinformação.

Além disso, a interconectividade entre as comunidades que discutem sobrevivência e ocultismo com outras teorias conspiratórias, como a Nova Ordem Mundial e o globalismo, deve ser mais bem compreendida. Estudos futuros poderiam explorar como essas redes se formam, se mantêm e interagem, oferecendo uma visão detalhada das dinâmicas internas dessas comunidades. Compreender essas interações pode ajudar a identificar pontos de intervenção eficazes para reduzir a disseminação de desinformação. Por fim, o desenvolvimento de ferramentas que permitam monitorar e identificar em tempo real os *hubs* de desinformação que emergem dentro dessas comunidades é uma área promissora para pesquisa. Em tempos de crises globais, como Pandemias ou desastres naturais, essas ferramentas poderiam ser utilizadas para identificar rapidamente as narrativas emergentes e implementar intervenções antes que elas ganhem tração e causem danos significativos.

## 5. Referências

Grootendorst, M. (2020). ***BERTopic:*** *Leveraging BERT and c-TF-IDF to create easily interpretable topics*. GitHub. https://github.com/MaartenGr/BERTopic

Loper, E., & Bird, S. (2002). ***NLTK:*** *The Natural Language Toolkit*. In *Proceedings of the ACL-02 Workshop on Effective Tools and Methodologies for Teaching Natural Language Processing and Computational Linguistics* - Volume 1 (pp. 63–70). Association for Computational Linguistics. https://doi.org/10.3115/1118108.1118117

McKinney, W. (2010). **Data structures for statistical computing in Python.** In *Proceedings of the 9th Python in Science Conference* (pp. 51–56). https://doi.org/10.25080/Majora-92bf1922-00a

Mueller, A. (2020). ***WordCloud.*** GitHub. https://github.com/amueller/word_cloud

Plotly Technologies Inc. (2015). ***Collaborative data science.*** Plotly Technologies Inc. https://plotly.com

Rocha, I., Silva, E. C. M., & Mielli, R. V. (2024). ***Catalytic conspiracism:*** *Exploring persistent homologies time series in the dissemination of disinformation in conspiracy theory communities on Telegram*. 14º Encontro da Associação Brasileira de Ciência Política. UFBA, Salvador, BA. https://www.abcp2024.sinteseeventos.com.br/trabalho/view?ID_TRABALHO=687

Silva, E. C. M. (2023, fevereiro). ***Web scraping Telegram posts and content.*** GitHub. https://github.com/ergoncugler/web-scraping-telegram/

Silva, E. C. M., & Sátiro, R. M. (2024). ***Conspiratorial convergence:*** *Comparing thematic agendas among conspiracy theory communities on Telegram using topic modeling*. 14º Encontro da Associação Brasileira de Ciência Política. UFBA, Salvador, BA. https://www.abcp2024.sinteseeventos.com.br/trabalho/view?ID_TRABALHO=903

## 6. Biografia do autor

**Ergon Cugler de Moraes Silva** possui mestrado em Administração Pública e Governo (FGV), MBA pós-graduação em Ciência de Dados e Análise (USP) e bacharelado em Gestão de Políticas Públicas (USP). Ele está associado ao Núcleo de Estudos da Burocracia (NEB FGV), colabora com o Observatório Interdisciplinar de Políticas Públicas (OIPP USP), com o Grupo de Estudos em Tecnologia e Inovações na Gestão Pública (GETIP USP), com o Monitor de Debate Político no Meio Digital (Monitor USP) e com o Grupo de Trabalho sobre Estratégia, Dados e Soberania do Grupo de Estudo e Pesquisa sobre Segurança Internacional do Instituto de Relações Internacionais da Universidade de Brasília (GEPSI UnB). É também pesquisador no Instituto Brasileiro de Informação em Ciência e Tecnologia (IBICT), onde trabalha para o Governo Federal em estratégias contra a desinformação. Brasília, Distrito Federal, Brasil. Site: https://ergoncugler.com/.